\documentclass[smallcondensed]{svjour3}

\usepackage{epsfig}
\usepackage{amsmath}
\usepackage{amssymb}
\usepackage{natbib}
\usepackage{threeparttable} 
\usepackage{xspace}
\usepackage{color}

\usepackage{epstopdf}

\newcommand{\astrosat}{\textit{ASTROSAT}\xspace}
\newcommand{\athena}{\textit{Athena}\xspace}
\newcommand{\beppo}{\textit{BeppoSAX}\xspace}
\newcommand{\chan}{\textit{Chandra}\xspace}

\newcommand{\extp}{\textit{eXTP}\xspace}
\newcommand{\ginga}{\textit{Ginga}\xspace}
\newcommand{\hitomi}{\textit{Hitomi}\xspace}
\newcommand{\hxmt}{\textit{Insight-HXMT}\xspace}
\newcommand{\inte}{\textit{INTEGRAL}\xspace}
\newcommand{\loft}{\textit{LOFT}\xspace}
\newcommand{\maxi}{\textit{MAXI}\xspace}
\newcommand{\nicer}{\textit{NICER}\xspace}
\newcommand{\nustar}{\textit{NuSTAR}\xspace}

\newcommand{\rxte}{\textit{RXTE}\xspace}

\newcommand{\swift}{\textit{Swift}\xspace}

\newcommand{\xmm}{\textit{XMM-Newton}\xspace}

\newcommand{\Msun}{\mathrm{M}_{\odot}}

\newcommand{\lum}{\mathrm{erg~s}^{-1}}

\newcommand{\flux}{\mathrm{erg~cm}^{-2}~\mathrm{s}^{-1}}

\newcommand{\gmc}{GM/c^2}

\newcommand{\cms}{\ensuremath{\mathrm{cm}^{2}}}

\def \mnras {MNRAS}
\def \apj {ApJ}
\def \apjs {ApJS}
\def \apss {Ap\&SS}
\def \apjl {ApJL}
\def \aap {A\&A}
\def \nat {Nature}
\def \nar {NAR}
\def \araa {ARAA}
\def \atel {ATel}
\def \pasp {PASP}
\def \aaps {AAPS}

\def \pasj {PASJ}
\def \ssr {Space Sci. Rev.}
\def \procspie {Proc. SPIE}
\def \physrep {Phys. Rev.}
\def \aapr {A\&A Rev.}

\journalname{Space Science Reviews}

\begin{document}

\title{Accretion disks and coronae in the X-ray flashlight}

\author{Nathalie Degenaar \and 
        David~R. Ballantyne \and
        Tomaso Belloni \and
        Manoneeta Chakraborty \and
        Yu-Peng Chen \and
        Long Ji \and
        Peter Kretschmar \and
        Erik Kuulkers \and
        Jian Li \and
        Thomas~J. Maccarone \and
        Julien Malzac \and
        Shu Zhang \and 
        Shuang-Nan Zhang
}
  
\authorrunning{N. Degenaar et al.}

\institute{%
N.~Degenaar \at
Institute of Astronomy, University of Cambridge, Madingley Road, Cambridge CB3 OHA, UK\\
Anton Pannekoek Institute, University of Amsterdam, Science Park 904, 1098 XH, Amsterdam, the Netherlands
\email{N.D.Degenaar@uva.nl}
\and 
D.R. Ballantyne \at
Center for Relativistic Astrophysics, School of Physics, Georgia Institute of Technology, Atlanta, GA 30332, USA
\and 
T. Belloni \at
Istituto Nazionale di Astrofisica, Osservatorio Astronomico di Brera, Via E. Bianchi 46, I-23807 Merate, Italy
\and 
M.~Chakraborty \at
Sabanc\i\ University, Faculty of Engineering and Natural Sciences, Orhanli Tuzla 34956, Istanbul, Turkey
\and 
Y.P. Chen \and S. Zhang \and S.N. Zhang \at
Laboratory for Particle Astrophysics, Institute of High Energy Physics, Beijing 100049, China
\and 
L. Ji \at
Institut f\"ur Astronomie und Astrophysik, Kepler Center for Astro and Particle Physics, Eberhard Karls Universit\"at T\"ubingen, Sand 1, D-72076, T\"ubingen, Germany
\and 
P. Kretschmar \at
European Space Astronomy Centre (ESA/ESAC), Science Operations Department, Villanueva de la CanÃÂada (Madrid), Spain
\and 
E. Kuulkers \at
European Space Research and Technology Centre (ESA/ESTEC), Keplerlaan 1, 2201 AZ Noordwijk, The Netherlands
\and 
J. Li \at
Institute of Space Sciences (IEECCSIC), Campus UAB, Carrer de Magrans s/n, E-08193 Barcelona, Spain
\and 
T.J.~Maccarone \at
Department of Physics and Astronomy, Texas Tech University, Box 41051, Lubbock, TX 79409-1051, USA
\and 
J. Malzac \at
Universit\'e de Toulouse, UPS-OMP, IRAP, 31000, Toulouse, France \\
CNRS, IRAP, 9 Av. colonel Roche, BP 44346, 31028, Toulouse Cedex 4, France
}

\date{Received: date / Accepted: date}

\maketitle

\begin{abstract} 
Plasma accreted onto the surface of a neutron star can ignite due to unstable thermonuclear burning and produce a bright flash of X-ray emission called a Type-I X-ray burst. Such events are very common; thousands have been observed to date from over a hundred accreting neutron stars. The intense, often Eddington-limited, radiation generated in these thermonuclear explosions can have a discernible effect on the surrounding accretion flow that consists of an accretion disk and a hot electron corona. Type-I X-ray bursts can therefore serve as direct, repeating probes of the internal dynamics of the accretion process. In this work we review and interpret the observational evidence for the impact that Type-I X-ray bursts have on accretion disks and coronae. We also provide an outlook of how to make further progress in this research field with prospective experiments and analysis techniques, and by exploiting the technical capabilities of the new and concept X-ray missions \astrosat, \nicer, \hxmt, \extp, and {\it STROBE-X}.

\end{abstract}


\section{Introduction}\label{sec:intro}
Astronomical bodies can gravitationally attract material from their
surroundings. This process, known as {\it accretion}, plays a fundamental role at all scales encountered in the universe and therefore forms a central theme in astrophysics. Low-mass X-ray binaries (LMXBs) are excellent laboratories to study accretion processes. In these systems a stellar remnant, either a black hole (BH) or a neutron star (NS), accretes the outer gaseous layers of a companion star that has a lower mass than the compact primary. Apart from the compact primary and the donor star, the main components involved in the accretion process in LMXBs are an accretion disk (Section~\ref{subsec:disk}) and a hot electron corona (Section~\ref{subsec:introcorona}). Moreover, accretion is universally linked to collimated outflows called jets (Section~\ref{subsec:jet}), which are likely also connected to coronae.

Although the theory of high-energy emission from LMXBs is generally well developed, the nature of the X-ray emitting corona remains poorly understood. The concept of a corona, a hot gas flow producing non-thermal radiation, has been widely used in connection to modeling spectral state transitions in LMXBs and the interplay between disks and jets (see Section~\ref{sec:accr}). The intrinsic properties of these coronae are difficult to obtain and quite different views are proposed in the literature. Interestingly, {\it Type-I X-ray bursts} (shortly X-ray bursts hereafter; Section~\ref{sec:bursts})\footnote{There are two Galactic NS LMXBs that exhibit so-called Type-II X-ray bursts; the ``rapid burster'' MXB 1730-335 and the ``bursting pulsar'' GRO J1744-28. Opposed to Type-I X-ray bursts, the Type-II events are thought to result from accretion instabilities rather than thermonuclear burning \citep[see e.g.][for a recent extensive discussion of the Type-II X-ray burst phenomena]{bagnoli2015}.} ignited on the surface of accreting NSs can offer a dynamic probe of the accretion flow in LMXBs, including coronae.

It has long been appreciated that there is an intricate connection between the properties of the accretion flow (e.g. the mass-accretion rate, accretion geometry, and composition of the accreted material), and the observable properties of X-ray bursts (e.g. their duration, repetition time, peak flux, and spectral evolution). Following early studies \citep[e.g.][]{fujimoto81,fujimoto1987,lapidus1985,vanparadijs1988}, there have been excellent reviews on this topic \citep[e.g.][]{lewin1993,bildsten1998,strohmayer2006,galloway2008}, and many new results have been obtained in recent years \citep[e.g.][]{cavecchi2011,zand2012,watts2012,chakraborty2014,kajava2014,chenevez2016,mahmoodifar2016,ootes2016}. 

The reverse side of the coin gained much attention in the past few years. Indeed, the intense radiation of an X-ray burst can have a discernible effect on the surrounding accretion flow; on both coronae and disks.\footnote{In this review we use the general term ``accretion flow'' to refer both to the accretion disk and the corona.} To investigate the properties of the corona it is useful to have a probe characterized by intense soft X-ray emission released in a short time interval, exactly the properties delivered by X-ray bursts (Section~\ref{sec:bursts}). The rapid increase of soft photons caused by an X-ray burst can interact with the corona via inverse Compton scattering, which can be expected to lead to cooling of the hot plasma. This mechanism is not unlike the coronal evolution inferred from transitions between different spectral states (see Section~\ref{subsec:disk}). Furthermore, the radiation pressure force caused by the thermonuclear detonations may blow away the corona and lead to a coronal ejection (see Section~\ref{sec:theory}). X-ray bursts occur frequently, every few hours for several NSs, and their X-ray emission can generally be well described by a black body with a color temperature of tens of millions of Kelvin (i.e $kT_{\mathrm{bb}}$ of a few keV). Since the corona dominates the emission above $\approx$40~keV, it can be disentangled from the X-ray burst spectrum. These explosions thus represent a repeating experiment to investigate corona cooling and recovery in response to an intense shower of soft X-ray photons.

Since jets appear to be connected to coronae, these collimated outflows may not escape the impact of X-ray bursts either. Jets are likely fed by large scale height magnetic fields (see Section~\ref{subsec:jet}) and the effects of the influx of soft photons into the corona may have important implications for jet production. If the corona is blown apart, then the region containing the large scale height magnetic fields which feed the jet is removed with it. Even if the corona is merely cooled off, if it collapses dynamically the jet power may be affected. Understanding how the jet flux responds to an X-ray burst may thus help us to understand how jets are powered.  

Apart from affecting coronae, the impulsive luminosity changes caused by X-ray bursts are also expected to influence the accretion disk, causing it to be heated, puffed up, blown away, or dragged inward (see Section~\ref{sec:theory}). Illumination of the accretion disk by an X-ray burst can cause reflection features; these can be a powerful tool to constrain the physical structure of the inner accretion disk and how it is affected by the violent explosion. Furthermore, quasi-periodic oscillations (QPOs) with kHz frequencies are considered to probe the innermost disk regions, and are observed to change in response to X-ray bursts. Like coronae, accretion disks are also connected to outflows, albeit in the form of (non-collimated) winds. These winds are launched relatively far from the compact object and may not be directly influenced by X-ray bursts. Nevertheless, a thermonuclear explosion may cause a (radiatively-driven) wind-like outflow from the inner accretion disk or cause the ejection of a nova-like shell of material from the NS (see Section~\ref{subsec:PRE}). All this indicates that studying X-ray burst-induced changes to an accretion disk is a promising, direct way to explore the internal dynamics of the accretion process. 

The purpose of the present work is not to review how the observational properties of X-ray bursts are connected to that of the accretion flow, but rather to scrutinize the reverse interaction: how X-ray bursts impact accretion disks and coronae. In particular, we will review to what extent the disruption and subsequent restoration due to a thermonuclear explosion can increase our knowledge about the geometry and dynamics of the accretion flow. After introducing the phenomenology of X-ray bursts (Section~\ref{sec:bursts}), accretion disks, coronae and jets (Section~\ref{sec:accr}), we give an overview of the observational evidence for X-ray bursts interacting with the accretion flow (Section~\ref{sec:obs}). We then lay out the current theoretical framework for interpreting these results (Section~\ref{sec:theory}). Having reviewed the current status of this research field, we next discuss the prospects of using X-ray bursts as a probe of the accretion flow in future studies, in particular by employing new and concept X-ray missions as well as multi-wavelength efforts (Section~\ref{sec:future}). We conclude with a summary in Section~\ref{sec:summary}.


\section{Type-I X-ray bursts}\label{sec:bursts}
X-ray bursts \citep[][]{belian1976,grindlay1976_bursts,hoffman1978} result from thermonuclear flashes on an accreting NS, which are caused by the unstable ignition of a layer of He and/or H-rich material supplied by the low-mass companion star (\citet{Hansen_vanHorn:1975,WoosleyTaam:1976,MaraschiCavaliere:1977,LambLamb:1978}; see \citet{lewin1993,strohmayer2006} for reviews). The first X-ray burst was, in hindsight, detected in 1969 by Vela 5b, from the nearby transient NS LMXB Cen X-4 \citep[][]{belian1972}. 
After nearly 5 decades of X-ray burst detections, this very first one still stands as the brightest ever detected, reaching an estimated bolometric flux of $\approx 2 \times 10^{-6}~\flux$ (see \cite{kuulkers2009} for a re-analysis of this event). To date, many thousands of X-ray bursts have been detected \citep[e.g.][for compilations using different instruments]{vanparadijs1988,cornelisse2003,sakamoto2003,intzand2004,chenevez2008,galloway2008,jenke2016}\footnote{See the Multi-INstrument Burst ARchive at http://burst.sci.monash.edu/minbar \citep[MINBAR;][]{galloway2010_minbar}.} from over a hundred NS LMXBs.\footnote{See https://personal.sron.nl/\textasciitilde jeanz/bursterlist.html (maintained by Jean in 't Zand).}

X-ray bursts generally appear as short transient events that have a characteristic light curve shape and spectral evolution. The X-ray intensity rises rapidly on a time scale of seconds, and decays back to the pre-burst level on a longer timescale \citep[e.g.][]{lewin1993}. Traditionally, X-ray burst decays have been described with an exponential function, but when high signal to noise ratio data are analyzed, a power-law decay is often a better description and sometimes complex shapes are observed \citep[e.g.][]{zand2014_cool,kuuttila2017}. The duration of X-ray bursts commonly ranges from several seconds up to several minutes, and the total radiated output is typically $\sim 10^{39}$~erg \citep[e.g.][]{galloway2008}, although longer and more energetic X-ray bursts are also observed (see Section~\ref{subsec:longbursts}). The time intervals between subsequent X-ray bursts from relatively bright X-ray sources are typically of the order of hours to days. This recurrence time is thought to be determined by the time required for the NS to accumulate enough fuel to power another X-ray burst \citep[e.g.][]{fujimoto81,ayasli1982}.\footnote{Rotation-induced mixing of chemical elements in the surface layers of the NS may also affect the recurrence time of X-ray bursts \citep[e.g.][]{piro2007,keek2009}.}

The X-ray burst spectra generally harden during the rise and soften during the decay. This has been attributed to the heating and cooling of the uppermost layers of the NS.\footnote{At high accretion rates, around the regime where thermonuclear burning may be stabilizing, there are not always clear signs of cooling during X-ray bursts \citep[e.g.][]{kuulkers2002,zand09,linares2010,linares2011}.} The spectra can generally be satisfactorily described by black-body emission from spherical regions with radii of $\approx$10~km at inferred temperatures up to $kT_{\mathrm{bb}} \approx 3$~keV \citep[e.g.][]{swank1977,hoffman1977}, although the underlying physics is complex and only captured by appropriate NS atmosphere models (e.g. \cite{london1986,lapidus1986}; see \citet{suleimanov2011,suleimanov2012,nattila2015,medin2016} for recent computations). In some cases rapid variability on millisecond time scales, so-called burst oscillations, are detected during X-ray bursts and these can be used to infer the spin period of the NS (\cite{strohmayer1996}; see \cite{strohmayer2006,watts2012} for reviews).

\subsection{Photospheric radius expansion (PRE)}\label{subsec:PRE}
Under certain conditions the local luminosity generated during an X-ray burst may reach or exceed the Eddington limit and hence matter may be pushed outward by radiation pressure. As a result, the NS photosphere expands and this causes the emitting area to increase while the observed black-body temperature drops. When the surge of energy release is over, the photosphere gradually returns to its original radius, causing the emitting area to decrease and the inferred temperature to increase again. During these expansion and contraction phases, the luminosity is expected to be close to the Eddington limit. After the photosphere recedes to its original radius (called ``touch-down'') cooling is typically observed. This phenomena is referred to as {\it photospheric radius-expansion}; PRE for short \citep[e.g.][]{Grindlay:1980,ebisuzaki1983,lewin1984,tawara1984_xb1715}. 

About 20\% of all X-ray bursts exhibit PRE and the typical amplitude of the radius expansion is a factor of $\approx 3$ \citep[][]{galloway2008}. On rare occasions, however, extreme-PRE X-ray bursts have been observed during which the emitting radius increases by a factor of $\sim 10^2$, i.e. out to $\sim 10^3$~km  \citep[e.g.][]{lewin1984,tawara1984,tawara1984_xb1715}. This phenomena occurs in only $\approx$1\% of X-ray bursts, has been termed ``superexpansion" and suggests the ejection of a nova-like shell of material \citep[e.g.][]{zand2010}. If the radiative flux is low enough, the atmosphere can achieve equilibrium by expanding (since the Eddington flux increases with increasing radius), but for X-ray bursts that (far) exceed the Eddington limit this equilibrium cannot be accomplished and gives rise to a radiative wind \citep[e.g.][]{hanawa1982,paczynski1986,fujimoto1987,woosley2004}. In the few cases where the expansion timescale could be measured, this was so short that the ejections must have been mildly relativistic \citep[$\approx 0.1-0.3 c$;][]{zand2014}.\footnote{The burst-driven winds are different from the disk winds that are typically seen for BH and NS LMXBs in their soft spectral states \citep[e.g.][]{miller2006_winds,miller2011,neilsen2009,ponti2012_winds,ponti2014_exo0748}; these soft-state winds appear to be thermally or magnetically driven, have typical outflow velocities of $\lesssim 0.1c$, and are launched at larger distances from the accreting compact object \citep[see e.g.][for a recent review]{diaztrigo2015}.}

\subsection{Long X-ray bursts}\label{subsec:longbursts}
There appears to be a separate class of so-called ``intermediate-duration bursts", which have a radiated energy output of $\sim 10^{40-41}$~erg and a typical duration of tens of minutes \citep[e.g.][]{swank1977,hoffman1978,zand05_ucxb,zand2011,molkov2005,chenevez2008,falanga08,kuulkers2010}. These events are thus significantly longer and more energetic than common (short) X-ray bursts. Many of the intermediate-duration bursts are detected from NSs that accrete at a low rate of $\approx$0.1--1\% of the Eddington limit, and are thought to result from the ignition of a particularly thick layer of He. 

A thick He layer may be build up if the companion is an evolved star and hence the accreted material is H-depleted \citep[e.g.][see also Section~\ref{subsubsec:classes}]{cumming06,zand07}. X-ray bursts with similar duration/energy output can, however, also occur when the accreted matter is H-rich \citep[e.g.][]{falanga09,degenaar2010_burst}: for instance, at low accretion rates H burns unstably but the energy release from these (weak and undetectable) H-flashes may not immediately trigger unstable He burning, allowing a thick layer of He to accumulate before it eventually ignites \citep[][]{cooper07,peng2007}. Regardless of the presence of H in the accreted matter, it takes a long time to build up a thick layer of He, particularly if the accretion rate is low. The recurrence time of the intermediate-duration bursts is therefore typically long; observational constraints suggests several weeks/months \citep[e.g.][]{zand07,linares09,degenaar2010_burst,degenaar2011_burst,keek2016}. A puzzling case is GX 17+2 (a ``Z-source''; see Section~\ref{subsubsec:classes}), which shows intermediate-duration bursts at very high mass-accretion rates \citep[][]{kuulkers2002}.  

About two dozen X-ray bursts have been detected to date that last for an extended period of hours to half a day and have an energy output of $\sim 10^{42}$~erg, i.e. about a thousand times higher than normal X-ray bursts and even 1--2 orders of magnitude higher than the intermediate-duration bursts \citep[e.g.][]{Cornelisse:2000,strohmayer2002}. These unusually long and energetic events have been dubbed ``superbursts'' \citep[][]{Wijnands:2001}. Their recurrence time is long, months to years, and superbursts are therefore rare. Indeed, only a few dozen of such events have been observed to date from about two dozen sources \citep[see][for a recent overview]{zand2017_superbursts}, compared to many thousands of short X-ray bursts. The origin of superbursts is different from the shorter X-ray bursts; rather than being due to the ignition of pure He or a H/He mixture, these are thought to be due to unstable burning of C,  deeper in the NS envelope \citep[][]{cumming2001_bursts,strohmayer2002}.

\section{Coronae, jets and disks}\label{sec:accr}


\subsection{Accretion disks and X-ray spectral states}\label{subsec:disk}
In LMXBs, the donor star is typically overflowing its Roche lobe and feeding matter into an accretion disk that surrounds the compact primary. Thermal-viscous instabilities within the accretion disk may cause LMXBs to be {\it transient}. In transient LMXBs, outburst episodes during which matter is rapidly accreted and hence the X-ray luminosity is high ($L_X\sim 10^{36-39}~\lum$), are alternated by quiescent episodes during which the accretion rate and the X-ray luminosity are reduced by several orders of magnitude \citep[e.g.][for an extensive review]{lasota01}. Out of $\approx$125 Galactic NS LMXBs currently known, about 43\% are persistently accreting, about 40\% are transient and the remaining $\approx$17\% fall more or less in between; these are formally transient but can persistently accrete for several years. Such sources are indicated as quasi-persistent transients \citep[][]{wijnands2004}.

Variations in the mass-accretion rate are accompanied by changes in the accretion geometry. This is borne out by the existence of distinct \textit{states} that have different X-ray spectral and timing properties. For BH LMXBs there are two main states, denoted as soft and hard according to the X-ray spectral shape, although several additional (intermediate) states can be defined \citep[e.g.][for a detailed overview]{homan2005_specstates,remillard2006}.\footnote{Since soft states often occur at higher X-ray luminosity than hard states, these states are also referred to as low/hard and high/soft states. However, in practice, hard and soft states can occur at similar $L_{\mathrm{X}}$ \citep[see e.g.][for a recent example of such behavior in an X-ray burst source]{chenevez2016}.} During {\it soft states}, an optically thick and geometrically thin accretion disk \citep[][]{shakura1973} is extending close to the compact primary. Indeed, in BH LMXBs it is clear that the thermal disk paradigm often provides a good description of the soft-state spectral data \citep[e.g.][]{Steiner:2010}. However, during {\it hard states} the X-ray spectra are not consistent with the thermal disk solution; these are instead dominated by a power-law emission component with a high-energy cutoff \citep[e.g.][]{sunyaev1991,tanaka1996,ibragimov2005,gilfanov2010}. This hard emission is typically ascribed to a corona (see Section~\ref{subsec:introcorona}).

\begin{figure}
\centering
\includegraphics[width=0.8\textwidth]{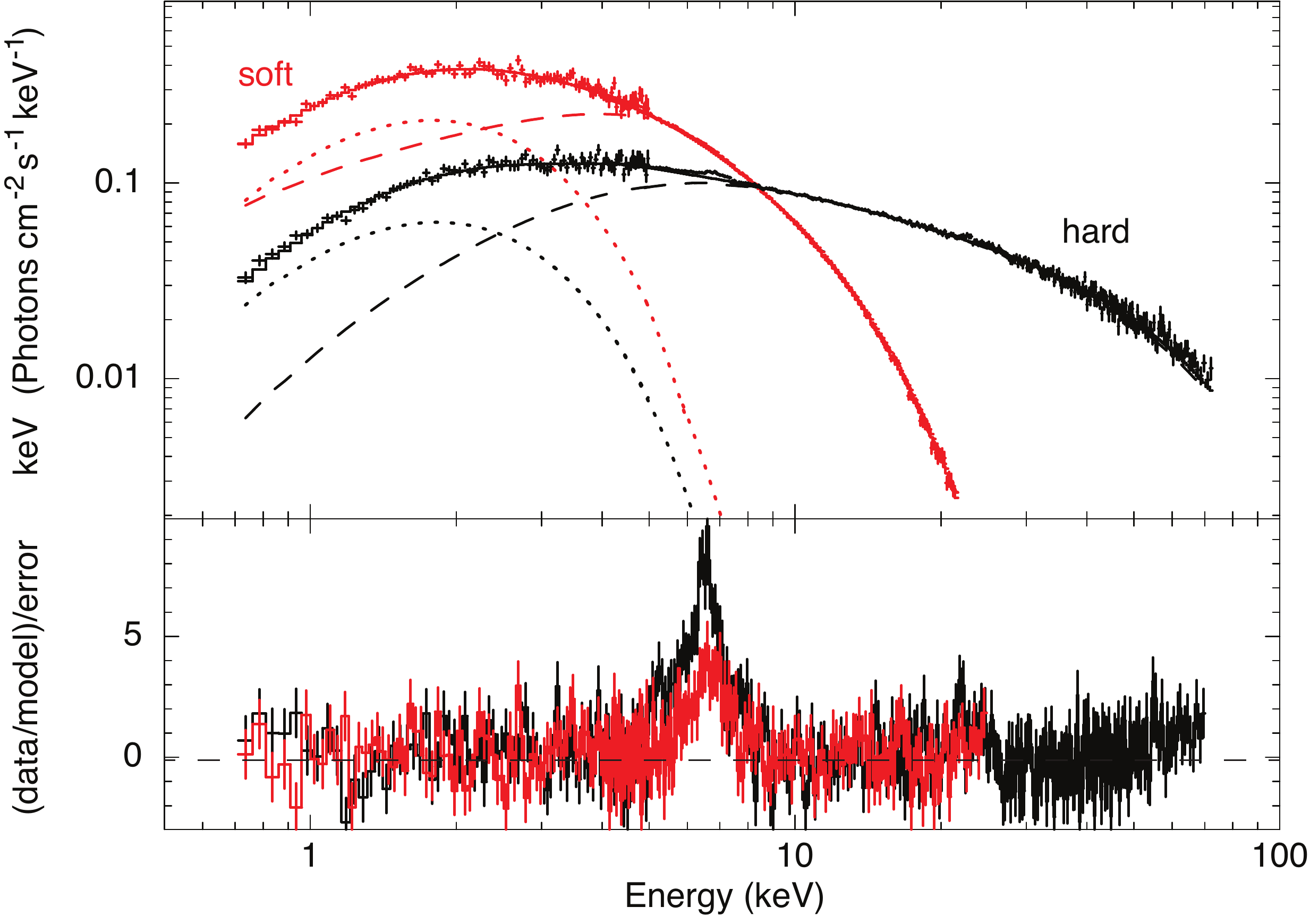}
\caption{Illustration of soft and hard spectral states in a NS LMXB. Shown is \swift\ (0.7--5 keV) and \nustar\ (3--79 keV) data of 1RXS J180408.9--34205 obtained during hard (black) and soft (red) spectral states of its 2015 outburst (reported by \cite{ludlam2016} and \cite{degenaar2016_rxsj1804}, respectively). For illustration purposes, both data sets were simply fitted to a combination of a black body (\textsc{bbodyrad}; dotted) and thermal Comptonization (\textsc{nthcomp}; dashed) model. The energy range between 5.0 and 8.0 keV was ignored during these fits to highlight the broad Fe-K$\alpha$ line that clearly stands out in the data to model ratio (bottom panel). Residuals present at other energies are due to the oversimplified spectral model used.}
\label{fig:states}
\end{figure}

The morphological behavior seen in BH LMXBs is also observed for NS systems, although these suffer from spectral degeneracy (e.g. \cite{barret2001,lin2007,burke2017} for extensive discussions).\footnote{The main hard and soft spectral states in the atoll class of NS LMXBs (see Section~\ref{subsubsec:classes}) are also referred to as ``island" and ``banana" states, respectively, based on the characteristic shape traced in the color-color diagram \citep[][]{hasinger1989}. These can be further classified into sub-states \citep[see][for a recent discussion]{wijnands2017}.} This is due to the presence of additional emission components that can influence the spectral shape; thermal emission directly from the NS surface itself, and/or that emerging from a boundary/spreading layer where the accretion flow impacts the stellar surface \citep[e.g.][]{sunyaev1986,inogamov1999,deufel2001,popham2001,grebenev2002,suleimanov2006}. Such a layer is formed because the gas must decelerate from the near-Keplerian velocities in the inner accretion disk ($\approx 0.2c-0.4c$ for typical inner disk radii; see below) to the slower rotational motion of the NS \citep[typically a few hundred Hz; see][for the most recent overview of spin rates measured for NS LMXBs]{patruno2017_spin}. Figure~\ref{fig:states} shows an illustrative example of hard and soft state spectra in a NS LMXB. A boundary/spreading layer is expected to be most prominent when the disk extends all the way down to the NS surface (i.e. typically during soft states). If the inner disk is truncated and replaced by a hot (optically thin) plasma (which likely happens at low $L_{\mathrm{X}}$, i.e. in hard states and quiescence), the accretion flow may smoothly connect to the NS surface through an optically-thin boundary layer \citep[][]{deufel2001}.

Super-imposed on the continuum X-ray emission, many BH and NS LMXBs show broad emission features (primarily in the Fe-K$\alpha$ band near 6.4--7.0 keV as shown in Figure~\ref{fig:states}), as well as absorption edges. The general interpretation of these features is that some fraction of the Comptonized emission is intercepted by the inner part of the disk and reflected into our line of sight by further Compton scattering and reflection \citep[e.g.][]{george1991,matt1991}. Since the gas in the disk moves in high-velocity (generally assumed to be Keplerian) orbits inside the gravitational well of the compact accretor, the shape of these reflection features is modified by Doppler and gravitational redshift effects. Modeling the reflection spectrum can thus be used to extract information about the accretion morphology, in particular the inner radial extent of the accretion disk \citep[e.g.][for a review]{fabian2010}, which is relevant in the context of the present work. 

The location of the inner disk is typically expressed in terms of the gravitational radius $r_g = \gmc$, where $G$ is the gravitational constant, $M$ is the mass of the compact accretor, and $c$ is the speed of light. For reference, for an $M=1.5~\Msun$ NS, the gravitational radius is $r_g \approx 2.2$~km and for a $M=10~\Msun$ BH we have $r_g \approx 14.8$~km. The location of the inner disk is often compared to that of the inner-most stable circular orbit (ISCO), i.e. the smallest orbit in which a test particle can stably orbit an object in the framework of General Relativity. In the Schwarzschild metric (i.e. zero spin), the ISCO is located at $r_{\mathrm{isco}}= 6~r_g$. For a maximally (prograde) rotating BH, on the other hand, we have $r_{\mathrm{isco}}=r_g$, and for the most rapidly spinning NSs, the ISCO is located at $r_{\mathrm{isco}} \approx 5~r_g$ \citep[e.g.][]{degenaar2017_igrj1706}.

Apart from broadened Fe-K$\alpha$ lines, there are several other means to infer inner disk radii such as the normalization of the disk black-body spectral component, rapid X-ray variability, and reverberation lags (see Section~\ref{subsec:theory:variability}). The fact that these methods often do not agree has led to a long and heated discussion about the location of the inner disk, particularly during the hard state of BH LMXBs. Whereas it is often assumed that the inner disk starts to recede upon entering the hard state, a number of papers used X-ray reflection to suggest that the disk keeps extending to the ISCO down to relatively low X-ray luminosity \citep[$\sim 10^{-3}$ of the Eddington limit; e.g.][and references therein]{reis2010}. These results have been debated in several other works \citep[e.g.][]{done2010,kolehmainen2014,plant2015,basak2016,basak2017}, but some of this criticism has been rebutted \citep[e.g.][]{miller2010,cackett2013,chiang2016}. X-ray timing studies of BH LMXBs may indicate that the inner disk radius varies between different spectral states (see Section~\ref{subsec:theory:variability}).

In case of NS LMXBs, the strong spectral degeneracy mentioned above prohibits using the disk normalization for reliable inferences about the location of the inner disk. Reflection studies lead to inner disk radius measurements of typically $\approx 5-20~r_g$ \citep[i.e. $\approx 10-45$~km from the NS; see e.g.][for a sample studies]{cackett2010_iron,ludlam2017}, with apparently no strong dependence on the spectral state \citep[e.g.][]{disalvo2015,mondal2017,wang2017}. Inner disk radii that extend a few $r_g$ away from the ISCO have often be ascribed to the presence of a geometrically thick boundary/spreading layer \citep[e.g.][]{dai2014_reflection,ludlam2017_rxj1709}. However, in a few cases the inferred inner disk radii are so large that other explanations have been invoked. For instance, the disk appears to be significantly truncated in the soft states of Aql X-1 and GRO J1744--28 \citep[][]{degenaar2014_groj1744,king2016,ludlam2017_aqlx1}. Since both of these sources show evidence for a dynamically active magnetic field (i.e. the detection of X-ray pulsations; see Section~\ref{subsubsec:classes}), it is suspected that their disks are truncated by the pressure exerted by the NS's magnetic field. This has also been proposed to cause the much larger inner disk radius of MXB 1730-335 during its hard state compared to other NSs \citep[][]{vandeneijnden2017}. In addition, there are 2 NSs for which truncated disks have been inferred at relatively low X-ray luminosity ($\sim 10^{-3}-10^{-2}$ of the Eddington limit); for these objects either magnetic truncation or disk evaporation has been considered \citep[][]{papitto2013_hete,degenaar2017_igrj1706}. The rapid variability properties of NSs can also be used to obtain information on the inner disk position (see Section~\ref{subsec:theory:variability}).

\subsubsection{Different sub-classes of neutron star LMXBs}\label{subsubsec:classes}
The accretion rate, accretion geometry and composition of the accreted material have an influence on X-ray burst properties (see Section~\ref{sec:intro}). It is therefore relevant for the present work to recap that there are different sub-classes of NS LMXBs that possibly have distinct accretion-flow properties.

Based on correlated X-ray spectral and (non-coherent) timing properties, NS LMXBs have been classified as Z or atoll sources \citep[][]{hasinger1989}. The small group of Z-sources are brighter and show, on average, softer X-ray spectra than the much more numerous atolls. The fact that some (transient) NS LMXBs can switch between Z and atoll behavior seems to suggests that the distinction between classes is related to the rate at which mass is accreted \citep[e.g][]{priedhorsky1986,shirey1998,shirey1999,homan2010,fridriksson2015}, although other models have been proposed \citep[e.g.][]{church2012}. 

The accreting millisecond X-ray pulsars (AMXPs) stand out as a sub-class of the atoll sources by displaying coherent X-ray pulsations in their persistent emission \citep[e.g.][]{wijnands1998}. It is believed that in these LMXBs the NS magnetic field is strong enough to channel the accretion flow on to the magnetic poles of the rapidly rotating NS (see \citet{patruno2012_review} for an overview up to 2012 and \cite{archibald2014,papitto2013_nature,papitto2015,sanna2017,sanna2017_igrj1659,strohmayer2017} for more recent discoveries). In 4 AMXPs the pulsations have been detection for only a fraction of the time; these are referred to as ``intermittent AMXPs'' \citep[][]{galloway2007,altamirano2008_amxp,casella2008,vandeneijnden2017_igr}. Among the AMXPs there are also 3 NSs that make transitions between an accretion powered X-ray pulsar state and a rotation-powered radio pulsar state; the ``transitional millisecond radio pulsars'' \citep[tMSRPs; e.g.][]{archibald2009,papitto2013_nature,bassa2014,stappers2014}. Anisotropic accretion such as occurring in AMXPs may affect the ignition conditions of X-ray bursts \citep[e.g.][]{cavecchi2011}.

Ultra-compact X-ray binaries \citep[UCXBs; e.g.][]{nelemans2006} form a sub-class of LMXBs that have very short orbital periods of $\lesssim$90~min. This implies that the donor star must be deprived of H in order for it to fit within the binary \citep[e.g.][]{nelson1986}. The different chemical composition of the accreted matter in UCXBs may influence the X-ray burst properties (see Section~\ref{subsec:longbursts}).

\subsection{Coronae}\label{subsec:introcorona}
Hard X-ray emission, in the form of a power-law extending up to tens and even to hundreds of keV, is commonly observed in NS and BH LMXBs, as well as accreting supermassive BHs (Active Galactic Nuclei; AGN). Different from an optically thick but geometrically thin accretion disk that produces the observed thermal X-ray emission in LMXBs, the power-law hard X-ray emission is believed to be produced from an optically thin but geometrically thick structure near the NS or BH, and is commonly called a corona \citep[e.g.][]{sunyaev1991,tanaka1996,ibragimov2005,gilfanov2010}. However, the formation and exact geometry of the corona is much less understood than that of the disk. 

\begin{figure}
\centering
\includegraphics[width=0.32\textwidth]{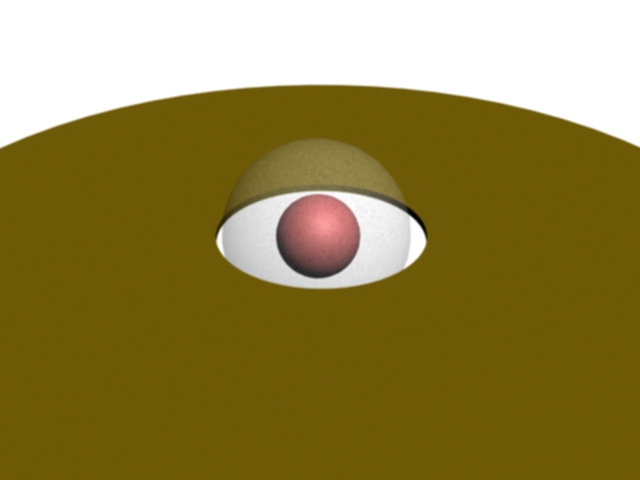}
\includegraphics[width=0.32\textwidth]{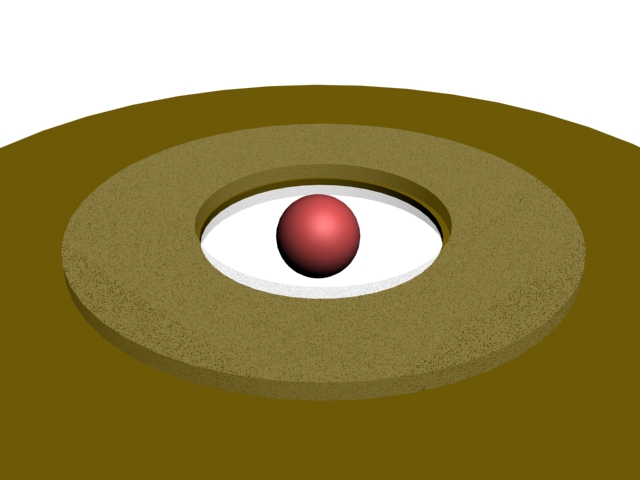}
\includegraphics[width=0.32\textwidth]{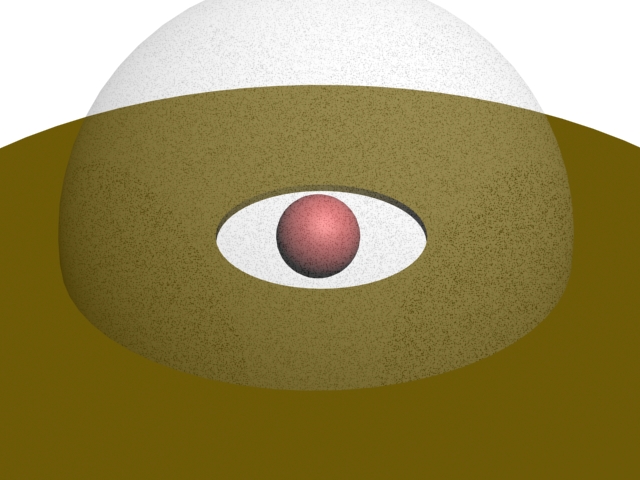}
\caption{Cartoon images of the possible location and geometry of the hot electron corona in LMXBs. The NS is presented as the red central sphere and the disk as the surrounding brown area. Left: A spherical corona that fills the inner region of the accretion flow. Middle: A corona that sandwiches the accretion disk. Right: An intermediate geometry in which a spherical hot flow is overlapping with the cold disk. }
\label{fig:corona_geometry}
\end{figure}

The hard X-ray spectral component seen in LMXBs, most prominently during hard X-ray spectral states, is commonly attributed to Compton up-scattering of soft disk photons by a population of hot electrons \citep[e.g.][]{thorneprice1975,sunyaev1980,poutanen1998,zdziarski2004}. 
Nevertheless, the specific geometry of this population of hot electrons is not completely clear. Figure~\ref{fig:corona_geometry} shows cartoon images of three envisioned coronal geometries \citep[see e.g.][for detailed discussions]{done2007,gilfanov2010}: the hot Comptonizing electrons may originate from some form of hot accretion flow that replaces the inner part of the disk \citep[Figure~\ref{fig:corona_geometry} left; e.g.][]{meyer1994,narayan1995,poutanen1997,malzac2009,veledina2013}, or the corona may be blanketing the disk, being fed by magneto-hydrodynamic (MHD) instabilities \citep[Figure~\ref{fig:corona_geometry} middle; e.g.][]{haardt1993}. Another possible geometry would be a somewhat intermediate case in which a spherical hot flow is overlapping with the cold disk \citep[Figure~\ref{fig:corona_geometry} right; e.g.][]{zdziarski1999}. In AGN, X-ray spectral-timing and microlensing studies have led to the consensus that the corona is very compact, located within $\lesssim 20~r_g$ from the supermassive BH \citep[e.g.][]{fabian2009,fabian2015,demarco2011,reis2013,cackett2014,uttley2014}. Indeed, in theoretical considerations the corona if often assumed to be compact \citep[e.g.][]{henri1997,reynolds1999,lu2001,miniutti2004,garcia2014}. The coronae in LMXBs may have a similar geometry (see below). 

It is not uncommon to adopt a somewhat ad hoc ``lamppost'' geometry, in which the compact emission source is located at a certain height along the rotation axis of the BH, although there are some physical objections to this simple representation \citep[e.g.][]{niedzwiecki2016,dovciak2015,dovciak2016}. The lamppost model has also been applied in reflection studies of weakly magnetized NSs \citep[e.g. 4U 1608--52; ][]{degenaar2015_4u1608}, leading to unphysically small values for the height of the illuminating X-ray source. This suggests that the X-ray source that shines on the disk, and causes the reflection spectrum, likely has a different geometry \citep[][]{degenaar2015_burst}. Indeed, in weakly accreting NSs it may be the boundary/spreading layer rather than the corona that is the prime source of X-rays illuminating the disk \citep[e.g.][]{cackett2010_iron}.

Efforts to explain the exact nature of coronae have primarily focussed on BHs. Three main models have been proposed, which differ in the distribution and location of the electrons: i) thermal electrons \citep[e.g.][]{poutanen1997}, ii) non-thermal electrons with a distribution close to a power-law \citep[proposed in the context of AGN; e.g.][]{ghisellini1993,henri1997,torricelli2005}, or iii) non-thermal electrons free-falling from the last stable orbit down to the horizon of the BH \citep{LaurentTitarchuk:1999}. Hybrid models consisting of a low-energy thermal population and an additional high-energy non-thermal population have also been developed, accounting for the fact that synchrotron emission from the magnetic corona may itself provide a source of seed photons for the Comptonization process \citep[e.g.][]{PoutanenCoppi:1998,coppi1999,merloni2001,veledina2011,delsanto2013,poutanen2014}. It is of note that the purely power-law models face several problems \citep[e.g.][]{malzac1998} and are currently not favored. The bulk-motion Comptonization model cannot explain the observed high maximum photon indices and is therefore also disfavored \citep[e.g.][]{niedzwiecki2006}. It seems therefore more likely that the electrons are mostly thermal with possibly a small fraction of non-thermal particles at higher energies (i.e. a thermal or hybrid model).

High-energy observations are vital in order to disentangle the thermal and
non-thermal components of the corona. Satellites such as the Compton Gamma Ray Observatory (\textit{CGRO}) and the INTErnational Gamma-Ray Astrophysics Laboratory \textit{(INTEGRAL)} have observed an additional emission component above 200~keV during the transition between spectral states of BH LMXBs, as well as in the hard state of both persistent and transient LMXBs \citep[e.g.][]{mcconnell1994,mcconnell2002,ling1997,DelSanto:2008,Bouchet:2009,jourdain2012}. The nature of this additional hard tail is usually 
attributed to the presence of a small fraction of non-thermal electrons in the hot-Comptonizing plasma \citep{Wardzinski:2002}. It could also arise from an additional thermal Comptonization component that results from
spatial/temporal variations in the coronal plasma parameters \citep{MalzacJourdain:2000}, although fluctuations of the thermal Compton parameter may not be able to explain emission extending up to 10 MeV such as detected for the BH Cyg X-1 with \textit{CGRO}/COMPTEL \citep[][]{mcconnell2002}. A weak high-energy tail is also observed
during soft states and is usually interpreted as non-thermal
emission as well, although there is no consensus regarding the origin of this tail. Overall, all the spectral components that are thought to originate in the corona are far from being understood. 

It is worth noting that the coronae in BH LMXBs and AGN appear to have strikingly similar properties despite that their disks are very different. Indeed, the electron temperatures are typically on the order of tens to hundreds keV, despite the fact that the disk temperatures, inner disk radii and variability time scales all scale with the mass of the BH, as predicted in the standard Shakura and Sunyaev accretion disk model \citep[][]{shakura1973}. This seems to suggest that the properties of coronae are scale independent, unlike the properties of the disk. Since the observed spectral and state transitions between some NS and BH LMXBs are quite similar \citep[e.g.][]{Zhang1996}, and because of the apparent scale-independence of coronae, it seems reasonable to assume that the basic coronal properties like its nature and geometry are the same in different accreting systems. Nevertheless, the presence of a solid surface in NS systems can have a direct impact on the properties of the corona. 

Thermal emission from either the NS surface or from the boundary layer should provide additional seed photons for the Comptonization process \citep[][]{sunyaev1989}. Indeed, the hard state of NSs is often softer than that of BHs, which can be attributed to a lower coronal temperature \citep[e.g.][]{sunyaev1991,gilfanov1993,barret1996,done2003}.\footnote{A few NS LMXBs have recently been identified to have exceptionally hard emission spectra, considerably harder than BHs \citep[][]{wijnands2015,wijnands2017,parikh2017}. There is no physical interpretation yet for such very hard spectra in NS LMXBs, although the observed very steep power-law spectral indices of $\Gamma \approx 1-1.3$ might be difficult to explain via Comptonization processes \citep[][]{haardt1993}, and perhaps require a different underlying mechanism.} A recent extensive study of \rxte\ data of a large sample of LMXBs in their hard states revealed a clear dichotomy in the strength of Comptonization between NSs and BHs \citep[][]{burke2017}. In particular, it was found that the electron temperature for the NS sample peaked at $\approx 15-25$~keV, whereas the BH sample showed higher temperatures that were distributed over a broader range of $\approx 30-200$~keV. Moreover, it was shown that at a given optical depth, the electron temperature is higher in BHs than in NSs. Another important implication of this work is that in the NS systems the accreted gas loses a fraction of $\sim 0.5-0.7$ of its energy through Comptonization in the corona, whereas the remaining energy is released when it impacts the stellar surface. The NS thus indeed provides an important source of soft photons that alter the properties of the corona. 

In a follow-up study, \citet{burke2017_spin} used \rxte\ data of hard state NS LMXBs to investigate the effect of spin on the key Comptonization properties. It was found that at a given (Eddington-scaled) accretion luminosity, the seed photon temperature is lower and the Comptonization stronger for a more rapidly rotating NS. This agrees with the theoretical idea that the energy liberated in the boundary layer decreases with increasing NS spin \citep[][]{shakura1988,sunyaev1989}. Since this results in weaker thermal emission and hence a lower supply of soft seed photons for the Comptonization process, the corona of more rapidly spinning NSs is expected to be hotter. The coronal properties of the fastest spinning NSs thus most closely resemble those of the BHs \citep[][]{burke2017_spin}. 

\begin{figure}
\centering
\includegraphics[width=0.85\textwidth]{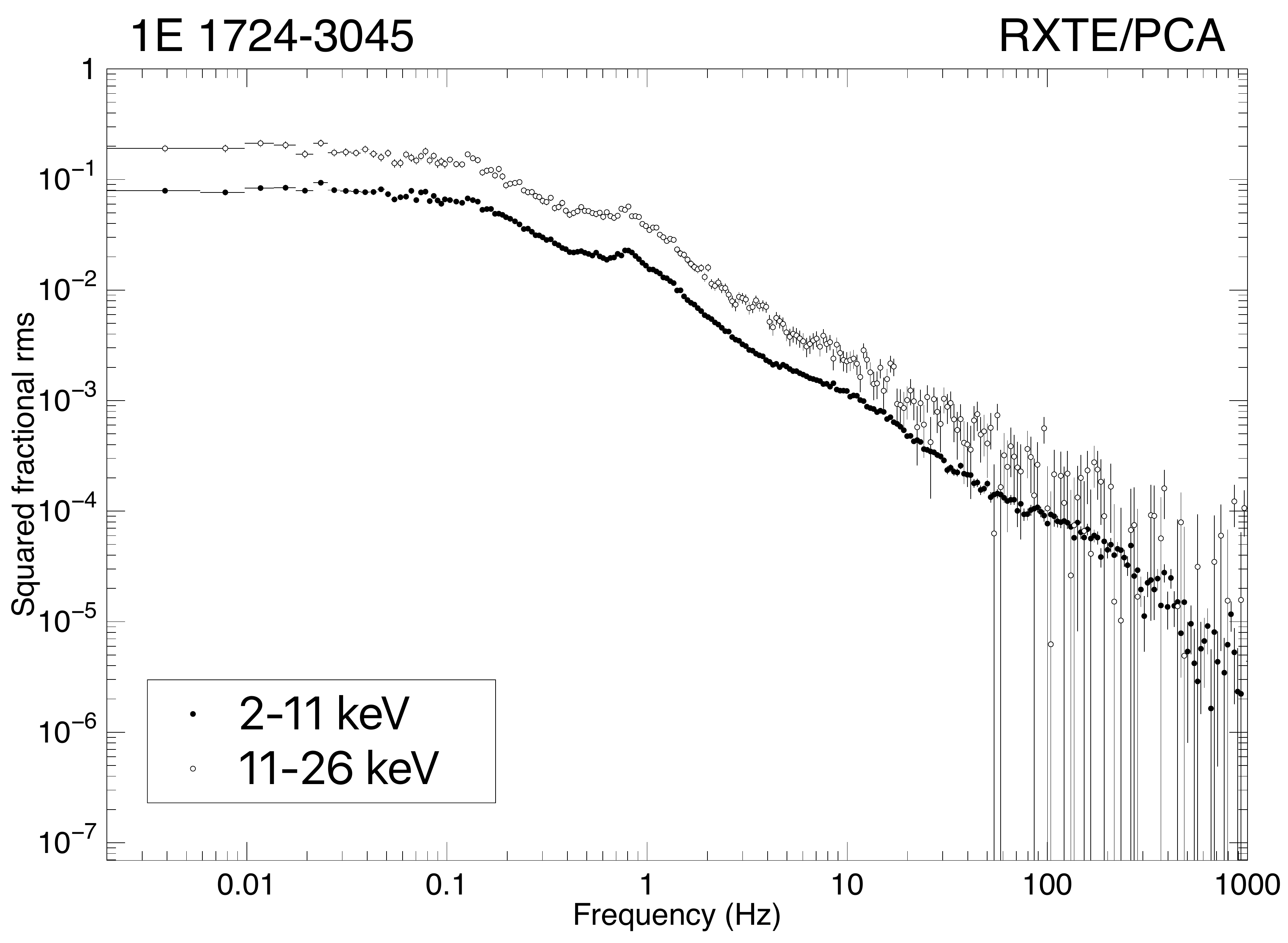}
\caption{Illustration of a PDS of the X-ray bursting NS LMXB 1E 1724--3045, using data from \rxte/PCA in two different energy bands (5 PCUs active). Without doing a component analysis, it is clear that there are multiple Lorentzian components and that they are all stronger in fractional rms at high energies \citep[for a detailed analysis of these data, see][]{olive1998}.}
\label{fig:pds}
\end{figure}

\subsubsection{Coronal geometry in LMXBs from fast time variability}\label{subsec:theory:variability}
The high-energy flux from LMXBs is variable on time scales that go down to the dynamical time scales very close to the compact object, which can be as fast as milliseconds \citep[see e.g.][for reviews]{vanderklis2006,vanderklis2010,bs2014,bm2016}. This fast time variability can reach levels of $\approx$40\% fractional rms and cannot be ignored for the development of theoretical models. This variability is complex and results in Power Density Spectra (PDS; see Figure~\ref{fig:pds}) that contain several frequency-dependent components such as band-limited noise and QPOs (which appear as broad peaks in the PDS). These components yield specific frequencies that are associated to the structure of the accretion flow. However, many different time scales could be connected to the observed features, both from the accretion process and from General Relativity, which makes it difficult to identify the exact nature of different components. Importantly, the observed fast variability is energy-dependent, which implies that the energy spectrum varies on short time scales. This is typically not accounted for in spectral modeling, which requires (much) longer integration times for sensitivity reasons \cite[e.g.][]{wu2010}, and thus averages over all rapid variations. 

Most components in the observed PDS of LMXBs have a hard spectrum, as illustrated in Figure~\ref{fig:pds}, which connects them to the corona. Indeed, in BH LMXBs the variability above 10 keV is always strong and at those energies the coronal emission dominates. Therefore, modeling the timing properties can give important information about the geometry of the corona and complement the information obtained from spectral analysis. Variability is very strong ($\approx$10-40\% fractional rms) in the hard states of both NS and BH LMXBs. The PDS can be decomposed into the sum of a few broad Lorentzian components (i.e. band-limited noise components), which yield characteristic frequencies \citep[][]{bpk2002}. These frequencies are correlated with spectral hardening: harder spectra correspond to higher frequencies \citep[with few exceptions; see][]{pottschmidt2003}. In transient BH LMXBs, the frequencies also correlate with source flux. As the energy spectrum in the hard states is dominated by a thermal (or hybrid) Comptonization component, the noise components must originate from the Comptonizing region. Their increase in characteristic frequency would be naturally explained by a decrease in the emission radius, since all accretion and relativistic frequencies are negatively correlated with distance from the compact object. Constraints on the size of the corona can in principle be derived from the rapid X-ray variability, provided that the observed frequencies are identified as physical time scales in the accretion flow, something for which there is no general agreement.

As mentioned above, relative changes in QPO frequencies can also be used to make inferences about variations in the accretion geometry. In case of NSs, however, there have been extensive discussions about evidence of the inner disk reaching the ISCO. For instance, \citet{barret2005} interpreted a drop in coherence as a function of frequency in the kHz QPOs from 4U~1636--536 as a sign of the accretion flow having reached ISCO, and extended this work to other sources \citep[][]{barret2005,barret2006}. However, \citet{mendez2006} studied the dependence of QPO coherence on source luminosity and concluded that the coherence drop is more likely caused by changes in the properties of the  region where the QPO is produced.

The technique of reverberation combines X-ray spectral and timing studies to make inferences about the accretion geometry (e.g. \cite{fabian1989,stella1990,reynolds1999}; see \cite{uttley2014} for a review). This method uses the fact that due to the finite light travel time, variations in the light reflected off the disk (i.e. the reflection spectrum) will be delayed compared to changes in the corona emission that illuminates the disk. Studies of these time delays in BH LMXBs indicate that there is a significant change when moving from the soft to the hard state, which may indicate that the inner disk moves outwards in the hard state \citep[e.g.][]{demarco2015,demarco2016,demarco2017}. Reverberation can possibly be observed in NS LMXBs as well \citep[][]{barret2013,cackett2016}, but needs to be further developed.


\begin{figure}
\centering
\includegraphics[width=1.0\textwidth]{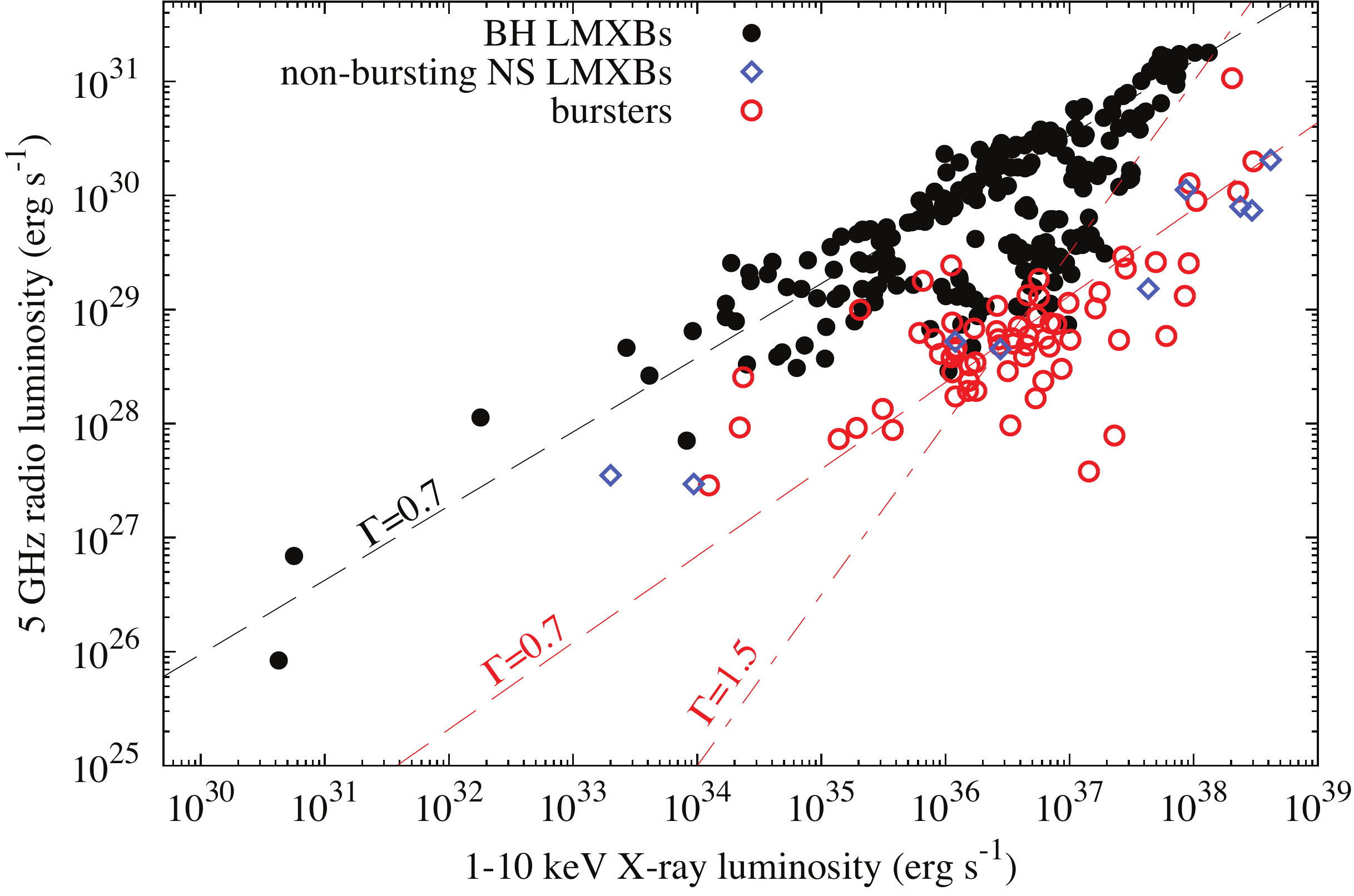}
\caption{
Illustration of the jet properties of BH and NS LMXBs. Shown is a compilation of radio and X-ray luminosities, where black filled circles are BHs, red open circles are bursting NSs and blue open squares are non-bursting NSs. The dashed and dashed-dotted lines highlight different source classes and different couplings between the radio and X-ray luminosity ($L_{\mathrm{R}} \propto L_{\mathrm{X}}^{\Gamma}$) that have been proposed for BHs (black line) and NSs (red lines). 
}
\label{fig:lxlr}
\end{figure}

\subsection{Jets}\label{subsec:jet}
Radio emission has been seen from LMXBs for decades, and it has been well understood since quite early on that there are clear correlations between the radio properties and the X-ray properties of LMXBs. In fact, the first strong evidence for the association of the X-ray source Cygnus X-1 with the object we now know to be its donor star came from the discovery of a sharp transition in the X-ray spectral shape in correlation with a sharp change in the radio luminosity \citep{tananbaum1972}. Radio and X-ray detections of a large number of LMXBs suggest that these different emission components are correlated as $L_{\mathrm{R}} \propto L_{\mathrm{X}}^{\Gamma}$, where $L_{\mathrm{R}}$ and $L_{\mathrm{X}}$ are the radio and X-ray luminosity, respectively, and $\Gamma$ is the coupling constant between the two. This is illustrated in Figure~\ref{fig:lxlr}, where the various curves indicate different $L_{\mathrm{R}}$-$L_{\mathrm{X}}$ couplings that have been proposed (see below). Both empirical \citep[e.g.][]{tananbaum1972} and theoretical \citep[e.g.][]{Meier:2001a} work suggest that jets are fed by large
scale height magnetic fields.  

\subsubsection{State dependence of jets and empirical correlations for BH LMXBs}\label{subsubsec:lrlx}
A broad range of correlations have emerged over time between the radio and X-ray properties of the accreting BH sources (black markers in Figure~\ref{fig:lxlr}). Broadly speaking, it is found that the radio luminosity, $L_{\mathrm{R}}$, scales as the X-ray luminosity, $L_{\mathrm{X}}$, to the 0.6--0.7 power in the hard state. The strong connection between the hard X-ray (corona) emission and the radio (jet) emission of LMXBs \citep[see][and references therein]{Corbel:2013} is sometimes suggested to indicate that the X-ray emission comes from the jet. Indeed, the $L_{\mathrm{R}} \propto L_{\mathrm{X}}^{0.7}$ relation can be explained in a straightforward manner if the X-ray emission comes from uncooled synchrotron emission from the jet \citep[e.g.][]{markoff01}. However, the correlation is equally well explained by producing the X-ray emission in a radiatively inefficient accretion flow with $L_{\mathrm{X}} \propto \dot{M}^2$, and having the kinetic power of the jet be a constant fraction of the gravitational power released in the accretion flow \citep[e.g.][]{heinzsunyaev03}. Ascribing all X-ray emission from a jet has therefore been questioned \cite[e.g.][]{poutanen2003,zdz03,maccarone2012}. During soft X-ray spectral states, on the other hand, the radio emission from the steady, compact jet is suppressed \citep[e.g.][for the deepest soft-state radio limits to date]{russell2011,rushton2016}. During the hard-to-soft state transition, some BH LMXBs release discrete ejecta that are referred to as ``ballistic jets" and the radio emission brightens substantially.

The radio spectral index of BH jets in hard states are normally seen to be flat (i.e. $\alpha \approx 0$, where the flux density is $f_\nu \propto \nu^{-\alpha})$.  This is most commonly explained as the result of having a compact, conical jet so that at each height up the jet the spectrum is strongly peaked due to synchrotron self-absorption below a critical frequency \citep{blandfordkonigl}.  During the ballistic ejections near state transitions, the dominant radio emission mechanism may be different; this is often modeled as coming from an expanding plasmon or set of plasmons. At high radio frequencies, the emission is generally optically thin synchrotron with $\nu^{-0.7}$ spectral indices, although the phenomenology can be considerably more complicated.

\subsubsection{Jets in NS LMXBs}\label{subsubsec:nsjets}
Early radio observations of NS LMXBs, which focussed on the bright Z-sources, also revealed a close correlation between the X-ray and radio emission \citep[][]{penninx1988,hjellming1990a,hjellming1990b,oosterbroek1994,berendsen2000}. However, our phenomenological understanding of the radio properties of NS LMXBs is far less developed than for the BH sources \citep[see e.g.][for reviews]{penninx1989,migliari11}. This is largely due to the fact that fewer NSs have been studied in the radio band and that the dynamical range in X-ray luminosity for which their jet properties are probed is much narrower than for BHs (see Figure~\ref{fig:lxlr} where X-ray bursters are indicated as red open circles and non-bursting NSs as blue open squares). Moreover it is clear that, on the whole, the NSs seem to be fainter in radio at a given $L_{\mathrm{X}}$ and are therefore more difficult to study at radio wavelengths \citep[e.g.][]{fender2001,migliarifender06}. There is some evidence that the correlation between radio and X-ray emission for NSs in the hard state has a steeper index, with $L_{\mathrm{R}}{\propto}L_{\mathrm{X}}^{1.5}$ or so \citep[][]{migliarifender06}, but such evidence is not based on a large number of measurements nor on a large number of sources (see Figure~\ref{fig:lxlr}). Indeed, there appears to be quite a diversity in behavior among the NS LMXBs \citep[e.g.][]{tetarenko2016,tudor2017}. One potential issue with comparing simply the total X-ray and radio luminosity of NSs and BHs is the spectral complexity of NSs (i.e. due to additional emission processes; see Section~\ref{subsec:disk}).

In contrast to BH LMXBs at least two (persistent) NS systems, Ser X-1 and 4U~1820--30, seem to show radio emission from a jet in their soft states \citep{migliari2004}. 
However, this is again not clearly established for a large sample of objects. Moreover, the (transient) NS LMXBs Aql~X-1 and 1RXS J180408.9--0342058 do show a significant downturn in their radio luminosity during the soft state \citep[][]{tudose09,gusinskaia2017}. At present there is thus no clear picture yet about jet suppression in NS LMXBs during their soft states.

Relatively little work has been done on the radio spectra of the NSs, especially in the hard states. The (moderate-$L_{\mathrm{X}}$) atoll sources generally show a flat radio spectrum when they are observed with enough sensitivity to measure a spectral index \citep[e.g.][]{hjellming1999,migliari2004,migliari2010,millerjones2010}. For the (high-$L_{\mathrm{X}}$) Z-sources there are also relatively few spectral measurements, but in these systems it is clear that sometimes flat radio spectra and sometimes spectra with $\alpha \approx 1$  are seen \citep[e.g.][]{grindlay1986,penninx1988,berendsen2000,migliari07}. For one source, 4U~0614+091, extrapolation of the jet spectrum above the break from optically thick to optically thin falls well below the observed X-ray flux. This argues against ascribing the bulk of the X-ray emission to a jet \citep[][]{migliari2010}.

A few key properties of the NS systems make understanding their jet properties especially attractive. First, many NSs have known spin rates, ranging from $\approx$1 to 600~Hz \citep[see][for the most recent overview]{patruno2017_spin}, allowing for a test of the effects of compact object rotation on jet production independent of the debated methods used to infer BH spins \citep[e.g.][for a recent discussion]{kolehmainen2014}. Second, NSs have magnetic fields. While it has long been thought that strong, dynamically important magnetic fields inhibit jet production \citep[e.g.][]{fender1997,fender2000,migliarifender06,massi2008,migliari11}, recent radio detections of the high-magnetic field NSs GX 1+4 and Her X-1 indicate that this idea may need to be revisited \citep[][]{vandeneijnden2017_gx,vandeneijnden2017_herx1}. Furthermore, it is not clear whether a somewhat weaker magnetic field might actually promote jet production. Strikingly, the tMSRPs appear to show a similar $L_R \propto L_X^{0.7}$ as BH LMXBs and are more radio-luminous than other NSs \citep[][]{deller15,bogdanov2017}. However, it is unclear whether for these objects the radio emission in these systems comes from a jet {\it per se} or some other type of less collimated outflow. Finally, the solid surface of NSs could potentially also play a role in jet production. For instance, it has been proposed that a boundary/spreading layer can act as a dynamo and support the jet mechanism \citep[][]{livio1999,maccarone2008}. This can possibly account for the lack of radio jet suppression in some NS LMXBs.


\section{X-ray bursts as a probe of the accretion flow}\label{sec:obs}
As described in Section~\ref{sec:intro}, the focus of the present work is not to review how the accretion flow properties determine the observational properties of X-ray bursts, but rather to scrutinize the reverse interaction: how X-ray bursts influence the accretion disk and corona. 

Observational examples of X-ray bursts interacting with the accretion flow have rapidly accumulated over recent years. Several different physical processes may be initiated in the accretion flow in response to an X-ray burst. The balance between these different processes determines how the accretion flow is affected. This may differ for hard and soft state accretion geometries (e.g., different locations of the corona and the inner accretion disk), and can also depend on the X-ray burst properties (e.g., peak flux, duration, total radiated energy and extent of photospheric radius expansion). Indeed, different but possibly linked types of behavior can be distinguished. These are summarized in Table~\ref{tab:overview} and discussed in more detail in Sections 4.2--4.6. Before describing these observations, we briefly introduce the X-ray missions most relevant to these type of studies in Section~\ref{subsec:currentmissions}.

\subsection{X-ray missions used most for X-ray burst/accretion flow interaction studies}\label{subsec:currentmissions}
The first promptly followed-on X-ray burst detection \citep[][]{grindlay1975} was achieved with the Astronomical Netherlands Satellite (ANS; operating between 1974 and 1976). X-ray burst studies then took a giant leap with NASA's Small Astronomy Satellite 3 ({\it SAS-3}; in service from 1975 till 1979), Japan's first X-ray astronomy satellite \textit{Hakucho} (in orbit between 1979 and 1985),  ESA's European X-ray Observatory Satellite ({\it EXOSAT}; operational from 1983 till 1985), and the Japanese mission \textit{Ginga} (operating between 1987 and 1991). Later missions like the Italian/Dutch satellite {\it BeppoSAX} (in use between 1996 and 2006) and the High Energy Transient Explorer 2 ({\it HETE-2}, an international project collaboration between the USA, Japan, Italy and France that was active between 2000 and 2008) collected a large number of X-ray bursts. 

Most studies of the impact of X-ray bursts on the accretion flow, i.e. the subject of the present work, have been performed with later missions. In particular, a wealth of X-ray burst data was collected with the high X-ray throughput Proportional Counter Array \citep[PCA;][]{jahoda2006} on board the Rossi X-ray Timing Explorer (\rxte). This NASA mission operated from 1995 till 2012 and detected thousands of X-ray bursts \citep[e.g.][]{galloway2008,galloway2010_minbar}. The PCA provided a large effective area\footnote{Quoted values for the effective area of various instruments refer to on-axis sources. For coded mask detectors we refer to that after the mask.} peaking at $\approx 6000~\mathrm{cm}^2$ near 10 keV and high time resolution (down to $\mu$s). It consisted of 5 proportional counter units (PCUs) behind $1\deg$ FWHM collimators, providing sensitivity in the 2--60 keV range. 

Another important tool for studying the response of the accretion flow to X-ray bursts is ESA's \inte\ \citep[][]{winkler2003}, which is still operating to date. \inte\ was launched in 2002 and the main two instruments used in X-ray burst studies are i) the \inte\ Soft Gamma-Ray Imager \citep[ISGRI;][]{lebrun2003} of the Imager on Board
the INTEGRAL Satellite \citep[IBIS;][]{ubertini2003} and ii) the Joint European X-ray Monitor \cite[JEM-X;][]{lund2003}. The ISGRI is sensitive from 15 keV to 1 MeV with a peak effective area of $\approx 1200~\mathrm{cm}^2$ near 60 keV. JEM-X consists of 2 identical units, both sensitive in the 3--35 keV band and with a an effective area peaking at $\approx 100~\mathrm{cm}^2$ near 10 keV.

The still active NASA missions \swift\ \citep{gehrels2004} and \nustar\ \citep{harrison2013} are also valuable tools for X-ray burst/accretion flow interaction studies. \swift\ is a multi-wavelength observatory that was launched in 2005. Its main instruments for X-ray burst studies are the Burst Alert Telescope \citep[BAT;][]{barthelmy2005} and the X-Ray Telescope \citep[XRT;][]{burrows2005}. The BAT is sensitive in the 15--150 keV range and has an effective area that peaks at $\approx 2000~\mathrm{cm}^2$ near 50 keV. X-ray bursts picked up with the wide-FOV BAT can cause an automated slew of the spacecraft that allows for rapid (within $\sim$ 1 minute) follow-up with the more sensitive narrow-field XRT. The XRT provides sensitivity in the 0.3--10 keV range with a maximum effective area of $\approx 130~\mathrm{cm}^2$ near 1.5 keV. \nustar\ is a hard X-ray imager that was launched in 2012. It consists of two co-aligned focal plane modules (FPM), A and B, that cover an energy range of 3--79~keV and provide a peak effective area of $\approx 1000~\mathrm{cm}^2$ near 10 keV. 

Albeit used on a smaller scale for X-ray burst studies, NASA's \chan\ and ESA's \xmm\ missions (both launched in 1999 and still active to date) provide high X-ray spectral resolution via their grating instruments. This allows to detect and study narrow spectral features, which is relevant in the context of the present work (see Section~\ref{subsec:specfeat}). The \chan\ observatory provides two different grating instruments; the Low-Energy Transmission Grating \citep[LETG, 0.1--10~keV;][]{brinkman2000} and the High-Energy Transmission Grating \citep[HETG, 0.4--10~keV;][]{marshall2004}. The \xmm\ satellite carries two Reflection Grating Spectrometers \citep[RGS1 and RGS2, 0.35--2.5~keV;][]{denherder2001}.

\begin{table*}
\caption{Different types of observable effects of X-ray bursts on the accretion flow.\label{tab:overview}}
\begin{threeparttable}
\begin{tabular*}{1.03\textwidth}{@{\extracolsep{\fill}}lccc@{\,}r}
\hline
Type of behavior & Sections & Figures & Sources & References  \\
\hline
\multicolumn{5}{c}{Clear evidence of interactions} \\
Variable accretion flux & \ref{subsubsec:enhanced}, \ref{subsubsec:changestate}, \ref{subsec:theory:persenhance} & \ref{fig:perschange} & Many sources & [1--5] \\ 
Variable accretion spectrum & \ref{subsubsec:changespec}, \ref{subsec:theory:persenhance}, \ref{subsubsec:extp} & \ref{fig:perschange}, \ref{fig:gs1826_comparison} & 4U 1636--536 & [6--8] \\ 
&  &  & 4U1608--52 & [9,39] \\ 
Discrete spectral features & \ref{subsec:specfeat}, \ref{subsec:theory:persenhance} & \ref{fig:reflection}, \ref{fig:4u1820_comparison} & 4U 1820--30 & [10--12] \\ 
 & & & 4U 1636--536 & [6,7] \\ 
 &  & & IGR J17062--6143 & [13,14] \\ 
 &  &  & 4U 2129+11  & [15] \\ 
 &  &  & GRS 1747--312  & [16] \\ 
 &  &  & 4U 0614+091 & [17] \\
  &  &  & 4U 1722--30 & [17] \\
Reduction $>$30~keV flux & \ref{subsec:corona}, \ref{subsec:theory:corona}, \ref{subsubsec:hmxt} 
& \ref{fig:hardXshort}, \ref{fig:hardxshortage_hxmt}, \ref{fig:hardxshortage_otherbursters} & Aql X-1 & [18,19] \\ 
 &  & & IGR J17473--2721 & [18,20] \\ 
 &  && 4U 1636--536 & [21] \\ 
 &  && GS 1826--238 & [22] \\ 
 &  && KS 1731--260 & [23] \\ 
  &  && 4U 1705--44 & [23] \\ 
    &  && 4U 1728--34 & [24] \\ 
kHz QPO suppression & \ref{subsec:qpo} & \ref{fig:qpofrechange} & 4U 1636--536 & [25] \\ 
kHz QPO frequency change & \ref{subsec:qpo} & \ref{fig:qpofrechange} & Aql X-1 & [26] \\ 
High amplitude variability & \ref{subsec:superexpansion} & \ref{fig:lcvar} & 2S 0918--549 & [27,28] \\ 
  & & & 4U 2129+11  & [22] \\ 
 & & & 4U 1820--30 & [17] \\ 
 & & & SLX 1735--269 & [29] \\ 
 & & & SAX J1712--3739 & [30] \\ 
  & & & IGR J17062--6143 & [20] \\ 
   & &  & GRS J1741.9--2853 & [31] \\ 
\hline
\multicolumn{5}{c}{Other possible indications of interactions} \\
Detection of pulsations & \ref{subsubsec:otherfast} & - & HETE J1900.1--2455 & [32] \\ 
 &  & - & SAX J1749.8--2021 & [33] \\ 
mHz QPO frequency & \ref{subsubsec:otherfast} & - & 4U 1636--53 & [34] \\ 
Trigger accretion outbursts & \ref{subsubsec:outbursts} & - & Cen X-4 & [32] \\ 
&  &  & Aql X-1 & [35] \\ 
 &  && 2S 1711--337 & [35] \\ 
  &  && XTE J1747--241 & [35] \\ 
 &  & & EXO 1745--248 & [36,37] \\
 &  & & IGR J17473--2721 & [38] \\ 
\hline
\multicolumn{5}{@{}p{1.0\textwidth}}{\rule[-.3\baselineskip]{0pt}{1.5\baselineskip}\footnotesize %
[1]~\cite{worpel2013}, 
[2]~\cite{worpel2015}, 
[3]~\cite{intzand2013}, 
[4]~\cite{ji2014_4u1608}, 
[5]~\cite{ji2015_gs1862},
[6]~\cite{keek2014}, 
[7]~\cite{keek2014_refl}, 
[8]~\cite{koljonen2016}, 
[9]~\cite{degenaar2015_burst},
[10]~\cite{strohmayer2002}, 
[11]~\cite{ballantyne2004}, 
[12]~\cite{ballantyne2005},
[13]~\cite{degenaar2013_igrj1706}, 
[14]~\cite{keek2016},
[15]~\cite{vanparadijs1990},
[16]~\cite{zand2003},
[17]~\cite{zand2010}
[18]~\cite{maccarone2003}, 
[19]~\cite{chen2013_aqlx1_bursts},
[20]~\cite{chen2012_xrbs_igrj1747},
[21]~\cite{ji2013_xrbs_4u1636},
[22]~\cite{ji2014_bursts},
[23]~\cite{ji2014_aa},
[24]~\cite{kajava2017_hardX},
[25]~\cite{peille2014},
[26]~\cite{yu1999},
[27]~\cite{zand05_ucxb}, 
[28]~\cite{zand2011}, 
[29]~\cite{molkov2005}, 
[30]~\cite{palmer2011}, 
[31]~\cite{Barriere2015},
[32]~\cite{galloway2007}
[33]~\cite{altamirano2008}
[34]~\cite{altamirano2008_amxp}
[35]~\cite{kuulkers2009},
[36]~\cite{serino2012}, 
[37]~\cite{altamirano2012},
[38]~\cite{chenevez2011},
[39]~\cite{kajava2017_4u1608}.
}
\end{tabular*}
\end{threeparttable}
\end{table*}


\subsection{Changing persistent emission}\label{subsec:change}
The standard approach in X-ray burst studies has long been to take a spectrum of the persistent emission and subtract that from the X-ray burst spectrum as background. This method implicitly assumes that the persistent emission is not changing during an X-ray burst. However, fitting X-ray burst spectra via this method can leave residuals, suggesting that the accretion emission may actually not be stable during X-ray bursts \citep[e.g.][]{vanparadijs1986,kuulkers2002,guver2012_bursts}. 

Possible changes in the persistent flux can be probed by not subtracting a pre-burst spectrum, as in the conventional method, but by rather modeling the persistent emission along with the burst. A relatively simple prescription to parametrize any change in the persistent \textit{flux} is to follow these steps: i) fit a pre-burst spectrum to obtain a model for the persistent emission, ii) in the X-ray burst spectral fits (using a black body or burst atmosphere model) include the model for the persistent emission with all parameters fixed to the values obtained from fitting the pre-burst spectrum, and iii) multiply the model for the persistent emission with a constant factor, $f_\mathrm{a}$, that is allowed to change during the fits. This approach assumes that the \textit{shape} of the persistent emission remains constant during the X-ray burst, but allows its flux to change. This simple method has now been applied to a large number of X-ray bursts from many different sources. 

Moving beyond this simple parametrization of the changing accretion flux during X-ray bursts, a few studies have recently applied a mathematical method called ``non-negative matrix factorization'' (NMF). This provides means to decompose the spectra of the X-ray burst and persistent emission and reduce the level of spectral degeneracies (see Section~\ref{subsec:techniques} for more details).

\subsubsection{Enhanced persistent flux}\label{subsubsec:enhanced}
There is indeed growing consensus that the accretion emission changes during the majority of X-ray bursts, regardless of the spectral state and whether or not the X-ray bursts exhibit PRE. A very clear example is the study of the brightest X-ray burst ever observed from the recurrent transient and AMXP SAX J1808.4--3658, which was caught by both \rxte/PCA and \chan/LETG. The X-ray burst spectra deviated from a black body both at high ($\gtrsim$15~keV) and low ($\lesssim$3~keV) energies, which could be explained as an increase of the persistent emission by a factor $\approx$20 \citep[][]{intzand2013}. 

The ``variable persistent emission approach'' described in Section~\ref{subsec:change} was also applied in a large and systematic sample study, which revealed that changes in the accretion emission during X-ray bursts are common \citep[][]{worpel2013}: For 332 PRE X-ray bursts detected with \rxte/PCA from 40 different NS LMXBs, the persistent emission was not subtracted but rather modeled along with the X-ray burst and allowed to vary by a factor $f_\mathrm{a}$. For the majority of X-ray bursts in the sample of \citet{worpel2013} it was found that $f_\mathrm{a}>1$  (with typical values up to $f_\mathrm{a}\approx 10$, but some reaching as high as $f_\mathrm{a}\approx 80$), thus implying an enhancement of the persistent emission. Interestingly, values of $f_\mathrm{a}<1$ (i.e. a reduction rather than an enhancement) were found for the few super-expansion X-ray bursts in the sample. This may be an effect of the X-ray burst emission moving out of the instrument bandpass due to severe expansion and associated cooling of the photosphere, or because these extreme X-ray bursts lead to the ejection of an optically thick shell that absorbs/scatters the emission from the inner accretion flow \citep[e.g.][see also Section~\ref{subsec:specfeat}]{zand2010}. Alternatively, values of $f_\mathrm{a}<1$ could indicate that the in-flow of gas is temporarily halted due to the intense radiation pressure of an X-ray burst (see Section~\ref{subsec:theory:diskradmech}). 

\begin{figure}
\centering
\includegraphics[width=0.8\textwidth]{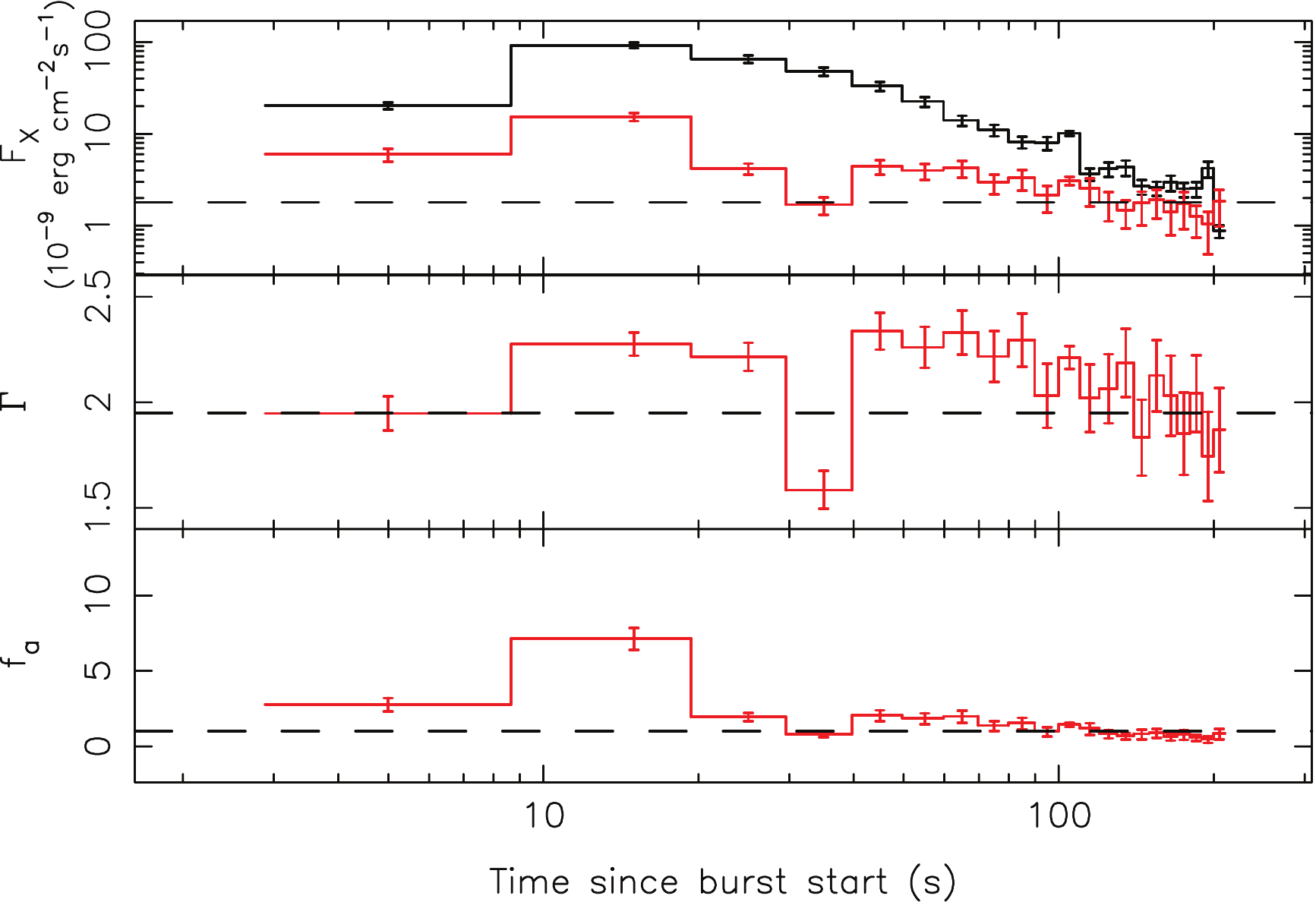}
\caption{Illustration of the changing persistent emission during an X-ray burst (Section~\ref{subsubsec:changespec}). Shown is \nustar\ data from 4U 1608--52 during its 2015 outburst when it was in a hard spectral state. Top: The 2--30 keV flux evolution of the X-ray burst (black) and persistent (red) emission. Middle: Changes in the power-law index $\Gamma$ of the persistent emission, suggesting a possible softening (i.e. an increase in $\Gamma$) along the X-ray burst. Bottom: The amplification factor of the persistent emission during the X-ray burst. Dashed horizontal lines indicate the pre-burst values of the different quantities. This analysis made use of the NMF technique to decompose the spectrum (see Section~\ref{subsec:change} and~\ref{subsec:techniques}). This figure was adapted from \citet{degenaar2015_burst}.}
\label{fig:perschange}
\end{figure}

The above described analysis approach was extended by \citet{worpel2015}, this time using a sample of 1759 X-ray bursts from 56 sources that included non-PRE X-ray bursts. Similar to the PRE sample  \citep[][]{worpel2013}, the persistent emission was generally found to be enhanced compared to the pre-burst level by a factor of several. Comparing the original PRE-burst sample with the (factor $\sim$4 larger) non-PRE sample, suggests that while the average peak $f_\mathrm{a}$ is similar, the PRE X-ray bursts show a larger spread and reach up to a higher maximum $f_\mathrm{a}$ than the non-PRE ones \citep[see figure~8 in][]{worpel2015}. For instance, while the majority of non-PRE X-ray bursts have $f_\mathrm{a} \sim 1-10$ at their peak and only a few have values reaching up to $f_\mathrm{a} \sim 10-30$, several of the X-ray bursts in the PRE sample show $f_\mathrm{a} \sim 30-80$. Figure~\ref{fig:perschange} shows an example of the changing persistent emission during an X-ray burst, which made use of NMF spectral decomposition technique (see Section~\ref{subsec:change} and~\ref{subsec:techniques}).

It is of note that in the two Z-sources Cyg X-2 and GX 17+2, which accrete near the Eddington rate, the enhancement factor is $f_\mathrm{a} \approx 1$ \citep[][]{worpel2015}. Indeed, previous work on GX 17+2 had shown that the persistent emission does need to be subtracted from the X-ray burst emission to obtain satisfactory fits \citep[][]{kuulkers2002}. Apparently the X-ray bursts have no measurable effect on the accretion disk in these systems, which might be related to their near-Eddington accretion rates. In Section~\ref{subsec:theory:persenhance} we discuss how the observations of a (lack of) varying persistent emission during X-ray bursts may be interpreted in a theoretical framework.

\subsubsection{Spectral state dependence of the change in persistent flux}\label{subsubsec:changestate}
It is instructive to investigate whether the enhancement in the persistent flux during X-ray bursts depends on the X-ray spectral states: Since the accretion geometry is expected to be different in hard and soft states (see Section~\ref{subsec:disk}), the response to an X-ray burst can potentially be different.

A study of the frequently active transient NS LMXB 4U 1608--52 revealed that the change in the persistent flux during an X-ray burst is indeed state-dependent \citep[][]{ji2014_4u1608}. A sample of 20 soft-state X-ray bursts (14 PRE), and 26 hard-state X-ray bursts (5 PRE) were examined, which revealed an average $f_\mathrm{a} \sim 3.2 \pm 0.1$ during hard state X-ray bursts and $f_\mathrm{a} \sim 6.8 \pm 0.3$ for soft state ones. Furthermore, whereas in soft states $f_\mathrm{a}$ was observed to increase with X-ray burst luminosity, for hard states $f_\mathrm{a}$ only increased with X-ray burst luminosity as long as $L_{\mathrm{burst}}\lesssim 1.8\times10^{38}~\lum$. For higher X-ray burst luminosity the trend reverses and $f_\mathrm{a}$ instead decreases with further increasing X-ray burst luminosity. It is of note that for very bright hard state X-ray bursts (both PRE and non-PRE) values of $f_\mathrm{a} < 1$ were obtained for 4U 1608--52, suggesting a reduction of the accretion emission rather than an enhancement. 

In a similar study, the steadiness of the accretion emission and X-ray burst profiles in the hard state of the persistent LMXB GS 1826--238, also known as the ``clocked burster'', was exploited to scrutinize the evolution of the persistent emission \citep[]{ji2015_gs1862}. Stacking 68 X-ray bursts and slicing this up in $\approx$3~s time steps revealed that i) $f_\mathrm{a}$ very rapidly increased during the rise and reached a maximum value of $f_\mathrm{a} \approx 2$ on a time scale of $\approx$10~s, ii) dropped to $f_\mathrm{a} \approx 1$ near the X-ray burst peak, and iii) gradually rose to the pre-peak value in $\approx$20~s after the X-ray burst peak and iv) gradually decayed again to $f_\mathrm{a} \approx 1$ in the tail of the X-ray burst ($\approx$130~s). This is the most detailed map of the evolution of $f_\mathrm{a}$ during normal X-ray bursts of a single source.

\subsubsection{Changes in the spectral shape of the persistent emission}\label{subsubsec:changespec}
In addition to probing how the flux of the persistent emission changes during an X-ray burst, it is also valuable to understand if, and how, the shape of the persistent emission spectrum is affected by an X-ray burst. Unfortunately, for most individual X-ray bursts the available data quality does not allow to investigate this. However, there are some exceptions and a few studies in this direction have been carried out. 

Due to the exceptionally long duration and good quality \rxte/PCA data, changes in the persistent emission could be investigated for the (non-PRE) superburst from the persistent LMXB 4U 1636--536 \citep[][]{keek2014}. It was found that the persistent emission increased by a factor $f_\mathrm{a}\approx$1.8, which is relatively low compared to the range of values found for (short) X-ray bursts in the sample studied by \citet{worpel2015}. This is likely not due to the fact that the superburst was sub-Eddington and non-PRE, since higher $f_\mathrm{a}$ values have been inferred for several short non-PRE X-ray bursts that had similar peak fluxes as the 4U 1636--536 superburst \citep[e.g.][]{ji2014_4u1608,worpel2015,degenaar2015_burst}. The high-energy part of the spectrum of 4U 1636--536 appeared to soften \citep[][see also Section~\ref{subsubsec:hardXspecchange}]{keek2014}.

Alternatively, a study of the same 4U 1636--536 superburst using the NMF technique (see Section~\ref{subsec:change} and~\ref{subsec:techniques}) suggested that the boundary/spreading layer significantly changed in response to the X-ray burst \citep[][]{koljonen2016}. The interpretation favored by this analysis was that the boundary/spreading layer moves out to higher latitude due to the radiation pressure of the X-ray burst. This increases the luminosity of the layer (since more X-ray burst photons are reprocessed in the layer) and can thus mimic an increase in the accretion rate. This scenario was previously speculated upon by \citet{suleimanov2011_eos} and \citet{kajava2014}. A similar study, using NMF, was performed for X-ray bursts occurring during the soft states of 4U 1608--52 and led to similar conclusions \citep[][]{kajava2017_4u1608}. 

Hints for spectral changes in the persistent emission were also found by applying an NMF analysis (Section~\ref{subsec:change} and~\ref{subsec:techniques}) to an X-ray burst of 4U 1608--52 that occurred during a hard spectral state \citep[][]{degenaar2015_burst}. Despite that the X-ray burst, detected with \nustar, was severely sub-Eddington ($\approx$0.35$L_{\mathrm{Edd}}$ at its peak), spectral analysis suggests that the persistent emission increased by a factor of $f_\mathrm{a}\approx$5 and possibly softened (Figure~\ref{fig:perschange}; see also Section~\ref{subsec:corona}). In Section~\ref{subsec:theory:corona} we touch upon possible mechanisms to induce changes in the persistent emission spectrum during X-ray bursts.

\begin{figure}
\centering
\includegraphics[width=0.7\textwidth]{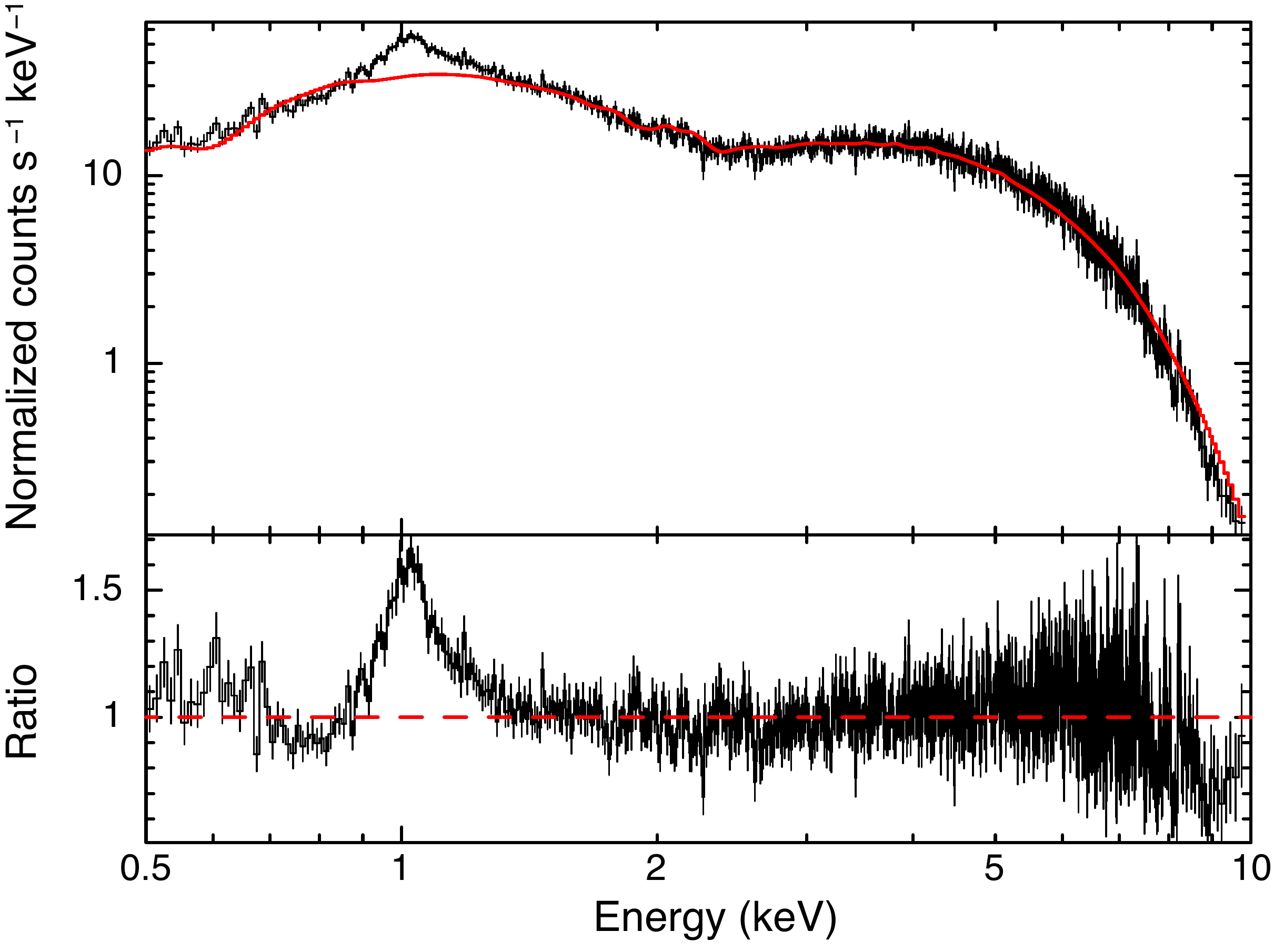}
\caption{Illustration of spectral features during an X-ray burst (Section~\ref{subsec:specfeat}). The top panel shows the average \swift/XRT count-rate spectrum of the intermediate-duration burst of IGR J17062--6143 detected in 2012 \citep[reported by][]{degenaar2013_igrj1706} together with a continuum fit. The residuals in the bottom panel clearly reveal a broad emission line near 1 keV, which possibly results from disk reflection. In addition, absorption features in the 7--9 keV range were reported. 
}
\label{fig:reflection}
\end{figure}


\subsection{Reflection and local absorption}\label{subsec:specfeat}
When studying NS LMXBs, there are several components that may cause discrete features in the X-ray spectra (e.g. narrow absorption lines and edges or broad Fe-K$\alpha$ lines). This includes the NS photosphere \citep[e.g.][]{paerels1997}, the accretion disk \citep[e.g.][]{cackett2010_iron}, an accretion disk wind \citep[e.g.][]{diaztrigo2015}, the interstellar matter \citep[e.g.][]{pinto2013}, and a wind or expanding shell from the NS launched by PRE X-ray bursts \citep[e.g.][]{zand2010}. If emission and absorption features are detected during an X-ray burst, and their origin can be determined, these may also give valuable information about how the accretion flow is responding.

There are a handful of solid cases of (evolving) emission and absorption features during X-ray bursts. This includes two superbursts from the persistent LMXBs 4U 1820--30 and 4U 1636--536,  two intermediate-duration bursts that were both detected from IGR J17062--6143, and two super-expansion X-ray bursts from the candidate UCXBs 4U 0614+091 and 4U 1722--30. Table~\ref{tab:absem} lists the emission and/or absorption features detected during these events. Given the limited number of X-ray bursts for which spectral features have been detected, it is not clear to what extent these are state-dependent. We discuss the possible interpretation of (evolving) reflection and absorption features during X-ray bursts in Section~\ref{subsec:theory:persenhance}.

\subsubsection{The 4U 1820--30 and 4U 1636--536 superbursts}\label{subsubsec:specfeatsuper}
The superbursts from 4U 1820--30 (PRE) and 4U 1636--536 (sub-Eddington and non-PRE) showed similar spectral features: a broad Fe-K$\alpha$ emission line near $\approx$6.4~keV and an absorption edge at $\approx$8--9~keV \citep[both were detected with \rxte;][]{strohmayer2002,ballantyne2004,keek2014,keek2014_refl,koljonen2016}. In both cases the energy and normalization of the reflection features (both the line and the edge) decreased during the superburst. The features were stronger in 4U 1820--30 than in 4U 1636--536, possibly due to the lower X-ray burst flux for the latter. 

In both superburst cases, the line and edge could be successfully modeled as the X-ray burst emission reflecting off the (inner) accretion disk \citep[][]{ballantyne2004,keek2014_refl}. Tracing the evolution of the reflection features for 4U 1820--30 suggested that the inner disk receded during the X-ray burst \citep[][]{ballantyne2004}. To explain this, it was proposed that the inner part of the accretion flow was evacuated during the X-ray burst \citep[][see also Section~\ref{subsec:theory:persenhance}]{ballantyne2005}. In case of the superburst from 4U 1636--536, the evolution of the ionization parameter suggested that the reflection initially occurred in the outer regions of the disk, but came from closer in (presumably the inner disk) during the tail of the X-ray burst, possibly due to a change in disk structure \citep[][]{keek2014_refl}. Interestingly, an increase in column density was seen during both superbursts, which persisted long after the X-ray burst peak \citep[][]{strohmayer2002,keek2014}. For 4U 1820--30, it was suggested that the increase in local absorption could be due to material being ejected from the photosphere during the PRE phase \citep[][]{ballantyne2004}. However, since the superburst of 4U 1636--536 was sub-Eddington and did not show PRE, this seems less likely. It was therefore proposed instead that the X-ray burst perhaps drew material off the disk in a wind \citep[][]{keek2014}. 

\subsubsection{The intermediate-duration bursts from IGR J17062--6143}\label{subsubsec:specfeatlong}
A broad emission line near $\approx$1~keV (which may be due to Fe-L, Ne or some combination of ionization states of the two elements) and possible absorption lines/edges in the $\approx8-9$~keV range were detected during an intermediate-duration burst from IGR J17062--6143 that was captured with \swift\ in 2012 \citep[see Table~\ref{tab:absem};][]{degenaar2013_igrj1706}. The average X-ray burst spectrum is shown in Figure~\ref{fig:reflection}, which illustrates the spectral features. A phase of strong, irregular light curve variability in the tail of the 2012 X-ray burst suggest that it may have exhibited superexpansion (see Section~\ref{subsec:superexpansion}). All emission/absorption features were seen throughout the X-ray burst, but the data quality was not good enough to track any possible changes in the properties of the features along the X-ray burst. Photo-ionization modeling of the blue-shifted absorption features pointed to material flowing out at a distance of $\sim 10^3$~km from the NS, broadly consistent with independent estimates from the width of the $\approx$1~keV emission line (assuming dynamical broadening by plasma executing Keplerian orbits) and the time scale of the light curve variations \citep[][]{degenaar2013_igrj1706}. It was therefore proposed that the irregular light curve variability and the discrete spectral features were all caused by material that was pushed out to a distance of $\sim 10^3$~km in response to the intense X-ray burst emission.

Another intermediate-duration burst was detected from IGR J17062--6143 with \maxi\ and \swift\ in 2015. Time-resolved X-ray spectral analysis suggested that this X-ray burst exhibited super-expansion. As for the 2012 event, an emission feature near 1 keV was seen and there were some hints that this feature evolved along the X-ray burst. Therefore, it was speculated upon that the illuminated gas (presumably the inner edge of the accretion disk) moved inward during the X-ray burst \citep[][]{keek2016}. The 2015 X-ray burst did not exhibit the wild light curve variability that was seen for the 2012 X-ray burst, although such a (short-lived) phase may have been missed due to non-continuous sampling of the 2015 X-ray burst. It is of note that an intermediate-duration burst detected from the persistent LMXB SAX J1712--3739 with \swift\ \citep[][]{palmer2011} showed similar light curve variability as the 2012 X-ray burst from IGR J17062--6143, but for SAX J1712--3739 no convincing emission/absorption features are seen in the spectrum. 

\subsubsection{The super-expansion X-ray bursts from 4U 0614+091 and 4U 1722--30}\label{subsubsec:specfeatsupexp}
Studying a large sample of 32 super-expansion X-ray bursts (from 8 different sources; see also Section~\ref{subsec:superexpansion}), 2 X-ray bursts were found to show significant deviations from black-body spectra \citep[in addition to the superburst from 4U 1820--30;][]{zand2010}: the super-expansion X-ray bursts detected with \rxte/PCA from the candidate UCXBs 4U 0614+091 and 4U 1722--30 both showed a broad emission line between 6--7 keV and absorption edges that evolved during the X-ray bursts. Although the limited spectral resolution of the PCA precluded an exact identification, the edges were interpreted as signatures of heavy element ashes in the expanding photosphere of the NS \citep[][]{zand2010}. It was previously suggested that such features should be detectable in the spectra of (PRE) X-ray bursts \citep[][]{weinberg2006}.

\subsubsection{Additional notes on reflection and spectral features}\label{subsubsec:specfeatnotes}
It is worth pointing out that targeted searches with the grating instruments of \chan\ and \xmm\ did not reveal any discrete spectral features from (normal) X-ray bursts of the peristent LMXBs GS 1826--24 \citep[][]{kong2007} and 4U 1728--34 \citep[][]{galloway2010_4u1728}, nor of the transient LMXBs SAX J1808.4--3658 \citep[][]{intzand2013} and the rapid burster MXB 1730--335 \citep[][]{intzand2017}. However, \citet{pinto2014} reported on the detection of blue-shifted absorption lines, indicative of outflowing plasma along the line of sight, right \textit{after} the particularly energetic X-ray burst from SAX J1808.4--3658 reported by \citet{intzand2013}. Those features were not present prior to or during the X-ray burst. This seems to suggest that the X-ray burst somehow ejected material (e.g. from the photosphere, disk, or corona), perhaps similar to the wind-like outflows that have been tentatively inferred from emission/absorption features in other X-ray bursts \citep[e.g.][]{ballantyne2005,zand2010,degenaar2013_igrj1706}. There is one reported case of the detection of narrow absorption features in the stacked X-ray spectrum of a number of X-ray bursts from the quasi-persistent LMXB EXO 0748--676, which were interpreted as gravitationally-redshifted lines from the neutron star photosphere \citep[][]{cottam2002}. However, the later discovered high rotation period of 552 Hz \citep[][]{galloway2010} firmly rules out a gravitational-redshift origin \citep[][]{lin2010}. Moreover, the putative lines were not observed when the experiment was repeated with new data \citep[][]{cottam2008}.

There are also cases reported in the literature of seemingly broad emission features between 4--8 keV that cannot clearly be interpreted as reflection or Comptonization. This includes intermediate-duration bursts from the (persistent) UCXB 4U 2129+11 in M15 \citep[][]{vanparadijs1990}, the Z-source GX 17+2 \citep[][]{kuulkers2002}, and the frequently active transient GRS 1747--312 in Terzan 6 \citep[][]{zand2003}. Such features have been ascribed to imperfect modeling of the X-ray burst emission \citep[i.e. deviations from a Planck spectrum; e.g.][]{vanparadijs1986}. Indeed, discrete spectral features may also be formed in the NS photosphere \citep[e.g.][]{weinberg2006,zand2010,nattila2015,kajava2017_edges} and are not necessarily a signature of disk reflection or a wind-like outflow.

Finally, it is worth noting that while emission features in X-ray spectra are direct evidence that (some of) the X-ray burst radiation is reflected by the disk, indirect evidence for reflection is provided by the difference in maximum peak fluxes observed in soft and hard spectral states (see figure 1 in \cite{kajava2014} and \cite{suleimanov2016} for the interpretation). It should also be kept in mind that Fe emission features are caused by gas that is only partly ionized; it is possible that the accretion flow extends further in, where Fe is fully ionized and hence does not leave any spectral features but mainly reflects through electron scattering radiation. 

\begin{table*}
\caption{Emission and/or absorption features detected during long/energetic X-ray bursts.\label{tab:absem}}
\begin{threeparttable}
\begin{tabular*}{1.0\textwidth}{@{\extracolsep{\fill}}lccr}
\hline
\hline
Source & Feature & $E_c$ (keV) & Reference \\
\hline 
4U 1820--30 & Fe-K$\alpha$ emission line & $5.8 -6.7$ & [1--3] \\ 
 & absorption edge & $8 -9$ &  \\
4U 1636--536 & Fe-K$\alpha$ emission line & $6.4$ & [4,5] \\ 
 & absorption edge & $8.1 -9.4$ & \\
IGR J17062--6143 (2012) & emission line (Fe-L or Ne?) & $1.0$ & [6] \\ 
 & absorption line & $7.7$ &  \\
 & absorption line & $8.1$ &  \\
 & absorption edge & $8.8$ (fixed) &  \\
 & absorption edge & $9.3$ (fixed) &  \\
(2015) & emission line  & $1.0$ & [7] \\ 
4U 0614+091 & Fe-K$\alpha$ emission line & $6-7$ & [8] \\ 
 & absorption edge & $4.6-8.5$ & \\
 4U 1722--30 & Fe-K$\alpha$ emission line & $6-7$ & [8] \\ 
 & absorption edge & $6-11$ & \\
\hline
\multicolumn{4}{@{}p{1.0\textwidth}}{\rule[-.3\baselineskip]{0pt}{1.5\baselineskip}\footnotesize%
[1]~\cite{strohmayer2002}, [2]~\cite{ballantyne2004}, [3]~\cite{ballantyne2005},
[4]~\cite{keek2014}, [5]~\cite{keek2014_refl},
[6]~\cite{degenaar2013_igrj1706},
[7]~\cite{keek2016}
[8]~\cite{zand2010}.
}
\end{tabular*}
\end{threeparttable}
\end{table*}

\begin{figure}
\centering
\includegraphics[width=0.49\textwidth]{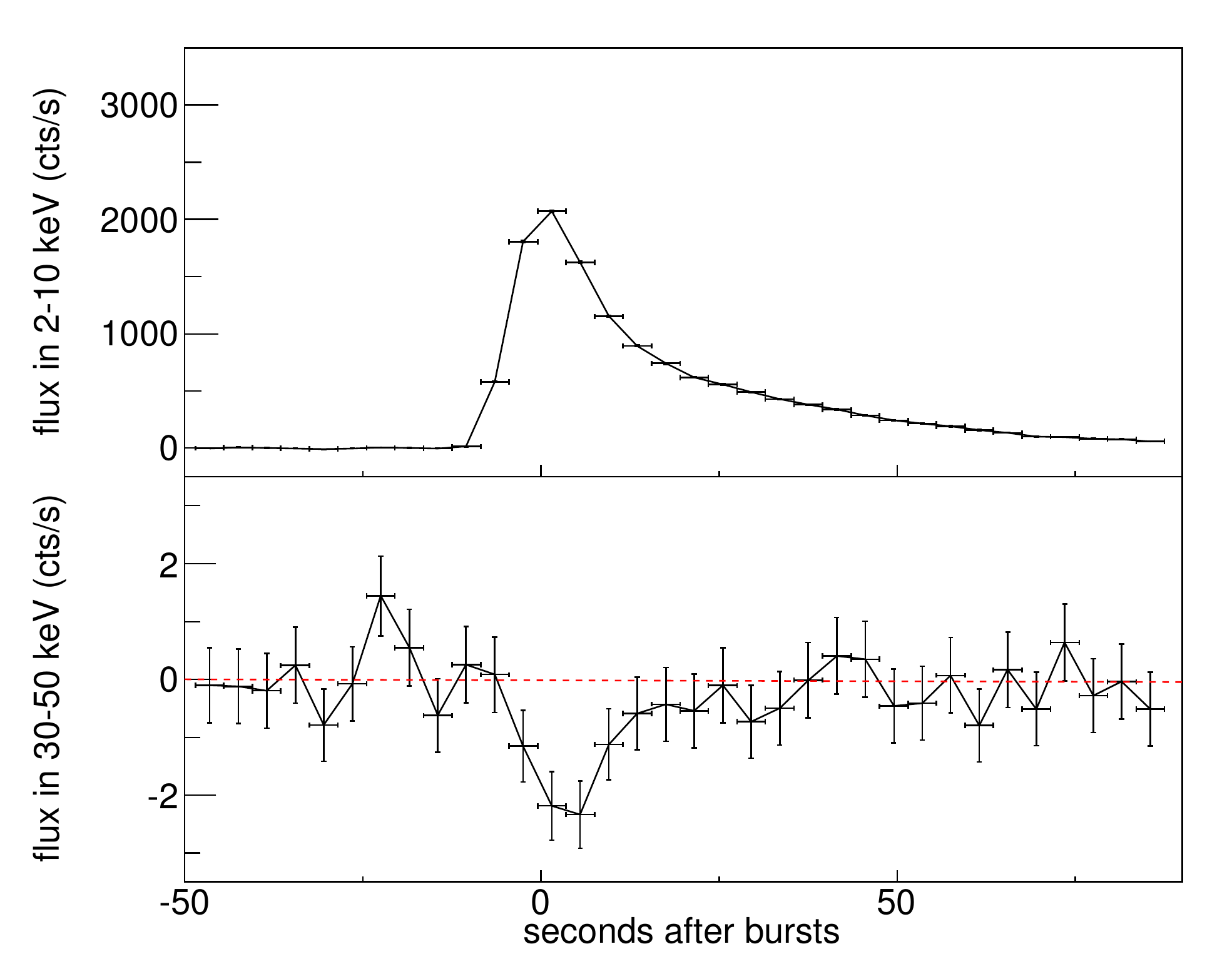}
\includegraphics[width=0.49\textwidth]{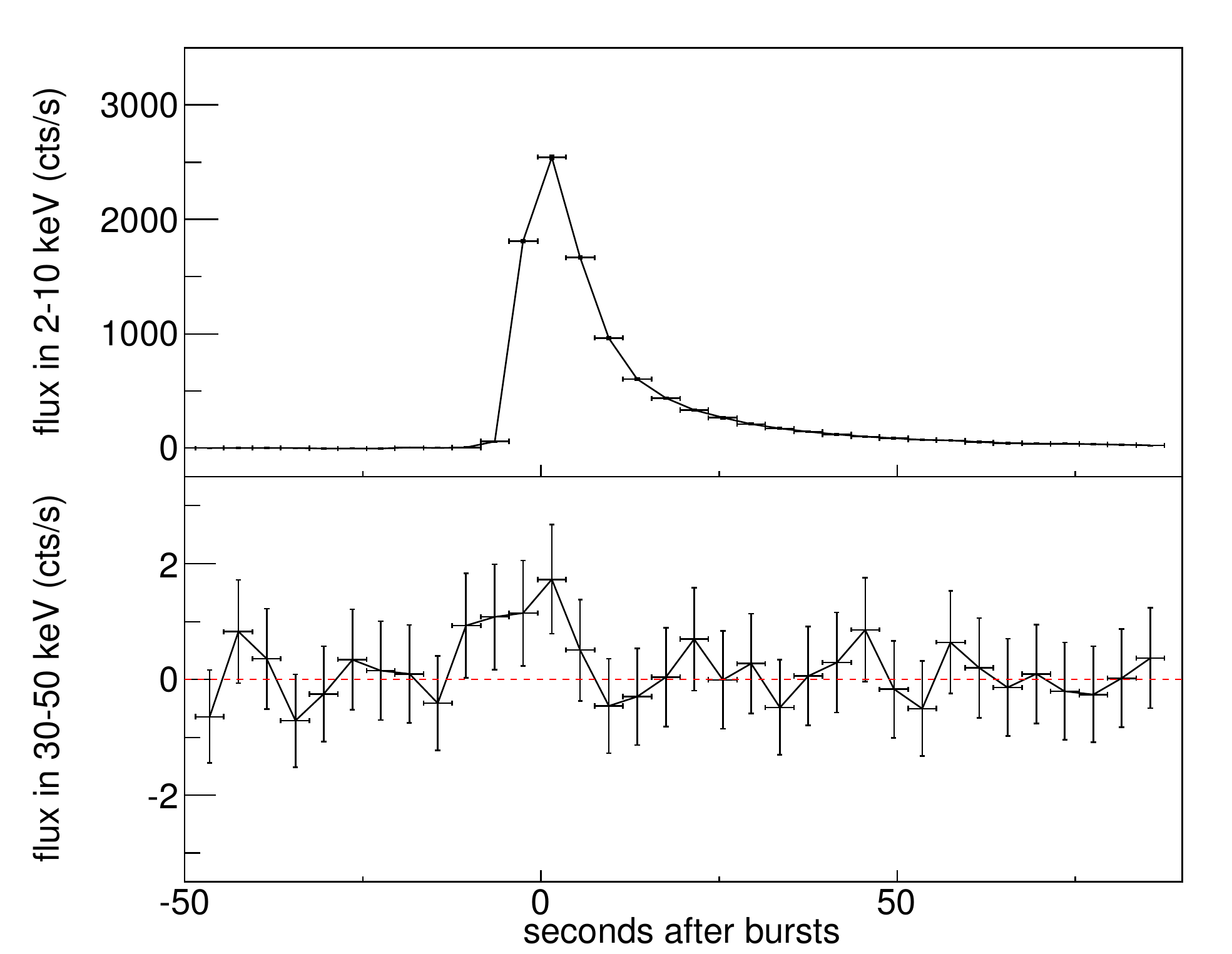}
\caption{Illustrative example of hard X-ray deficits during X-ray bursts. Shown are data obtained with \rxte\ during the 2008 outburst of IGR 17473--2721. In this outburst the source moved from a hard state to a soft state and then through a hard state back to quiescence. The top panels show the 2--10 keV count rate, whereas the bottom panels show the hard, 30--50 keV count rate. Each data point is the sum over the spectral residual after subtracting the persistent emission in the relevant energy band. Left: 32 stacked X-ray bursts detected during the initial hard state that preceded the soft state. A hard X-ray deficit during the X-ray burst peak is evident. Right: 7 stacked X-ray bursts detected during the decaying hard state that followed the soft state. In this case, the hard X-ray count rate remains stable during the X-ray burst. These plots were adapated from \citet{chen2012_xrbs_igrj1747}.}
\label{fig:hardXshort}
\end{figure}


\subsection{Effect on the high-energy persistent emission tracing the corona}\label{subsec:corona}
It was first noted by \citet{maccarone2003} that there was a factor $\approx$2 reduction of the high-energy flux (30--60 keV) during the peak of an X-ray burst from the frequently active transient Aql X-1 (detected with \rxte). This X-ray burst occurred during a hard X-ray spectral state and was not particularly long or luminous. Based on energetic considerations, the observed high-energy flux reduction was interpreted as cooling of the corona due to the injection of soft X-ray burst photons. We explore this idea further in Section~\ref{subsec:theory:corona}.

There are several sources now for which hard X-ray ($>$30 keV) deficits have been reported during X-ray bursts. An example of this is shown in Figure~\ref{fig:hardXshort}. For six of those sources, attempts have been made to measure the lag between the detection of the hard X-ray deficit and the peak of the X-ray burst: these are on the order of $\approx$1--4~s, although not all these measurements are significant (see Table~\ref{tab:hardshort}). These sources are quite different in terms of their accretion behavior (e.g., average accretion luminosity or persistent versus transient nature). Note that the enhancement in the persistent emission (Section~\ref{subsubsec:enhanced}) is typically measured at lower energies of $\lesssim30$~keV, so this does not conflict with the deficits seen at higher ($>30$~keV) energies. 
In exceptional cases it may be possible to study variations in the  shape of the persistent high-energy spectrum during X-ray bursts \citep[e.g.][see Section~\ref{sec:future} for a future outlook]{keek2014,degenaar2015_burst,koljonen2016,kajava2017_4u1608}.

\subsubsection{State dependence of hard X-ray deficits}\label{subsubsec:hardXstatedepend}
Studying a larger sample of X-ray bursts from Aql X-1 revealed similar hard X-ray deficits \citep[][]{chen2013_aqlx1_bursts}. Interestingly, however, despite the similar X-ray burst profiles seen for hard and soft state X-ray bursts, the reduction of the $>30$~keV X-rays was only present during hard-state X-ray bursts, and not during the soft-state ones. This underlines the importance of the accretion geometry and/or physical properties of the corona in the interaction with X-ray bursts \citep[][]{chen2013_aqlx1_bursts}. 

A similar state-dependence of the hard X-ray reduction has been seen for the persistent LMXB 4U 1636--536 \citep[][]{ji2013_xrbs_4u1636}, and in a sample of X-ray bursts collected from the transient LMXB IGR J17473--2721 \citep[][]{chen2012_xrbs_igrj1747}. Results for latter source are shown in Figure~\ref{fig:hardXshort}: Most X-ray bursts were detected during the rising hard state at the beginning of its 2008 outburst, and stacking these revealed a clear deficit of 30--50~keV flux around the X-ray burst peaks. Such a deficit is not detected, however, in the falling hard state at the end of the outburst (i.e., after the source had gone through a soft state). 

Prompted by the discoveries of hard X-ray deficits during X-ray bursts, \citet{ji2014_aa} investigated a sample of 17 NS LMXBs (that all had $>$5 X-ray bursts detected) to search for similar behavior. A hard X-ray deficit was found only during the hard state X-ray bursts of the (quasi-)persistent LMXBs KS 1731--260 and 4U 1705--44, but not during their soft state X-ray bursts nor in any of the X-ray bursts of the other 15 sources. It was therefore proposed that the lack/occurrence of hard X-ray deficits may be related to the accretion geometry (e.g. the height of the corona) or the properties of the X-ray burst \citep[e.g. if its temperature is high, the X-ray burst will contribute significant hard X-ray flux that may dilute any possible deficit due to coronal cooling;][]{ji2014_aa}. 
It is of note, however, that a study with \inte\ of the persistent accretor 4U 1728--34, stacking 123 individual hard-state X-ray bursts, revealed a significant deficit of 40--80 keV photons compared to the persistent emission \citep[][]{kajava2017_hardX}, whereas such a hard X-ray reduction was not found for this source in the previous \rxte\ study \citep[][]{ji2014_aa}. This may have to do with the different responses of the PCA and ISGRI instruments, as well as their different levels of background contamination \citep{kajava2017_hardX}. 

\subsubsection{X-ray burst-induced changes in the hard X-ray accretion spectrum}\label{subsubsec:hardXspecchange}
Stacking X-ray bursts of the clocked burster (GS 1826--238) allowed to investigate the underlying mechanism of the reduction of the high energy photons during X-ray burst peaks \citep[][]{ji2014_bursts}. Modeling the persistent emission during the X-ray bursts suggested that the deficit could possibly be caused by cooling the corona from $T_e \approx 21$ to $16$~keV (see also Section~\ref{subsec:theory:corona}). 

Similar inferences about coronal cooling can be made from other studies. For instance, NMF analysis of a hard state (non-PRE and sub-Eddington) X-ray burst detected from 4U 1608--52 in 2015 with \nustar, revealed hints of a softening of the persistent emission spectrum were found \citep[Figure~\ref{fig:perschange};][]{degenaar2015_burst}. This can possibly be explained as cooling of the corona. However, no deficit of the hard ($>$30~keV) emission was observed for this particular burst \citep[][]{degenaar2015_burst}, nor in a stacked sample of \rxte-detected X-ray bursts from 4U 1608--52 \citep[][]{ji2014_aa}. 

The high quality data of the superburst from 4U 1636--536 also allowed to track possible changes in the shape of the persistent emission during the X-ray burst. It was found that the pre-burst emission could be described by a cutoff power-law and that the cutoff energy decreased around the X-ray burst peak, recovering to its original value in the X-ray burst tail \citep[][]{keek2014}. This was also interpreted as cooling of the corona by the injection of the soft burst photons. We note, however, that \citet{koljonen2016} used the NMF technique to instead argue that the accretion emission remained unaltered and that only the radiation of the boundary/spreading layer changed during the X-ray burst (see also Section~\ref{subsubsec:changespec}).

\begin{table*}
\caption{Lags for hard X-ray ($>$30~keV) deficits with respect to the X-ray burst peak.\label{tab:hardshort}}
\begin{threeparttable}
\begin{tabular*}{1.0\textwidth}{@{\extracolsep{\fill}}lcc}
\hline
\hline
Source & Lag (s) & Reference \\
\hline
Aql X-1          & $1.8 \pm 1.5 $ & \cite{chen2013_aqlx1_bursts} \\
IGR J17473--2721 & $0.7 \pm 0.5$ &  \cite{chen2012_xrbs_igrj1747} \\
4U 1636--536     & $2.4 \pm 1.5$ & \cite{ji2013_xrbs_4u1636} \\
GS 1826--238     & $3.6 \pm 1.2$ & \cite{ji2014_bursts} \\
KS 1731--260     & $0.9 \pm 2.1$ & \cite{ji2014_aa} \\
4U 1705--44      & $2.5 \pm 2.0$ & \cite{ji2014_aa} \\
\hline
\end{tabular*}
\end{threeparttable}
\end{table*}


\subsection{Effect on kHz quasi-periodic oscillations}\label{subsec:qpo}
The physical origin of QPOs at kHz frequencies has not been established, but it is generally assumed that these signals represent the dynamical timescale of the inner part of the accretion flow, close to the NS. Since this region may be subject to the intense radiation of X-ray bursts, there could be observable effects on the properties of kHz QPOs. Indeed such effects have been reported for the frequently active transient Aql X-1 and the persistent source 4U 1636--536 \citep[][]{yu1999,peille2014}.

\begin{figure}
\centering
\includegraphics[width=0.4\textwidth]{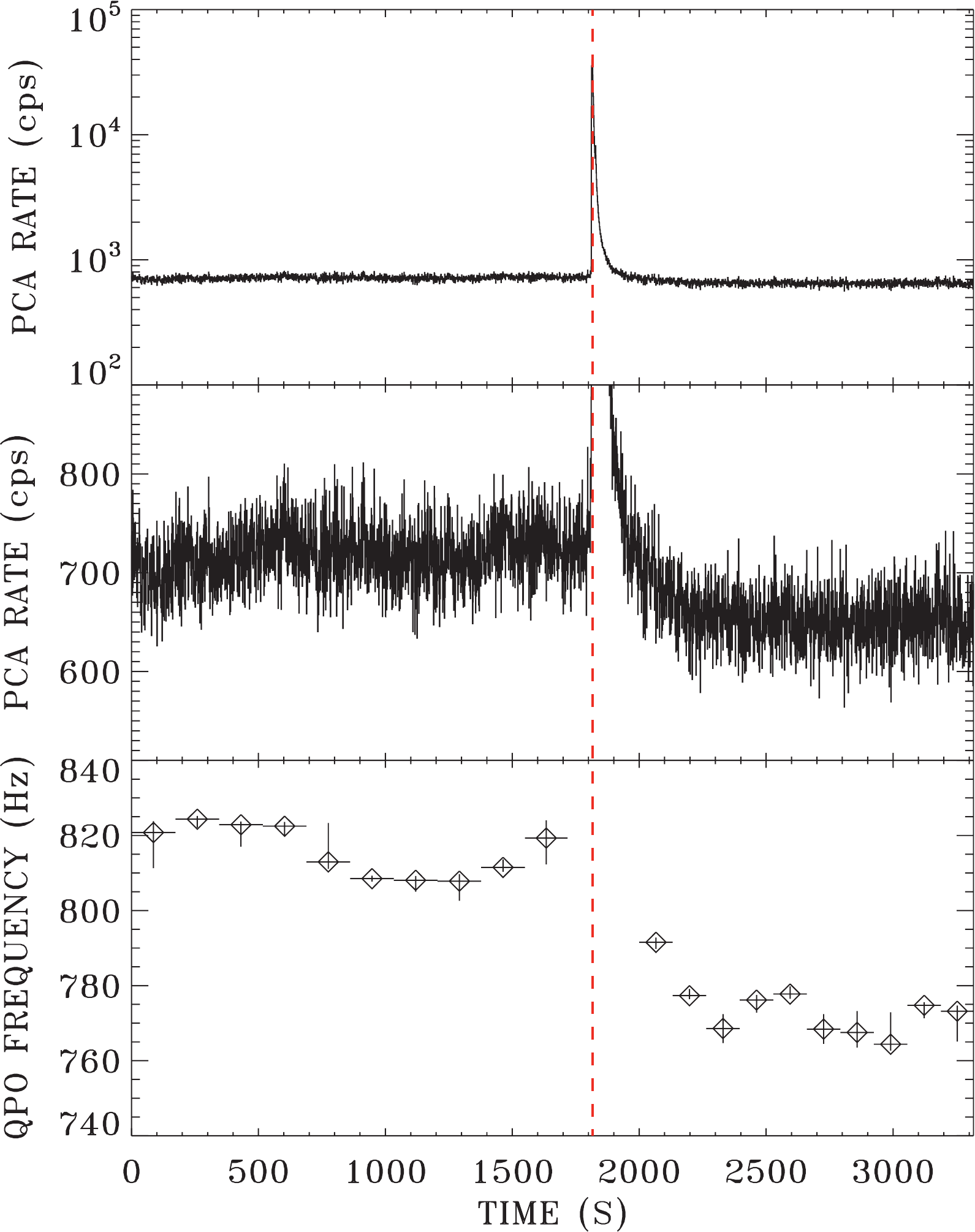}\hspace{+0.3cm}
\includegraphics[width=0.55\textwidth]{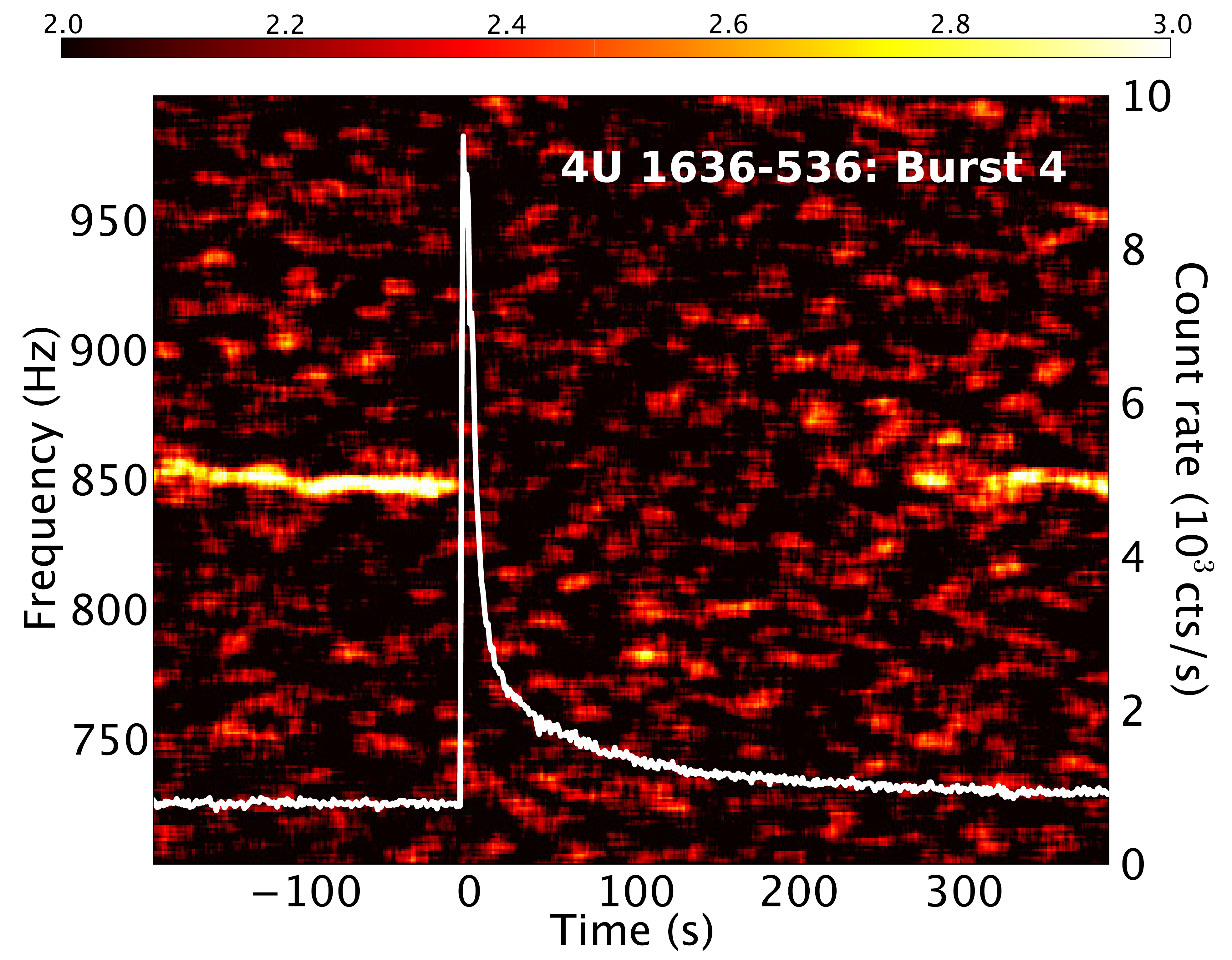}\vspace{+0.5cm}
\caption{Detectable effects of X-ray bursts on kHz QPOs. Left: Change in the QPO frequency around an X-ray burst of Aql X-1 detected with \rxte. Top: Instrument count rate. Middle: zoom to illustrate the change in persistent flux around the time of the X-ray burst. Bottom: QPO frequency. The red dashed line marks the peak of the X-ray burst. This plot was kindly provided by Wenfei Yu as adapated from \citet{yu1999}. Right: Dynamical PDS and light curve (white, overlaid) of an X-ray burst from 4U 1636--536 detected with \rxte. The non-detection of the QPO is significant in a time interval of 80--248~s after the X-ray burst peak. This plot was kindly provided by Philippe Peille as adapated from \citet{peille2014}. }
\label{fig:qpofrechange}
\end{figure}

A significant reduction of the kHz QPO frequency was detected with \rxte\ in Aql X-1 from $813\pm3$~Hz before an X-ray burst to $776\pm4$~Hz thereafter \citep[][]{yu1999}. This was accompanied by a $\approx$10\% decrease in the X-ray flux, as shown in Figure~\ref{fig:qpofrechange} (left). Since the QPO frequency is expected to scale with the radial distance from the NS (i.e., higher frequencies occur further in), it was proposed that the X-ray burst blew away the inner accretion flow. There was nothing unusual about this X-ray burst in terms of peak luminosity, duration or energetics. 

A sample of 15 X-ray bursts detected from two sources, 4U 1636--536 (10 X-ray bursts) and 4U 1608--52 (5 X-ray bursts), was analyzed by \citet{peille2014} to track the kHz QPO on a timescale commensurable with the X-ray burst duration (tens of seconds) in the dynamical PDS. For the majority of X-ray bursts in the sample (12), the QPO went undetected for an episode of $\approx$20--30~s after the onset of the X-ray burst, but this is plausibly due to reduced sensitivity (i.e. the injection of non-modulated burst photons causes a strong drop in the significance of the QPO signal). However, in the 3 remaining X-ray bursts from 4U 1636--536, the QPO signal remained undetected up to several hundreds seconds after the onset of the X-ray burst. It is of note that the X-ray burst for which the suppression of the QPO was most significant, which is shown in Figure~\ref{fig:qpofrechange} (right), had a longer duration and higher total energy output than all other X-ray bursts in the sample. Nevertheless, spectral analysis did not yield an unusually large amplification factor for the persistent emission during these X-ray bursts (maximum values of $f_\mathrm{a}\approx6$, whereas values up to $f_\mathrm{a}\approx20$ were reached for other X-ray bursts in the studied sample). The disappearance of the QPO signal could possibly be due to draining of the accretion disk (accompanied by a temporary enhancement in the accretion rate up to a factor of $\approx6$), but \citet{peille2014} pointed out that the (viscous) time to restore a depleted disk is $>100$~s, whereas for most X-ray bursts the QPO was detected already $\approx$20--30~s after the X-ray burst onset. Therefore, these authors instead favored an interpretation in which the X-ray burst emission irradiates and heats the inner accretion disk, causing it to puff up \citep[see also][]{ballantyne2005}. In Section~\ref{subsec:theory:persenhance} we discuss these different mechanisms in more detail.

\subsection{Superexpansion and large-amplitude light curve variability}\label{subsec:superexpansion}
\begin{figure}
\centering
\includegraphics[width=1.0\textwidth]{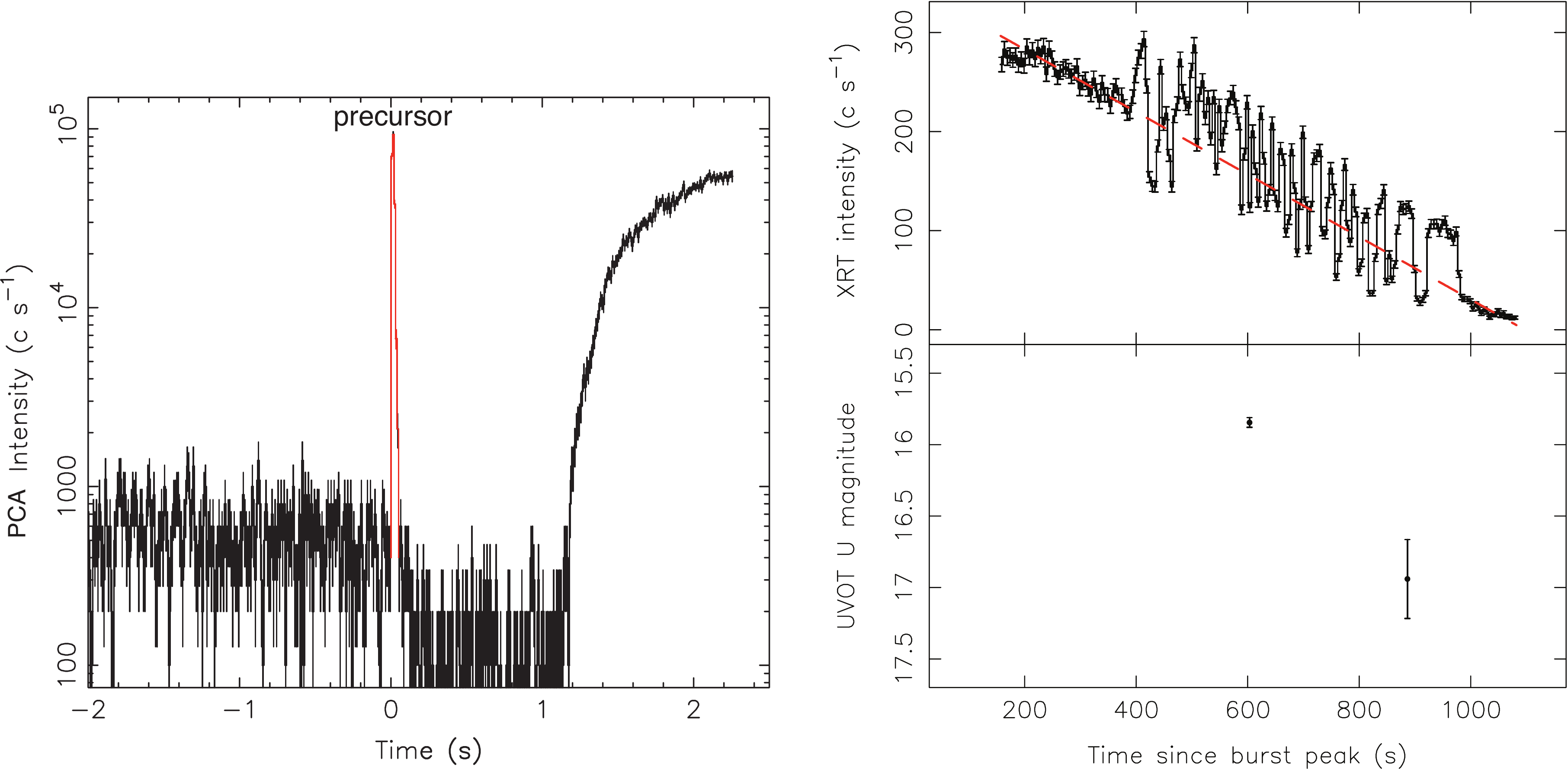}
\caption{Observable manifestations of superexpansion. Left: A precursor burst for 4U 0614+091 (\rxte\ data, 2--60 keV, 10 ms time resolution). This plot is adapted from
\cite{kuulkers2010}. Right: Wild light curve variability during an intermediate-duration burst of IGR J17062--6143 \citep[observed with \swift;][]{degenaar2013_igrj1706}. The top panel shows the evolution of the 0.3-10 keV count rate detected with the XRT (5 s time resolution). The dashed line represents the overall decay trend in the X-ray burst tail and illustrates the amplitude of the fluctuations. The bottom panel shows the evolution of the magnitude in the U band (central wavelength 3465~\AA, Vega system), which is likely due to reprocessing of the X-ray burst emission. }
\label{fig:lcvar}
\end{figure}

Extreme PRE, i.e. superexpansion (see Section~\ref{subsec:PRE}), is manifested as a dramatic drop in burst emission. It is plausible that the extreme expansion causes the temperature to decrease to such an extent that the emission moves out of the X-ray band. This would then lead to the observation of an X-ray precursor to the main burst event \citep[first reported by][an example is shown in Figure~\ref{fig:lcvar} left]{hoffman1978_precursor}. A compilation of 39 X-ray bursts with precursor events (i.e., indicating a superexpansion phase) from 9 different sources was recently given by \citet{zand2014}. It is worth noting that all of the sources showing precursors are either confirmed or strong candidate UCXBs \citep[see][and references therein]{zand07,zand2010}. 

During superexpansion X-ray bursts, the X-ray emission may actually decrease below the pre-burst emission level: an example is shown in Figure~\ref{fig:lcvar} (right). It has been proposed that this is caused by a geometrically thin but optically thick gas shell that is driven off the NS during the extreme PRE burst \citep[][]{zand2010}. Such an expanding gas shell can engulf the accretion flow so that it is temporarily blocked from our view, and may explain why the X-ray emission falls below the pre-burst level during this phase. This ejection of matter, moving at mildly relativistic speeds \citep[][]{zand2014}, is bound to affect the accretion flow. Indeed, there are a few clear examples where superexpansion seems to disrupt the disk.

On a few occasions, particularly energetic X-ray bursts have shown large amplitude, erratic flux variability during a phase of the X-ray burst tail. An example of this is shown in Figure~\ref{fig:lcvar} (right). It appears that these variations are related to superexpansion \citep[][]{zand2011} and it has been observed for 8 X-ray bursts from 7 different sources (listed in Table~\ref{tab:lcvar}). Seven of these X-ray bursts were of intermediate duration, whereas one was a superburst. The start time and duration of the strong light curve variations varies from source to source (see Table~\ref{tab:lcvar}). In two sources with these strong variations, 4U 1820--30 and IGR J17062--6143, emission/absorption features were seen in the X-ray burst spectra, which may be related to the unusual flux variations \citep[][see Section~\ref{subsec:specfeat}]{strohmayer2002,ballantyne2004,ballantyne2005,degenaar2013_igrj1706}. The large-amplitude flux variability is likely linked to the material that is ejected from the photosphere during the superexpansion \citep[][]{paczynski1986,ballantyne2005,weinberg2006,zand2011,degenaar2013_igrj1706}. This is further discussed in Section~\ref{subsec:theory:persenhance}. It is worth pointing out that for instruments with no energy coverage below $\approx$3~keV (e.g. \rxte) the large amplitude variations may appear achromatic \citep[][]{zand2011}, but when lower-energy coverage is available (e.g. with \swift) the variations do show an energy dependence \citep[][]{degenaar2013_igrj1706}.

\begin{table*}
\caption{Large amplitude light curve variability during X-ray bursts. These X-ray bursts show dramatic time variability, with strong deviations from the FRED-like envelopes of the X-ray bursts, as described in the text and shown in Figure~\ref{fig:lcvar} (right). The start time is given with respect to the X-ray burst peak. \label{tab:lcvar}}
\begin{threeparttable}
\begin{tabular*}{1.0\textwidth}{@{\extracolsep{\fill}}llrrcc}
\hline
\hline
Source & Instrument & Start (s) & Duration (s) & Remark & Ref. \\
\hline
2S 0918--549    & \beppo\  & $\approx 130$  & $\approx 100$  &            & [1] \\ 
                & \rxte\   & $\approx 120$  & $\approx 80$   &            & [2] \\ 
4U 2129+11      & \ginga   & $\approx 0$    & $\approx 30$   & M15        & [3] \\ 
4U 1820--30     & \rxte\   & $\approx 6000$ & $\approx 1500$ & superburst & [4] \\ 
SLX 1735--269   & \inte\   & $\approx 750$  & $\approx 500$  &            & [5] \\ 
SAX J1712--3739 & \swift\  & $\approx 100$  & $\approx 500$  &            & [6] \\ 
IGR J17062--6143  & \swift\ & $\approx 400$ & $\approx 600$ &   2012     & [7] \\ 
GRS J1741.9--2853 & \nustar\ & $\approx 50$ & $\approx 50$ & Galactic center & [8] \\ 
\hline
\multicolumn{6}{@{}p{1.0\textwidth}}{\rule[-.3\baselineskip]{0pt}{1.5\baselineskip}\footnotesize%
[1]~\cite{zand05_ucxb},
[2]~\cite{zand2011},
[3]~\cite{vanparadijs1990},
[4]~\cite{strohmayer2002},
[5]~\cite{molkov2005},
[6]~\cite{palmer2011},
[7]~\cite{degenaar2013_igrj1706},
[8]~\cite{Barriere2015}
}
\end{tabular*}
\end{threeparttable}
\end{table*}

\subsection{Other possible examples of the effect of X-ray bursts on the accretion flow}\label{subsec:other}
Sections~4.2--4.6 describe comprehensive and convincing cases of how X-ray bursts affect the surrounding accretion flow. In this Section we describe a few other types of observations that can possibly also be indicative of such interactions but are somewhat more ambiguous. 


\subsubsection{Changes in fast variability properties related to X-ray bursts}\label{subsubsec:otherfast}
\citet{altamirano2008} report on a decrease in frequency and eventual disappearance of the mHz QPO in the frequent burster 4U 1636--53, which happened between the transition from the hard to soft state. This behavior is somewhat reminiscent of the changes in the kHz QPOs observed for Aql X-1 and 4U 1636--53 \citep[see Section~\ref{subsec:qpo};][]{yu1999,peille2014}. However, in the case of 4U 1636--53 the X-ray burst happened after (rather than before) the observed change in the mHz QPO properties. This different causal connection suggests that the physical mechanism is likely different. Indeed, unlike kHz QPOs that are associated to phenomena happening in the inner accretion flow, the current paradigm is that mHz QPOs are instead related to marginally stable thermonuclear burning on the surface of the NS \citep[e.g.][]{revnivtsev2001,yu2002,heger2007,altamirano2008,linares2012,keek2014,lyu2014,lyu2016}. Nevertheless, for 4U 1608--52 an anti-correlation between the occurrence of mHz QPOs and the kHz QPO frequency was observed, which suggests that mHz QPOs can perhaps cause (small) perturbations in the accretion flow \citep[][]{yu2002}.

Another intriguing phenomena is that for the intermittent AMXPs HETE J1900.1--2455 (377 Hz) and SAX J1749.8--2021 (442 Hz), the sporadic detection of X-ray pulsations might be triggered by the occurrence of an X-ray burst \citep[][]{galloway2007,altamirano2008_amxp}. However, the detection of intermittent pulsations in Aql X-1 (550 Hz) did not have any connection with X-ray bursts \citep[][]{casella2008} and later detections of pulsations from HETE J1900.1--2455 were not related to X-ray bursts either \citep[][]{patruno2012_hete}.

It may be argued that both mHZ QPOs and pulsations are arising from the NS surface rather than accretion flow and that these phenomena are therefore different from the ones discussed in Section 4.2--4.6. However, the matter arriving on the NS surface is of course coming from the accretion flow and it is plausible that the properties of the flow (and their response to an X-ray burst) play a role.

\subsubsection{X-ray bursts inducing accretion outbursts?}\label{subsubsec:outbursts}
Particularly with older X-ray missions, several X-ray bursts were discovered while no persistent accretion emission could be detected above the instrument background (implying $L_X\lesssim 10^{36}~\lum$). These NS LMXBs were therefore referred to as ``burst-only sources"  \citep[e.g.][see also Section~\ref{subsec:targets}]{zand1999_burstonly,cocchi2001,cornelisse2002,kuulkers2009}. A few of these X-ray bursts were followed by an accretion outburst within a few days (see Table~\ref{tab:overview}).  The X-ray burst from the transient LMXB EXO 1745--248 (in Terzan 5) that occurred before its faint 2011 accretion outburst was likely a superburst \citep[][]{serino2012,altamirano2012}, whereas the X-ray burst detected from Cen X-4 prior to its 1969 accretion outburst was of intermediate duration \citep[][]{belian1972,kuulkers2009}. All other cases concern normal, short X-ray bursts. 

\citet{kuulkers2009} addressed whether the intense X-ray burst emission could ionize the cold, neutral (i.e. quiescent) accretion disk so that matter can flow more rapidly onto the NS and an accretion outburst is generated. For all but the superburst of EXO 1745--248 \citep[][]{altamirano2012}, however, the energy output from the X-ray burst is insufficient to ionize a cold disk. For most of these cases it therefore seems more likely that the accretion rate was already enhanced before the outburst, which is a prediction of the disk-instability model, and that this facilitates the ignition of an X-ray burst rather than vice versa \citep[][]{kuulkers2009}.


\section{Theoretical framework}\label{sec:theory}
As described in detail in Section~\ref{sec:obs}, there exists a vast body of evidence that X-ray bursts can affect the structure of the accretion disk and corona in a variety of ways. Although the individual topics of accretion disks and corona have a rich theoretical history, little attention has been paid to date to the processes that may arise in these systems when subject to an intense burst of radiation. In this Section, we review the current theoretical framework for interpreting the influence of X-ray bursts on accretion disks and corona. However, we emphasize that much more work needs to be done before a coherent understanding of the relevant physics emerges. 

\subsection{The influence of X-ray bursts on accretion disks and the accretion rate}\label{subsec:theory:persenhance}
The standard method of subtracting a pre-burst spectrum as background when modeling the X-ray burst emission reveals spectral residuals when high signal to noise data is available (e.g. when a large sample of X-ray bursts is analyzed in aggregate, data is taken with high sensitivity instruments, or one can integrate over long times such as for long X-ray bursts). Allowing for variations in the persistent emission during the X-ray burst significantly improves the fits (see Section~\ref{subsec:change}). We here summarize the main findings of applying this analysis approach to a large number of X-ray bursts:

\begin{itemize}
\item Allowing the persistent flux to vary by a factor of $f_{\mathrm{a}}$ suggests that it increases for the majority of X-ray bursts/sources, independent of the presence of PRE or the peak flux of the X-ray burst. The exception are Z-sources for which $f_{\mathrm{a}} \approx 1$, and super-expansion X-ray bursts for which $f_{\mathrm{a}} < 1$.

\item Studying a large sample of X-ray bursts suggest that while $f_{\mathrm{a}}$ is on average similar for PRE and non-PRE X-ray bursts, the former shows a larger spread that extends to higher maximum values.

\item The amplification factor of the persistent emission $f_{\mathrm{a}}$ appears to be state-dependent; it is higher in soft states than in hard states.

\item Using high-quality data of very long X-ray bursts or employing the NMF spectral decomposition technique allows to detect changes in the shape of the persistent spectrum; this generally suggests that the persistent emission softens during an X-ray burst.
\end{itemize}

Most of these studies have been performed using data up to $\approx$30~keV. These findings suggest that the accretion disks in NS LMXBs respond to X-ray bursts. In the truncated disk paradigm, in which the accretion disk recedes when moving from the soft state into the the hard state (see Section~\ref{subsec:disk}), the impact on the accretion disk should be most noticeable during soft states. It is worth noting that apart from its effect on the disk, an X-ray burst can also temporarily change the accretion rate (see Section~\ref{subsec:theory:diskradmech}), which can affect both the disk and the corona.

Moving beyond the simple parametrization discussed above, the detection of (evolving) X-ray reflection features during a superburst from 4U 1820--30 has been particularly valuable to study the dynamics of the accretion disk in response to an X-ray burst. The intense radiation field generated by X-ray bursts will irradiate the accretion disk surrounding the NS. The X-rays will be scattered and thermalized by the gas in the disk with a large fraction re-emerging in the X-ray band. This backscattered, or reflected, radiation can potentially be imprinted with spectral features that provide important diagnostics of the accretion flow such as its dynamics, density and abundances \citep[e.g.][]{fabian2010}. The possibility that reprocessing from the accretion disk may be detected in the X-ray spectra of X-ray bursts was initially suggested by \citet[][]{dd91}, but the first clear example was the detection of the Fe-K$\alpha$ line and edge in the \textit{RXTE} observation of the 4U 1820--30 superburst \citep[][]{strohmayer2002,ballantyne2004}. These spectral features are known to be produced by the reprocessing of X-rays in a dense and relatively cool environment that lies out of the line-of-sight \citep[e.g.][]{george1991}, and are commonly observed in a wide variety of accreting systems \citep[e.g.][]{miller07}.

Modeling the time evolution of the reflection features with sophisticated ionized reflection models allowed for time-dependent measurements of the ionization state of the reflecting region of the disk during the superburst of 4U 1820--30, as well as the total reflection strength in the spectrum (which is related to the covering fraction of the reflector). The Fe-K$\alpha$ features observed in the spectra of 4U 1820--30 were also observed to undergo a variable redshift during the superburst. Interpreting this redshift as gravitational, i.e. due to the potential well of the NS, also allowed constraints to  be placed on the radius of the reflecting region during the superburst. Combining all this information, \cite{ballantyne2004} found the striking result that the reflecting region in the accretion disk around 4U 1820--30 moved from a distance of $\approx 20~r_g$ to $\approx 100~r_g$ (corresponding to $\approx 44$ to $220$~km for a $1.5~\Msun$ NS) over the first 1000~s of the X-ray burst, before returning to $\approx 20~r_g$ over the following $\approx 1000$~s. 

\subsubsection{Radiative mechanisms that can affect the disk and accretion rate}\label{subsec:theory:diskradmech}
With the information obtained from the reflection analysis of the 4U 1820--30 superburst one can scrutinize the response of the accretion disk to the X-ray burst radiation. In addition to simply photo-ionizing the surface of the accretion disk, causing the above mentioned reflection features, the X-ray burst photons can  impact the disk in the following ways:

\begin{enumerate}
\item Launch a radiatively-driven mass outflow (see also Section~\ref{subsec:theory:diskmech}). Although a thermally-driven disk wind should be ineffective in the inner disk region due to the strong gravity of the NS, the case of a radiatively-driven outflow from a disk during an X-ray burst was investigated by \citet{ballantyne2005}. The computations suggested that the mass-outflow rate through the wind did not exceed the accretion rate through the disk, therefore the inner $\approx 100~r_g$ ($\approx 220$~km) of the disk around 4U 1820--30 could not be emptied simply by a wind. Moreover, significant column densities could only be driven for super-Eddington luminosities, which were only obtained at the beginning of the X-ray burst. 

\item Increase the mass inflow as a result of radiation drag. The intense burst emission can cause a radiation drag, the so-called Poynting-Robertson effect, that could drain the inner disk onto the NS.\footnote{Although this affects the electrons, the general assumption is that Coulomb-coupling is efficient enough that the material can be treated as a single fluid, which is valid expect for very tenuous accretion flows \citep[][]{walker1992}.} The competing mechanisms of radiation pressure and Poynting-Robertson drag on disk particles were simulated by \citet{stahl2013} and \citet{mishra2014}. These studies demonstrated that the influence of both processes decreases with increasing distance from the NS, but that radiation drag does so faster. Therefore, the minimum burst luminosity that is required to eject particles will increase with distance so that the net effect on the accretion rate depends on the position of the inner disk radius. This seems to be qualitatively consistent with the observations: during soft states the disk is thought to be close in, hence an enhancement in the persistent emission may be expected, even for bright X-ray bursts. However, the timescale for this effect is very fast \citep[][]{ballantyne2005}, so this mechanism can not explain on its own why the reflecting region observed in the 4U 1820--30 burst evolved out to $\approx 100~r_g$. During hard states, on the other hand, the disk is thought to be receded and radiation pressure may win the tug of war when an X-ray burst is intense enough, hence a reduction of the persistent emission would be expected. This can be tested with X-ray reflection observations of (long) hard-state X-ray bursts \citep[e.g.][]{degenaar2013_igrj1706,keek2016}. We note that this mechanism can operate both on (optically thick) disks and (optically thin) coronae, but possibly with different efficiencies.

\item A change in disk structure due to X-ray heating. The intense burst emission also heats the disk, which could increase its scale height and change its surface density. These effects should typically occur on the viscous timescale. For reasonable parameters, this timescale is in agreement with the $\approx 1000$~s variation in the reflecting region observed in the 4U 1820--30 superburst \citep{ballantyne2005}. This simple estimate does imply that such a large change in the reflecting region might only be observed during superbursts or during intermediate-duration bursts from low-luminosity bursters (which can be observed up to hours due to the very low accretion flux). 
\end{enumerate}

Ultimately, all three of these processes (outflow, inflow and heating) will be operating on an accretion disk during an X-ray burst, and it is unlikely that a single one dominates over the others \citep[see also][for earlier discussions]{fukue1983,walker1989,walker1992,fukue1995}. Also, the effect on the corona cannot be fully decoupled: heating and outflows may truncate the disk, which can change the accretion rate and in turn affect both the disk and the corona. We note that \citet{degenaar2015_burst} considered whether cooling and condensation of coronal gas onto the disk could account for the enhanced mass-accretion rate inferred for the \nustar-detected hard state X-ray burst of 4U 1608--52. However, their simple estimates suggested that there was not sufficient material in the corona to account for that. Numerical simulations are required to fully untangle the physical interaction between the X-ray burst and the surrounding accretion flow. 

Additional observations of X-ray reflection features during X-ray bursts are also needed to further understand the range of interactions and phenomena that are possible. The reflection signatures should be commonly observed in the spectra of X-ray bursts if enough counts are obtained. As noted by \citet{zand2010}, the grating instruments on board of \textit{Chandra} and \textit{XMM-Newton} in principle have the required spectral resolution and sensitivity to achieve this, although it would require catching a (rare) long burst, or otherwise being able to stack many X-ray bursts. Fortunately, there are several new and concept X-ray detectors that can provide the data necessary to unravel the physics of the interaction between X-ray bursts and disks \citep[][see also Section~\ref{sec:future}]{keek2016_future}.

\subsubsection{Mechanical impact on the accretion disk}\label{subsec:theory:diskmech}
In addition to the above described radiative processes, the mechanical interaction between a radiatively-driven outflow or an expanding shell can possibly also disrupt the accretion disk. \citet{ballantyne2005} found that for the superburst from 4U 1820--30 the radiatively-driven wind carried insufficient kinetic energy to overcome the potential energy of the wind. This argument is likely true for the majority of X-ray bursts. However, it was pointed out by \citet{zand2011} that in case of superexpansion X-ray bursts, a nova-like shell is ejected that carries much more density than the radiatively-driven winds considered previously. 

\citet{zand2011} investigated the impact on the disk for a superexpansion X-ray burst from the UCXB 2S 0918--549 detected with \rxte/PCA. This X-ray burst had a total duration of a few minutes and featured a brief (40 ms) precursor as well as large-amplitude variations (staring $\approx$120~s after the X-ray burst onset and lasting for $\approx$60~s). Making basic assumptions for the  shell ejected by the X-ray burst and the accretion disk in this LMXB, \citet{zand2011} indeed found that the expanding gas shell likely had sufficient momentum to sweep up the inner part of the disk, out to a few hundred km, and ablate the disk surface over much larger distances of $\sim 10^{3}-10^{4}$~km. Since the donor star continues to feed the accretion disk, once the effect of the super-Eddington radiation subsides, the disk should settle back at its original thin structure and extend down to the vicinity of the NS again.

Based on the analysis of the 2S 0918--549 X-ray burst, \citet{zand2011} proposed that the large-amplitude variations seen in a number of superexpansion X-ray bursts (see Section~\ref{subsec:superexpansion}) are due to the resettlement of the accretion disk after it is swept up and puffed up by the expanding shell and radiation pressure. The total duration of the variability phase may then be related to the radial extent over which the obscuring gas is located when it starts to fall back. This idea is supported by the apparent correlation between the duration of the Eddington-limited phase of the X-ray burst and the total time over which the variability is observed \citep[i.e. the longer the Eddington-phase lasts, the further out disk material can be pushed;][]{zand2011}.  

\subsection{The influence of X-ray bursts on coronae}\label{subsec:theory:corona}

\subsubsection{Interpretation of the hard X-ray deficits during X-ray bursts}\label{subsec:theory:hardXshortage}
As described in Section~\ref{subsec:corona}, there is vast evidence that X-ray bursts affect the hard ($\approx$30--50~keV) X-ray emission component in NS LMXBs, which is usually ascribed to the corona. Summarizing, the observational facts are:

\begin{itemize}
\item The hard X-ray deficit during X-ray bursts is dependent on the spectral state; the suppression of hard X-rays is only seen during hard states and not during soft states.

\item Most NS LMXBs in their hard spectral states show suppression of the hard X-ray emission during X-ray bursts, but in a few sources such a hard X-ray deficit is not seen despite being in the hard state.

\item The hard X-ray deficits are less strong during the hard state at the end of an outburst (i.e., when a source is transitioning to quiescence) than during the hard state at the beginning of an outburst (when the deficits are seen during X-ray bursts), even if the hard X-ray fluxes are similar.
\end{itemize}

The hard X-ray deficits during X-ray bursts can in principle understood as Compton cooling of the hot corona by the soft X-ray shower of an X-ray burst. There are a few requirements to achieve this. Firstly, the corona should have a relatively large opening angle with respect to the NS, so that the hot corona receives a sufficient amount of soft X-ray photons to cause cooling. There are several factors that determine whether this condition is met. One is the location of the corona, which can be either along the surface of the accretion disk, or concentrated in the inner part of the disk (see Figure~\ref{fig:corona_geometry}). A corona of the latter geometry would be cool more easily than the former, even if the distance to the NS is similar. 

The fact that in the decaying phase of the outburst the observed hard X-ray deficit is less strong, and that for some NSs the hard X-ray deficits are not observed at all, can then possibly be explained in terms of disk evaporation (see also Section~\ref{subsubsec:correform}). In this model there is no inner disk during the initial hard state, when the source is coming out of quiescence, so that the corona subtends a large solid angle with respect to the NS. Once a soft state is entered, the disk extends close to the NS. When the system then transitions into a hard state in the decaying phase of the outburst, the disk may possibly be extending further in than during the initial hard state, so that the corona forms (partly) along the disk. In such a configuration, the corona would have a large enough disk photon supply to produce relatively strong hard X-rays. However, if the corona is now (in part) stretched out along the disk, it has relatively smaller opening angle to the NS than during the initial hard state. The corona may therefore be cooled to a lesser extend by an X-ray burst. In this picture the corona is thus cooled more efficiently during the initial hard state, causing a strong reduction in the hard X-rays, than during the later later state (i.e. causing a less strong reduction).

An alternative explanation could involve the distance to the NS. Even if the corona is always located in the inner region of the disk, the inner disk radius may be further away from the NS in the initial hard state than during the hard state at the end of an outburst. When the inner disk is further away from the NS, the number of soft photons injected in the corona by the X-ray burst is smaller due to the relatively smaller opening angle of the corona with respect to the NS. This would then cause different amplitudes for the hard X-ray deficits in the initial and later (i.e. post-soft state) hard spectral states. Nevertheless, in this case some hard X-ray deficit during the later hard state is still expected. It is also possible that the corona has a large radial extent (e.g. when it is the jet) and only part of the corona facing the NS is cooled off, while the remaining part faced to the disk, where the hard X-ray emission is observed, is influenced to a lesser extent. Regardless of the exact underlying mechanism, the spectral state dependence of the hard X-ray deficits may simply be due to the fact that the hard X-ray emission is generally much weaker during the soft state.

\subsubsection{Lessons from X-ray bursts on coronal (re)formation}\label{subsubsec:correform}
One idea that explains the formation and evolution of the corona in a BH LMXB or AGN is that the mass accretion rate $\dot{M}$ (in units of the Eddington ratio) drives the variations of the complex accretion flows by the interaction between the cold standard disk and the hot corona \citep[e.g.][]{Liu1999,Liu2002,Liu2007,Liu2011,Meyer2000a,Meyer2000}. Specifically, the coupling between the hot corona and the cold disk leads to mass exchange between these two components of the accretion flow. The gas in the thin disk is heated up and evaporates into the corona as a consequence of thermal conduction from the hot corona, or the coronal gas condenses into the disk as a result of overcooling by, for example, external inverse Compton scattering. If $\dot{M}$ is low, evaporation occurs and can completely remove the thin disk, leaving only the hot corona in the inner region and a truncated thin disk in outer region. If $\dot{M}$ is high, the gas in the corona partially condenses to the disk due to strong Compton cooling, resulting in disk-dominated accretion. This model naturally explains the different structures of accretion flow in different spectral states \citep[e.g. ][]{Liu1999,Liu2002,Liu2006,Liu2007,Liu2011,Meyer2007,Meyer2000,Taam2008,Qiao2009,Qiao2012}.

It remains challenging to determine the structure of the corona for stellar-mass and supermassive accreting BHs from observations. The potential of turning to NS LMXBs is demonstrated by a recent study of the X-ray bursts from IGR~J17473--2721. It was found that the X-ray burst emission was well-described by a black body, suggesting that the X-ray burst emission itself experiences negligible Comptonization. Furthermore, it was shown that the corona cools quickly during the rising phase of X-ray bursts, while it is heated rapidly again during the decaying phase of an X-ray burst \citep[][]{chen2012_xrbs_igrj1747}. These results suggest that the corona cannot cover the central compact object completely and that the (partial) destruction and formation time scales of the corona are as short as seconds. Such a quick evolution is quite difficult to understand in the above discussed evaporation model, where the time scales involved are related to the viscous time scale of the accretion disk. Perhaps the corona  is not destroyed and just cools quickly but has no time to condensate into the disk before it heats up again when the X-ray burst radiation drops. Alternatively, this short time scale is consistent with a corona produced by magnetic reconnections in the accretion disk, in a similar way as the Solar corona is heated \citep{Zhang2000}.

\subsubsection{Simulations of Compton cooling due to an X-ray burst}\label{subsec:theory:comptoncool}
As described above, Compton cooling of a corona by the injection of soft X-ray burst photons can possibly explain the hard X-ray deficits and/or spectral softening observed during some X-ray bursts. To test whether this is indeed a viable explanation, we here derive simple estimates of this effect.

Let us first consider the energetics of the persistent emission. The total accretion luminosity is:
\begin{equation}
L=GM_{\star}\dot{M}/R_{\star}
\end{equation}
where $M_{\star}$ is the mass of the NS and $R_{\star}$ is its radius.
This total luminosity is the sum of three components:
\begin{equation}
L=L_\mathrm{d}+L_\mathrm{c}+L_\mathrm{ns}
\end{equation}
where,
\begin{equation}
L_\mathrm{d}=(1-f)L/2
\end{equation}
is the disk luminosity, with $f$ the fraction of the accretion luminosity going to the corona. Furthermore,
\begin{equation}
L_\mathrm{c}=(1-f_\mathrm{c}) fL/2
\end{equation}
is the power fed to electrons of the coronal plasma and $f_\mathrm{c}$ is the fraction of coronal accretion power that is not radiated but advected onto the NS. The rest of the power is dissipated at the boundary/spreading layer where the accreting material is stopped by the surface of the NS. Finally,
\begin{equation}
L_\mathrm{ns}=L-L_\mathrm{d}-L_\mathrm{c}=L\left(1+f_\mathrm{c}f\right)/2
\end{equation}
is the luminosity of the NS surface and/or boundary/spreading layer.

Depending on the geometry, a fraction $f_\mathrm{d}$ and $f_\mathrm{ns}$ respectively of the disk and NS luminosity intercepts the corona. In addition, the disk is illuminated by the coronal X-rays and the radiation impinging on the disk is reprocessed and thermally re-emitted (we neglect Compton reflection). The reprocessed disk luminosity can be written as:
\begin{equation}
L_\mathrm{rep}=\alpha \left[L_\mathrm{s}(1-e^{-\tau})+L_\mathrm{c}\right]
\end{equation} 
 where $L_\mathrm{s}$ is the soft photon luminosity entering the corona and $\alpha$ is the fraction of the coronal luminosity illuminating the disk. The latter is directly related to the amplitude of the reflection component $R\simeq2\alpha$.  The Thomson depth of the corona $\tau$ can be measured from spectral fits of the X-ray spectrum with Comptonization models. 

When accreting but not X-ray bursting, the soft photon luminosity entering the corona is therefore:
\begin{equation}
L_{\mathrm{s}0}=f_\mathrm{d}L_\mathrm{d}+f_\mathrm{d}L_\mathrm{rep}+f_\mathrm{ns}L_\mathrm{ns}
\end{equation}
or equivalently:
\begin{equation}
L_{\mathrm{s}0}=\frac{L}{2}\frac{f_\mathrm{d}+f_\mathrm{ns}+f\left[f_\mathrm{ns}f_\mathrm{a}-f_\mathrm{d}\left(1-\alpha+\alpha f_\mathrm{a}\right)\right]}{1-\alpha f_\mathrm{d}(1-e^{-\tau})}
\end{equation}

During X-ray burst phases, however, the soft photon flux entering the corona increases with the instantaneous X-ray burst luminosity $L_\mathrm{b}$ as:
\begin{equation}
L_\mathrm{s}=L_{\mathrm{s}0}+K_\mathrm{b} L_\mathrm{b}
\label{eq:ls}
\end{equation}
where 
\begin{equation}
K_\mathrm{b}=\frac{f_\mathrm{b}}{1-\alpha f_\mathrm{d}(1-e^{-\tau})}
\end{equation}
 and $f_\mathrm{b}$ is the fraction of the X-ray burst luminosity that intercepts the corona. 
The temperature of the electrons in the corona can be estimated from Compton cooling equilibrium \citep[using equations 13 and 14 of][for stellar-mass systems]{beloborov1999}:

\begin{equation}
kT_\mathrm{e}\sim \frac{m_\mathrm{e} c^2}{4 (\tau+\tau^2)}\left(\frac{L_\mathrm{c}}{L_\mathrm{s}}\right)^{4/3}
\label{eq:kte}
\end{equation}

Assuming that the coronal power $L_\mathrm{c}$ remains constant, Equations~\ref{eq:ls} and~\ref{eq:kte} show that the coronal temperature will decrease with the X-ray burst luminosity. The X-ray burst will have a significant effect on the energetics of the corona as soon as $K_\mathrm{b}L_\mathrm{b}$, the contribution of the X-ray burst to the coronal soft cooling, becomes comparable to  $L_{\mathrm{s}0}$. We therefore expect to see a significant softening of the coronal X-ray spectrum when the X-ray burst luminosity becomes a significant fraction of the critical luminosity $L_\mathrm{b,crit}= L_{\mathrm{s}0}/K_\mathrm{b}$. This critical luminosity can be written as:
\begin{equation}
L_\mathrm{b, crit}= \frac{f_\mathrm{d}+f_\mathrm{ns}+f\left[f_\mathrm{ns}f_\mathrm{a}-f_\mathrm{d}\left(1-\alpha+\alpha f_\mathrm{a}\right)\right]}{2f_\mathrm{b}} \,L
\label{eq:lcrit}
\end{equation}

Based on this basic mathematical framework, we can start to interpret the observations. It is of note that Equation~\ref{eq:lcrit} shows  that  the critical luminosity $L_\mathrm{b, crit}$ depends on the geometry of the system.  Some of the parameters can be constrained by the observations of the accretion emission outside X-ray bursts. The ratio $L_\mathrm{c}/L_{\mathrm{s}0}=L_\mathrm{c}/K_\mathrm{b}L_{\mathrm{b,crit}}$ is directly related to the slope of the hard X-ray non-thermal emission and can be inferred from spectral fits with the Comptonization model \textsc{eqpair} \citep[][]{coppi1992}, for instance, which takes into account the energy balance in the corona. This adds further constraints on the other parameters. In addition, one can assume a specific geometry and estimate the parameters expected for this specific geometrical model. 

Let us consider a model where the inner disk is truncated (because it is replaced by a hot flow), which is the typically assumed geometry of hard states in LMXBs. In this geometry the solid angle subtended by the corona as seen for the disk is small, we assume  $f_\mathrm{d}=0.05$ and that reprocessing is not important; we set $\alpha\simeq 0.1$ \citep[based on reflection measurements; see e.g.][]{bob00,zdz03}. The corona/hot flow is geometrically thick. We assume that the hot flow has a constant aspect ratio $\epsilon=h/r=0.2$ and extends from the NS surface to the inner disk radius, which we set at $R_\mathrm{in}=10~R_{\mathrm{\star}}$ (where $R_{\mathrm{\star}}$ is the NS radius). Therefore, $f=1-R_{\mathrm{\star}}/R_\mathrm{in}=0.9$. We also assume that the boundary/spreading layer and the X-ray burst can be considered as isotropic point sources: $f_\mathrm{ns} =f_\mathrm{b}\simeq \epsilon/\sqrt{1+\epsilon^2}\simeq 0.2$. The advection parameter is set to an intermediate value of $f_\mathrm{c}=0.2$, and we use $\tau=1.5$. 
For these parameters, $L_\mathrm{b,crit}\simeq 0.6 L$.  More generally,  $L_\mathrm{b,crit}$ is not sensitive to the parameters and remains comparable to $L$ (within a factor of a few) as long as $f_\mathrm{ns} \sim f_\mathrm{b} \geq f_\mathrm{d}$. {\it Therefore it is a prediction of the truncated disk model that the corona should start to respond to the X-ray burst as soon as the X-ray burst luminosity represents a sizable fraction of the accretion luminosity.} 

Keeping the same parameters and geometry, we can use the \textsc{eqpair} model to simulate the spectral evolution of the system during an X-ray burst. For simplicity we assume that both the disk emission and the NS surface emissions are all emitted with a unique black-body spectrum of temperature $T_\mathrm{BB}$.
The total thermal emission of the system, $L_\mathrm{BB}$, varies linearly with the X-ray burst luminosity $L_\mathrm{b}$:
\begin{equation}
L_\mathrm{BB}=L_\mathrm{d}+L_\mathrm{ns}+L_\mathrm{b}+L_\mathrm{rep}=L_\mathrm{BB0}+K_\mathrm{BB} L_\mathrm{b}
\end{equation}

The constants $L_{\mathrm{BB}0}$ and $K_\mathrm{BB}$ can be easily expressed as a complicated function of the geometrical parameters.
We assume that the accretion luminosity and the geometry of the system do not change during the whole duration of the outburst (radius expansion is neglected). The black-body temperature therefore varies as $T_\mathrm{BB}= T_{\mathrm{BB}0} \left(L_\mathrm{BB}/L_{\mathrm{BB}0}\right)^{1/4}$, where the pre-burst temperature is set to $T_\mathrm{BB0}$=0.3 keV.  

\begin{figure}
 \centering
 \includegraphics[width=0.7\columnwidth]{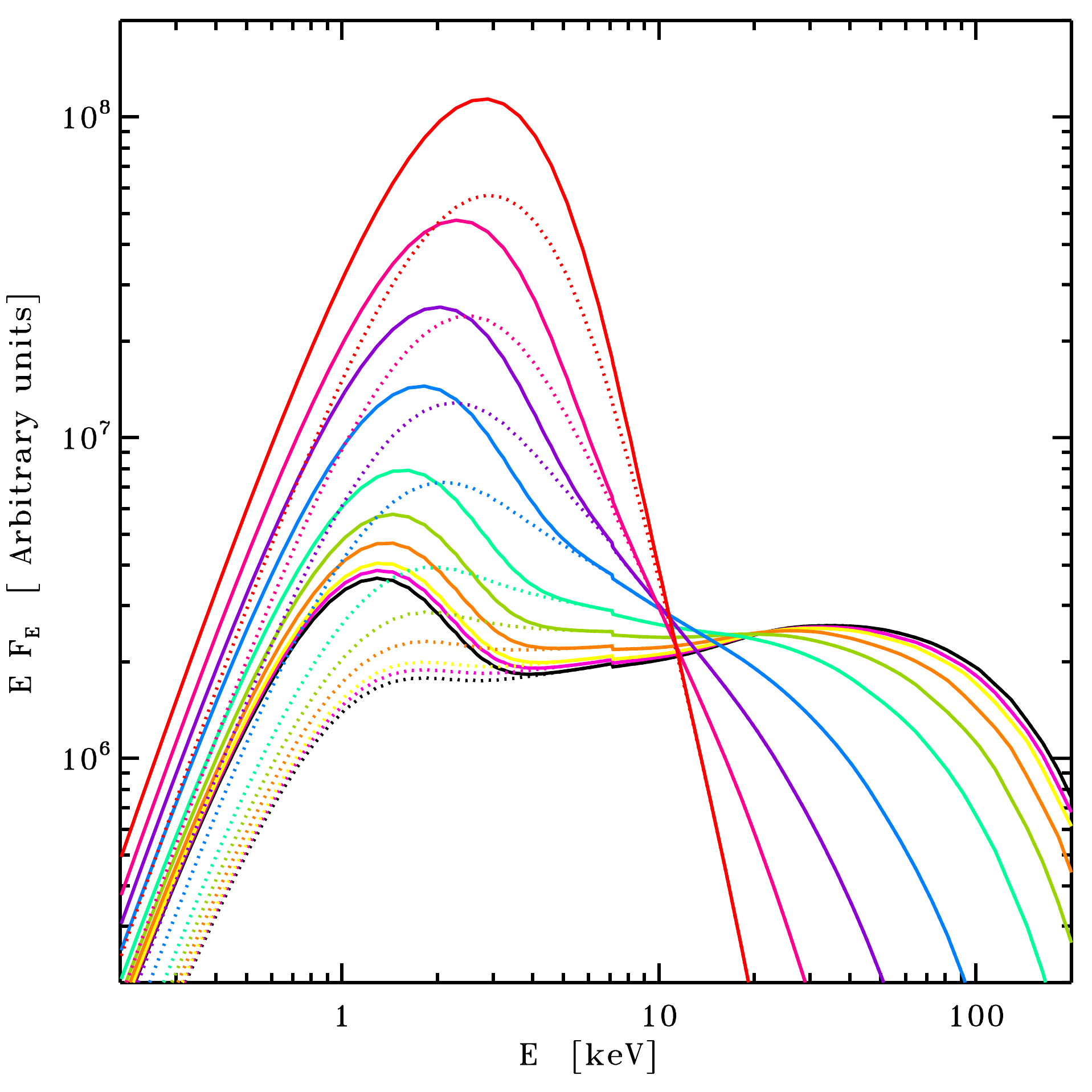}
 \caption{Basic simulations of the spectral evolution and coronal cooling as the X-ray burst luminosity increases, using the \textsc{eqpair} model with the range of parameters listed in Table~\ref{tab:parsim}. The dotted curves show the \textsc{eqpair} Comptonization model, the full curves also include the thermal emission which does not cross the corona. The different colors reflect the series of spectra resulting from varying the burst to accretion luminosity ratio $L_\mathrm{b}/L$ between 0 and 20 (as listed in the first column of Table~\ref{tab:parsim}). These simulations show that as $L_\mathrm{b}$ increases, the corona gradually cools down and the X-ray spectrum becomes increasingly softer while the high-energy cutoff decreases, hence causing a strong suppression of the hard ($\gtrsim$30~keV) X-ray emission.}
 \label{fig:spectra}
\end{figure}

For a given ratio of the X-ray burst to accretion luminosities, $L_\mathrm{b}/L$, we estimate the soft photon temperature $T_\mathrm{BB}$, the soft coronal luminosity $L_\mathrm{s}$ and the compactness ratio $L_\mathrm{c}/L_\mathrm{s}$, which are used as input parameters for the \textsc{eqpair} model. The \textsc{eqpair} code calculates the resulting equilibrium temperature of the electron plasma assuming a purely Maxwellian electron energy distribution, as well as the resulting spectrum (shown in Figure~\ref{fig:spectra}). We also add a second thermal component of luminosity $L_\mathrm{BB}-L_\mathrm{s}$ corresponding approximately to the thermal emission that does not intersect the corona and is observed directly. The \textsc{eqpair} model also includes neutral disk reflection; the reflection amplitude was set to $2\alpha$. Figure~\ref{fig:spectra} shows the resulting sequence of spectra when $L_\mathrm{b}$ is varied between 0 and 20 times the accretion luminosity. The exact values of $L_\mathrm{b}/L$ used for each simulation are given in Table~\ref{tab:parsim}, together with the corresponding temperatures of the thermal and Comptonized components and the resulting photon index in the 5 to 30 keV band. {\it Our simulations show that as $L_\mathrm{b}$ increases the coronal electrons are gradually cooled down, the X-ray spectrum becomes softer and softer, while the high energy cutoff decreases and causes a dramatic suppression of the hard X-ray emission.} At the brightest X-ray burst luminosities the emission is essentially thermal. This qualitatively agrees with the behavior of the persistent emission during X-ray bursts (e.g. a softening of the accretion spectrum and a reduction of the hard X-ray flux; Sections~\ref{subsec:change} and~\ref{subsec:corona}).
 
\begin{table*}
\caption{Spectral parameters of the simulated \textsc{eqpair} spectra shown in Figure~\ref{fig:spectra}.\label{tab:parsim}} 
\begin{threeparttable}
\begin{tabular*}{1.0\textwidth}{@{\extracolsep{\fill}}ccccccc}
\hline\hline
 $L_{b}/L$  & $L_\mathrm{BB}/L$ & $L_\mathrm{s}/L$ & $L_\mathrm{c}/L_\mathrm{s}$ & $kT_\mathrm{BB}$  (keV) & $\Gamma_{5-30}$ &  $kT_\mathrm{e}$ (keV) \\
\hline
      0.000 &       0.731 &      0.125 &       2.88 & 
      0.300 &       1.78 &       50.1 \\
    0.040 &       0.773 &      0.133 &       2.70 & 
      0.304 &       1.82 &       48.7 \\
    0.080 &       0.814 &      0.141 &       2.55& 
      0.308 &       1.85 &       47.4 \\
     0.200 &       0.938 &      0.165 &       2.18 & 
      0.319 &       1.91 &       44.1 \\
     0.400 &        1.14 &      0.206 &       1.75 & 
      0.335 &       2.02 &       39.6 \\
     0.800 &        1.56 &      0.286 &       1.26 & 
      0.361 &       2.20 &       33.5 \\
      2.00 &        2.79 &      0.528 &      0.682 & 
      0.417 &       2.66 &       23.7 \\
      4.00 &        4.86 &      0.931 &      0.387 & 
      0.478 &       3.32 &       16.4 \\
      8.00 &        8.98 &       1.74 &      0.207 & 
      0.558 &       4.41 &       10.5 \\
      20.0 &        21.4 &       4.16 &     0.087 & 
      0.692 &       6.56 &       5.38 \\
\hline
\end{tabular*}
\end{threeparttable}
\end{table*}


\section{Future prospects for X-ray burst/accretion interactions studies}\label{sec:future}
Having reviewed the current observational evidence for X-ray burst/accretion flow interaction and the current theoretical framework for their interpretation, we now explore the future direction of this emerging research field. The observable phenomena to study the effect of X-ray bursts on accretion flows as discussed in the present work fall into four categories: i) changes in the persistent flux/spectrum at energies $<$30~keV, ii) reduction of the hard ($>$30~keV) X-ray emission, iii) evolving reflection features and iv) changes in the rapid time variability. The missions that have been used most for these particular kind of studies up till now are \rxte, \inte, \swift, and \nustar. However, within the next decade several new tools are available: The recently (2015) launched mission \astrosat\ (see Section~\ref{subsec:astrosat}) has open observing time from 2017 onward, and in 2017 both \hxmt\ (Section~\ref{subsubsec:hmxt}) and \nicer\ (see Section~\ref{subsubsec:nicer}) were successfully deployed.\footnote{We note that further in the future, currently planned for the late 2020s, \athena\ will be launched and there are plans for a \hitomi\ recovery mission (named the X-ray Astronomy Recovery Mission, {\it XARM}). However, as demonstrated by \citet{zand2015_loft} and \citet{keek2016_future}, these two missions are not expected to make a particularly large contribution to X-ray burst studies. Therefore, we do not discuss these missions here any further.} In addition, the mission concepts \extp\ (Section~\ref{subsubsec:extp}) and \textit{STROBE-X} (based on the design of \loft; Section~\ref{subsubsec:loft}) are in development.

We here discuss how this new and concept missions can contribute to further develop studies of the impact of X-ray bursts on accretion flows. \citet{keek2016_future} present a thorough study of the capabilities of various missions (\loft\ and \nicer, as well as \athena\ and \hitomi) to study reflection features during X-ray bursts (i.e. category iii listed above). Furthermore, \citet{peille2014} perform simulations for studying kHz QPOs during X-ray bursts (category iv listed above) with the \loft\ concept, whereas \citet{zand2015_loft} lay out the general importance of a \loft-like mission (such as \textit{STROBE-X} concept) for X-ray burst science. We briefly summarize the main results here, but do not repeat or expand those studies. We have rather chosen to put emphasis on instruments that have not gained a lot of attention from the perspective of X-ray burst studies yet: \astrosat, \hxmt\ and \extp. Furthermore, we focus on the interaction categories i and ii listed above (i.e. studying changes in the persistent emission during X-ray bursts and in particular the response of the hard X-rays), since these have not been discussed yet in the context of new X-ray missions. In Table~\ref{tab:missions} we list the main characteristics of several new and concept missions in comparison with previous/current satellites used for studying X-ray burst/accretion flow interactions (see Section~\ref{subsec:currentmissions}). In Table~\ref{tab:prospects} we summarize the expectations of the new missions with respect to the different types of interactions that can be studied (as reviewed in Sections~\ref{sec:obs} and~\ref{sec:theory}). 

Following the discussion of new X-ray instruments, we also point out how coordinated multi-wavelength observations can yield more insight into the interaction between X-ray bursts and disks/coronae as well as jets, how analysis techniques can be further developed, and which specific targets might be most promising for future studies (Section~\ref{subsec:studies}).

\begin{table*}
\caption{Relevant properties of various missions for X-ray burst studies discussed in this work.\label{tab:missions}}
\begin{threeparttable}
\begin{tabular*}{1.0\textwidth}{@{\extracolsep{\fill}}rccc}
\hline
Mission (instrument) & energy range (keV) & $A_{\mathrm{eff}}$ 7 keV (cm$^2$) & $A_{\mathrm{eff}}$ 30 keV (cm$^2$) \\
\hline
\rxte\ (PCA) & 2--60 & $\approx$5\,000 & $\approx$450 \\
\inte\ (JEM-X)  & 3--35 & $\approx$200 & $\approx$60 \\
 (IBIS/ISGRI)  & 15--10\,000 & - & $\approx$800 \\
\swift\ (XRT) & 0.5--10 & $\approx$50 & - \\
 (BAT) & 15--150 & - & $\approx$2000 \\
\nustar\ (FPMA/B) & 3--79 & $\approx$900 & $\approx$300 \\
\hline
\astrosat (SXT)  & 0.3--8 & $\approx$50 & - \\
(LAXPC)  &  3--80 & $\approx$5\,500 & $\approx$5\,500 \\
 (CZTI)  & 10--50 & - & $\approx$400 \\
\nicer\ (XTI) & 0.2--10 & $\approx$300 & - \\
\hxmt\ (LE) & 1--15 & $\approx$200 & - \\
(ME) & 5--30 & $\approx$200 & $\approx$200 \\
(HE) & 20--250 & - & $\approx$4\,500 \\
\extp\ (SFA)  & 0.5--20 & $\approx$6\,000 & - \\
(LAD) & 2--30 & $\approx$40\,000 & $\approx$2\,000 \\
{\it STROBE-X} & 0.5--30 & $\approx$80\,000 & $\approx$7\,000 \\
\hline
\end{tabular*}
\begin{tablenotes}
\item[]Note. -- $A_{\mathrm{eff}}$ gives the approximate maximum effective area (for all detectors combined). For comparison we also list the characteristic of some recent and current missions that have been used most to study the impact of X-ray bursts on the accretion flow. We do not list \chan\ and \xmm, but note that the low-energy coverage and good sensitivity of these instruments combined with the high spectral resolution of their grating instruments can be valuable tools to study narrow absorption and emission features in the X-ray spectra of X-ray bursts \citep[see ][for a discussion; see also Section~\ref{subsec:specfeat}]{zand2010}.
\end{tablenotes}
\end{threeparttable}
\end{table*}

\begin{table*}
\caption{Summary of future prospects for X-ray bursts-accretion interaction studies.\label{tab:prospects}}
\begin{threeparttable}
\begin{tabular*}{1.0\textwidth}{@{\extracolsep{\fill}}lccccc}
\hline
\rule[-.3\baselineskip]{0pt}{1.5\baselineskip}%
Science case & \astrosat\ & \nicer\ & \textit{HXMT} & \extp\ & {\it STROBE-X} \\
\hline
Changing persistent emission & + & + & $\sim$ & + & + \\
Hard X-ray deficits & + & $-$ & + & + & + \\
Reflection features & $\sim$ & + & $\sim$ & + & + \\
Changing kHz QPOs & + & + & + & + & +  \\
\hline
\end{tabular*}
\begin{tablenotes}
\item[]Note. -- The different signs indicate the performance of a mission compared to previous/current missions to detect the X-ray burst effects listed in the first column: `$-$' $=$ not achievable, '$\sim$' = similar, `+' = expected to be better. 
\end{tablenotes}
\end{threeparttable}
\end{table*}

\subsection{New X-ray missions for X-ray burst studies}\label{subsec:newmissions}

\subsubsection{Recently launched missions: \astrosat}\label{subsec:astrosat}
India's first dedicated astronomy satellite, \astrosat\ \citep[][]{singh2014}, was successfully launched on 2015 September 28 and is operating according to expectations. It provides broad-band X-ray coverage via three instruments: the Soft X-ray Telescope \citep[SXT, 0.3--8 keV;][]{singh2016_sxt}, the Large Area Xenon Proportional Counters \citep[LAXPC, 3--80 keV;][]{yadav2016_laxpc}, and the Cadmium-Zinc-Telluride Imager \citep[CZTI, 10--150 keV;][]{vadawale2016_czti}. In addition, it carries an X-ray monitor, the Scanning Sky Monitor (SSM, 2--10 keV), and twin Ultraviolet Imaging Telescopes \citep[UVIT, 130--320~nm;][]{subramaniam2016_uvit} that provide far-UV to optical coverage using different narrow and broad-band filters.

The technical specifications of the LAXPC are comparable to that of \rxte's PCA (see Table~\ref{tab:missions}). Indeed, the LAXPCs were designed to allow high-resolution X-ray timing as to extend the rich legacy of \rxte. The \astrosat/LAXPC may therefore have similar value for X-ray burst studies as \rxte/PCA. Variations in the flux or spectral shape of the persistent emission (Sections~\ref{subsec:change} and~\ref{subsec:specfeat}) and hard X-ray deficits (Section~\ref{subsec:corona}) can thus likely be studied by stacking multiple X-ray bursts. The much higher effective area of LAXPC above 10~keV compared to PCA, will be particularly important to study coronae. Furthermore, as discussed by \citet{keek2016_future}, reflection features may be detectable with the LAXPC during very long and energetic X-ray bursts (superbursts and perhaps also intermediate-duration bursts; Section~\ref{subsec:specfeat}). Changes in the kHz QPO frequency (Section~\ref{subsec:qpo}) can possibly be detected for single X-ray bursts (Table~\ref{tab:prospects}).

Another interesting prospect of \astrosat, and advantage over other missions, is the ability to detect X-ray bursts simultaneous at soft and hard X-rays with the SXT, LAXPC, and the CZTI. The SXT has similar specifications as \swift's XRT (with a nominal timing resolution of 2.5~s; see Table~\ref{tab:missions}), and adds value by allowing the detection of X-ray burst photons below 3 keV (down to 0.3 keV). The CZTI might be regarded as a better version of \rxte's High Energy X-ray Timing Experiment (HEXTE) instrument, as it allows for better treatment of the X-ray background, provides spatial resolution, and is likely better suitable for timing analysis due to its lower deadtime than HEXTE. 

\astrosat\ also allows for simultaneous detection of an X-ray burst with three different optical and UV filters through the UVIT. X-ray bursts lead to optical/UV bursts that are likely due to reprocessing of the X-ray burst emission in the accretion flow \citep[e.g.][]{hackwell1979,lawrence1983,matsuoka1984,kong2000,hynes2006,mescheryakov2011}. Simultaneous measurements of the direct X-ray burst emission and its reprocessed optical/UV signature could be a powerful tool to understand the accretion geometry and disk structure (see Section~\ref{subsec:xopt}). Furthermore, being able to simultaneously observe an X-ray burst at X-ray and UV/optical wavelengths provides the exciting prospect of also studying the direct X-ray burst emission at longer wavelengths, especially during superexpansion when the peak of the X-ray burst emission completely drops out of the X-ray band towards longer wavelengths (see Section~\ref{subsec:superexpansion}). 

The UVOT on board \swift\ has detected the optical/UV counterparts of a few intermediate-duration bursts. An example of that is shown in the bottom panel of Figure~\ref{fig:lcvar} left, which shows that the brightness in the U filter increased by $> 1$~mag from $\approx 17$ to 15.7~mag (Vega system, central wavelength $\approx 3465$~\AA) in response to the energetic (superexpansion) X-ray burst from IGR J17062--6143. Also for another intermediate-duration burst, from 1RXH J173523.7--354013, \swift/UVOT detected an UV/optical counterpart with the broad-band white filter (1592--7532~\AA) that brightened from $\approx 21$ to 19 mag during the X-ray burst \citep[][]{degenaar2010_burst}. \astrosat's UVIT has a larger effective area than \swift's UVOT and has the advantage that it can observe in three rather than one optical/UV filter at the same time. It is therefore expected that \astrosat\ will be able to detect optical/UV counterparts of X-ray bursts too. The UVIT can reach a sensitivity of $\approx 18.2$~mag for a 1 hr integration time, which implies that it will be of value mostly for long and bright X-ray bursts or relatively nearby sources.

\subsubsection{Recently launched missions: \hxmt}\label{subsubsec:hmxt}
\noindent
The Hard X-ray Modulation Telescope (\hxmt) is a Chinese astronomical satellite that was designed to reconstruct images from data obtained in a scanning mode based on the Direct Demodulation Method \citep[DDM;][]{liwu1994}. The DDM technique allows a celestial object to be imaged with a positional precision of a few arc-minutes. The concept of \hxmt\ was proposed $\sim 20$~years ago and eventually evolved into a mission covering an energy range of 1--250 keV. The \hxmt\ was successfully launched in 2017 July. The nominal mission life time is 4~years.

There are three main payloads on board \hxmt, all of which are slat-collimated instruments: the High Energy X-ray Telescope (HE), the Medium Energy X-ray Telescope (ME), and the Low Energy X-ray Telescope (LE). The HE consists of 18 NaI/CsI phoswich modules (main detectors) with a maximum effective area of 4\,500~\cms\ between $\approx 20-70$~keV, and covering a total energy range of 20--250 keV. The ME uses 1728 Si-PIN detectors that cover an energy range of 5--30 keV and provide a peak effective area of 900~\cms\ near 10 keV. The LE uses a swept charge device (SCD) as its detector, which is sensitive in 1--15 keV with an effective area that peaks at 300~\cms\ near 5 keV. The time resolution of HE, ME and LE is designed to be 25~$\mu$s, 1~ms, and 1~ms, respectively.

Using the DDM and scanning observations, \hxmt\ can obtain X-ray images with reasonable (few arcminutes) spatial resolution. The large detection areas of these telescopes also allow pointed observations with good statistics and good signal to noise ratio. Among the main scientific objectives of the \hxmt\ are to: i) scan the Galactic Plane to find new transient sources and to monitor the known variable sources, and ii) observe X-ray binaries to study the dynamics and emission mechanism in strong gravitational or magnetic fields. It is expected that \hxmt\ will discover a large number of new transient X-ray sources and allows for detailed studies of the temporal and spectral properties of accreting BH and NS LMXBs.\\

\noindent
{\it Probing X-ray burst/accretion flow interactions with \hxmt}\\
With the quoted specifications, \hxmt\ will be a promising new tool to probe the interaction between an X-ray burst and the surrounding disk/corona. In particular, \hxmt\ should allow to obtain a good characterization of the hard X-ray continuum  prior to an X-ray burst and enhance the statistics of hard X-ray deficits during X-ray bursts. It is therefore expected that this mission will increase the sample of X-ray bursts/sources that can be used to study the hard X-ray deficits during X-ray bursts and perhaps even to investigate how the persistent hard X-ray emission evolves during an X-ray burst (see Section~\ref{subsec:corona}, which could be compared to our basic simulations presented in Section~\ref{subsec:theory:comptoncool}). 

Here we take the study of \cite{chen2012_xrbs_igrj1747} on IGR J17473--2721 as an example to show what can be achieved with \hxmt\ with respect to \rxte\ (PCA). For our simulations we used the most up-to-date predicted \hxmt\ response files (released on 2016 May 2). Based on \citet{chen2012_xrbs_igrj1747}, we assumed an absorbed cutoff power-law spectral shape for the persistent emission. From \rxte\ data, the model parameter $E_\mathrm{cut}$ was measured to be 30--50 keV \citep[][]{chen2012_xrbs_igrj1747}, which should represent roughly twice the corona temperature. We assume that the X-ray burst light curve has a rapid rise and an exponential decay and that the corona temperature is anti-correlated with the X-ray burst strength. We consider a non-PRE burst with a constant emitting radius, a peak temperature of $kT_{\mathrm{bb}}=2.5$~keV, and a 2--10 keV peak flux of $F^{\mathrm{peak}}_{\mathrm{X}}=3.8\times10^{-8}~\flux$. Finally, in the simulations we used an exposure time of 1~s to produce light curves at 1-s resolution. These light curves were then rebinned to 16~s to illustrate the results. We take the template described above to recover the hard X-ray deficit as observed in IGR J17473--2721. 

\begin{figure}
\centering
\includegraphics[width=0.8\textwidth]{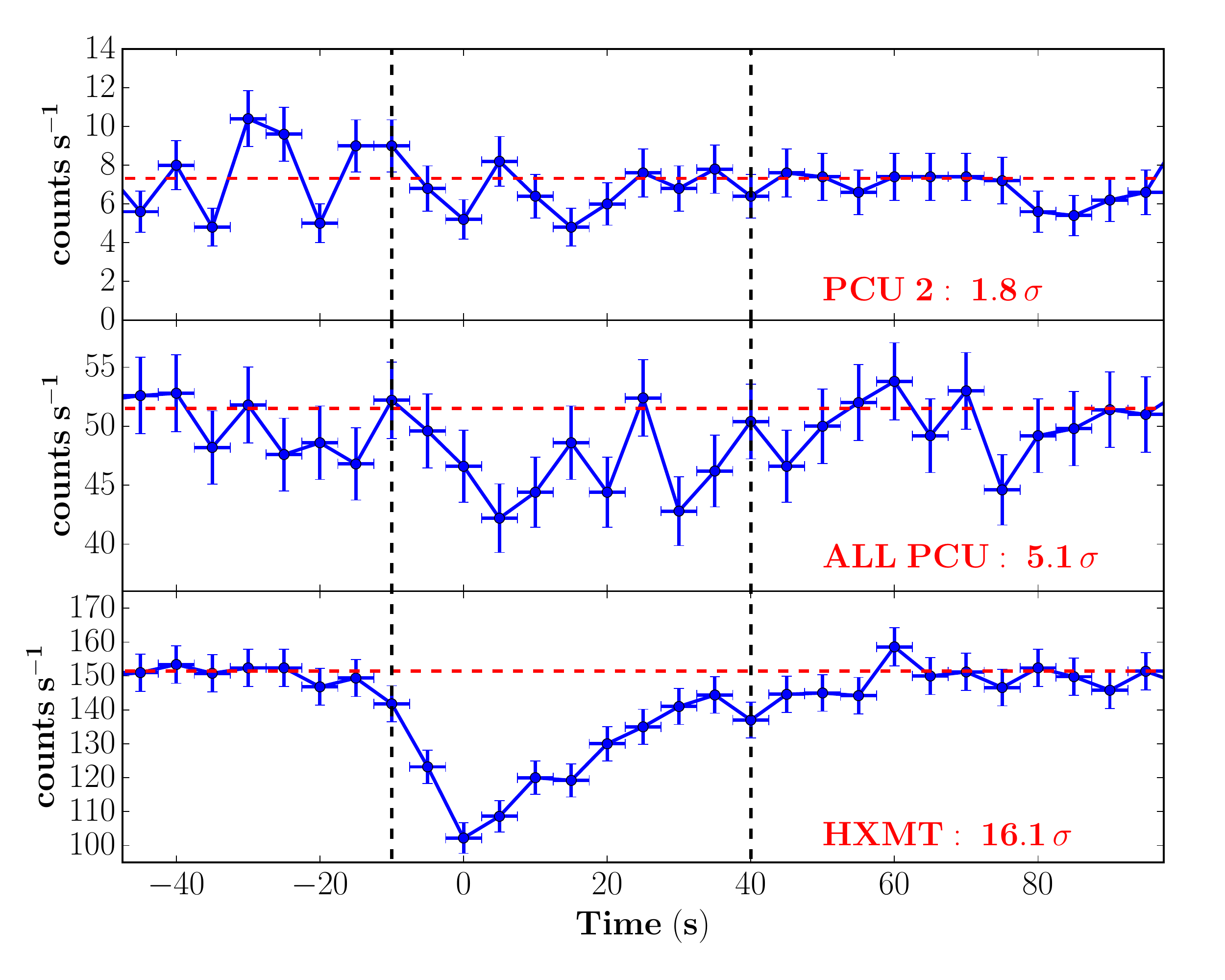}
\caption{Hard X-ray deficit as expected from one PCU of \rxte/PCA (top panel), all 5 PCUs of \rxte/PCA (middle panel) and \hxmt/HE (bottom panel), during a single burst event from IGR J17473--2721 \citep[based on data reported by][]{chen2012_xrbs_igrj1747}. Count rates refer to the 30--50 keV band. The averaged effective area of \hxmt/HE and \rxte/PCA (5 PCUs) are $\approx 3500$ and $1600~\mathrm{cm}^2$, respectively. After considering the spectral shape, the HE will receive a factor $\approx 2-3$ more photons than the PCA (not considering background).}
\label{fig:hardxshortage_hxmt}
\end{figure}

We next took these simulations as input to investigate how a hard X-ray deficit can be observed from \textit{individual} bursts by \hxmt\ and how this compares to what was possible with \rxte. The results are shown in Figure~\ref{fig:hardxshortage_hxmt}. We estimated the significance level of detecting a hard X-ray deficit by calculating the deficit area and its error within the black dashed lines in Figure~\ref{fig:hardxshortage_hxmt}. We find that a hard X-ray deficit during a single burst, presumably due to coronal cooling, can be detected at $\approx 16 \sigma$ significance with the \hxmt/HE. This is a large improvement with respect to the previous \rxte\ observation that resulted in only marginal evidence for a hard X-ray deficit \citep[][]{chen2012_xrbs_igrj1747}.

\begin{figure}
\centering
\includegraphics[width=0.7\textwidth]{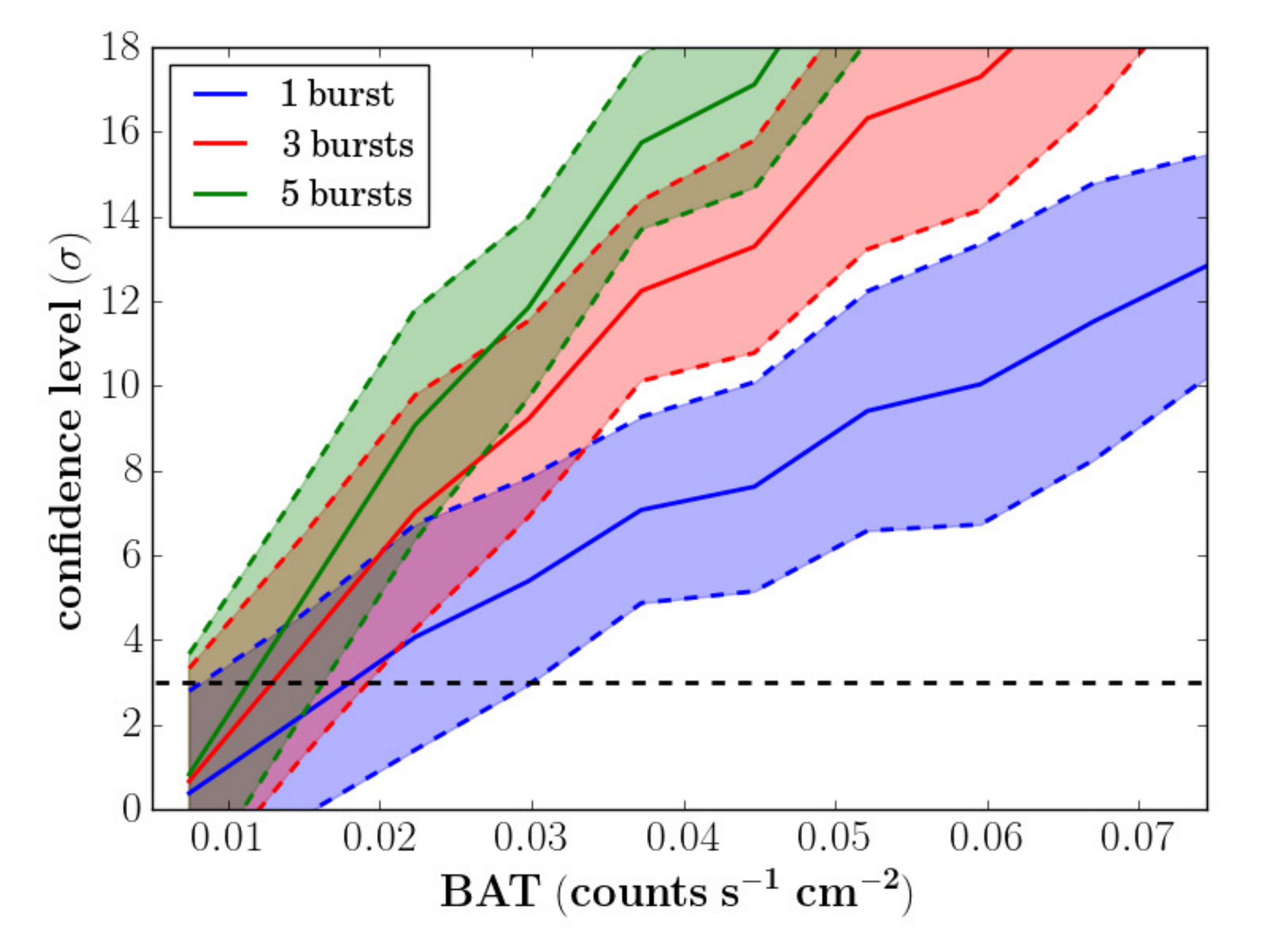}
\caption{Relation between the significance of a hard X-ray deficit achieved by \hxmt\ as a function of hard X-ray flux (represented by the \swift/BAT 15--150 keV count rate), and the number of bursts that are stacked (a single burst in blue, 3 bursts in red, and 5 bursts in green). As expected, the significance is positively correlated with the hard X-ray flux and a higher significance is also achieved as more bursts are stacked.}
\label{fig:hardxshortage_otherbursters}
\end{figure}

It is of note that the persistent 30--50 keV flux of IGR J17473--2721 is very strong compared to most other bursters. For a weaker corona (i.e. a lower hard X-ray flux), we would also need to stack bursts with \hxmt\ to investigate any possible hard X-ray deficit. This is illustrated in Figure~\ref{fig:hardxshortage_otherbursters}, where we used \swift/BAT (15--150 keV) count rates as an example to represent the hard X-ray flux. Obviously, as Figure~\ref{fig:hardxshortage_otherbursters} shows, a hard X-ray deficit is more clearly detected if more bursts are stacked or if the persistent hard X-ray emission is more luminous.

\subsubsection{Recently launched missions: \nicer}\label{subsubsec:nicer}
The NS Interior Composition ExploreR (\nicer) is a scientific payload that was successfully installed at the International Space Station in 2017 June \citep[][]{Gendreau:2012,Arzoumanian:2014}. Its prime scientific
goal is to achieve precise ($\pm$5\%) mass and radius measurements for a few selected NSs from detailed
measurements of X-ray pulsar lightcurves. In addition, the instrument will be used for a test of space
navigation using X-ray pulsars, under the label of Station Explorer for X-ray Timing and Navigation Technology 
\cite[\textit{SEXTANT};][]{Winternitz:2015}. The primary mission duration is 18 months, while 6 additional months are foreseen for a Guest Observer program.

\nicer\ covers the energy range from 0.2--12\,keV using concentrator optics (non-imaging, 5\,arcmin diameter FOV)
and silicon-drift detectors in order to achieve almost twice the effective area of \xmm\ around 1.5\,keV 
\citep[see figure~2 of][]{Arzoumanian:2014} with a similar energy resolution. The mission has
very strict requirements for absolute time-tagging ($<$300\,ns), which also require a built-in GPS receiver to 
determine its precise position with respect to the Solar System barycentre. Overall, \nicer\ is expected to
achieve a sensitivity $\approx 4\times$ better than \xmm\ and $\approx 30\times$ better than \rxte\ below 10~keV.

A key aspect of \nicer\ for burst studies is that it will be especially useful to allow for reflection modeling of selected sources with low absorption (see Sections~\ref{subsec:specfeat} and~\ref{subsec:theory:persenhance}), due to the high effective area at soft X-ray energies. The prospects of using \nicer\ for burst reflection studies were investigated and quantified by \citet{keek2016_future}. This shows that even with exposures of a few seconds, clear deviations between simulated reflection spectra and the black-body models commonly used in X-ray burst analysis can be detected \citep[see for instance figures~3, 4 and 5 in][]{keek2016_future}. Although \nicer\ does not cover hard X-rays, its high sensitivity at $<10$~keV will likely also facilitate detecting changes in the soft persistent X-ray flux and spectrum during bursts. Furthermore, its high timing resolution may allow for studying QPOs and how these are affected by the blast of an X-ray burst (Table~\ref{tab:prospects}), but we do not explore that here.

\subsubsection{Mission concepts: \extp}\label{subsubsec:extp}
\noindent
The enhanced X-ray Timing and Polarization (\extp) telescope is a next generation space-born X-ray telescope currently in development jointly in China and Europe \citep[][]{zhang2016_extp}. A launch in the early 2020s is currently envisioned. The exact specifications of the \extp\ mission are subject to change, but the design includes the following detectors: the large area spectroscopic focusing array (SFA), the large area collimated array detector (LAD), the Polarimetry Focusing array (PFA), and the Wide Field Monitor (WFM). The SFA and PFA are located at a central platform, while the LAD is set as two wings at each side of the satellite. The current baseline is that LAD has a maximum effective area of $\approx 3.5$~m$^2$ near 6 keV and covers a total energy band of 2--30 keV. The SFA works at 0.5--20 keV, and has 11 individual telescopes that add up to a peak effective area of $\approx$5\,000~\cms\ at 6 keV. Furthermore, the PFA has an effective area of $\approx$500~\cms\ at 3 keV and covers a total energy range of 2--10 keV. Finally, the WFM consists of a few pairs of coded-mask detectors, each covering a 90 by 90 degrees field of view, and will be employed to study X-ray variability in an energy range of 2--50 keV.\\

\noindent
{\it Probing X-ray bursts with eXTP}\\
Already through its WFM, \extp\ will play a major role in X-ray burst studies as it is expected to collect huge numbers of bursts, including rare phenomena such as intermediate-duration and superbursts \citep[the WFM of \extp\ is similar to that of the \loft\ design, so we refer to][to illustrate the impact of this]{zand2015_loft}. Furthermore, as shown by \citet{zhang2016_extp}, \extp\ will also be a powerful tool to detect and study QPOs as it provides more than an order of magnitude improvement in sensitivity over \rxte/PCA. It is therefore expected that \extp\ will also be of value to study possible changes in the QPO frequency or amplitude in response to X-ray bursts (see Section~\ref{subsec:qpo}). 

In the scope of the present work, the advantages of using \extp\ to investigate the X-ray burst/accretion flow interaction reside in its unprecedented large effective area, which is essential for improving the statistics for short bursts. Furthermore, the \extp/SFA is not very susceptible to pile-up and can therefore observe very bright sources (a factor $\gtrsim 10$ brighter than the Crab) without any issues. Therefore, we here explore simulations of how the persistent energy spectrum and the disk reflection component can be measured during X-ray bursts with \extp. 

\begin{figure}
\centering
\mbox{
\includegraphics[width=1.0\textwidth]{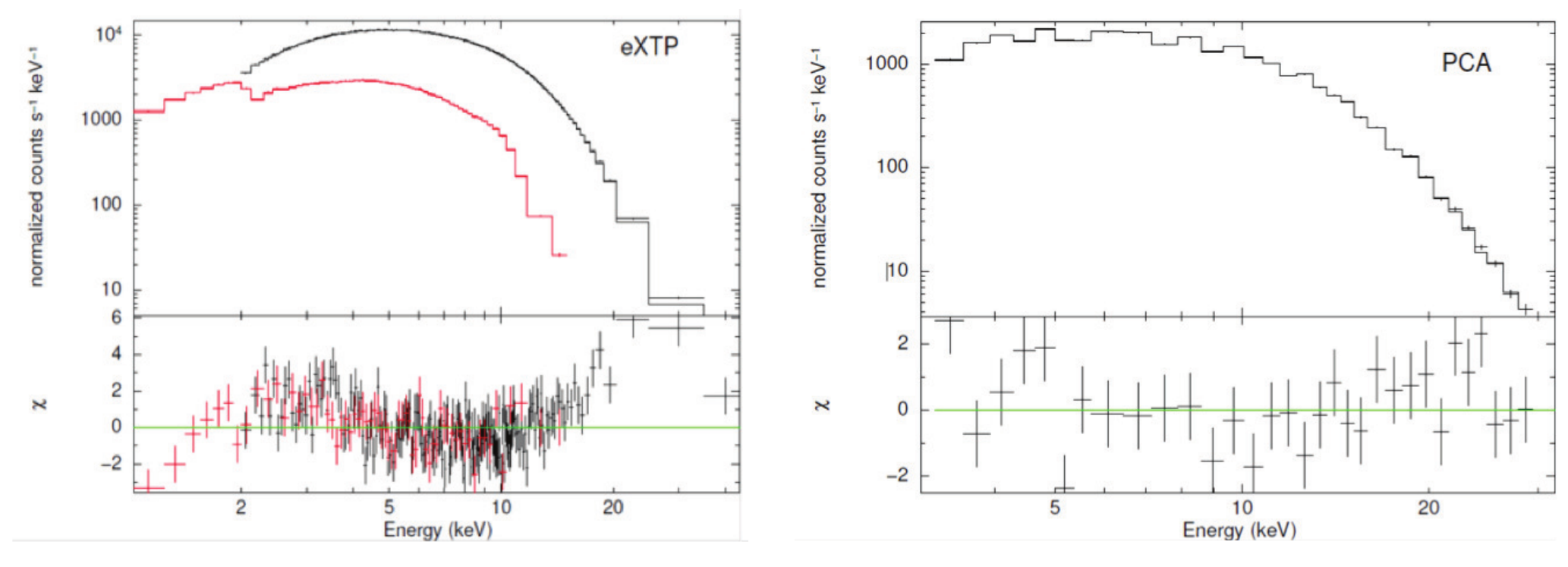}
}
\caption{Simulation of the X-ray burst influence upon the persistent emission as possible with \extp\ (left panel; SFA in red and LAD in black) and \rxte/PCA (right panel). The bottom panels show the model residuals when the pre-burst spectrum is subtracted from the X-ray burst spectrum as background. As a template for these simulations we used an X-ray burst detected from GS 1826--238 \citep[][]{ji2015_gs1862}. The large residuals seen for the \extp\ show that changes in the persistent spectrum during an X-ray burst can be readily identified.}
\label{fig:gs1826_comparison}
\end{figure}

When the persistent spectrum is dominated by the corona, as was observed in GS~1826--238, the persistent spectrum is well described by a cutoff power law. As described in \citet{ji2015_gs1862}, in GS 1826--238 the noticeable effects of an X-ray burst on the persistent spectrum are: i) the temperature of the corona drops from $\approx 20$ to $13$~keV, and ii) the optical depth of the corona changes (see also Section~\ref{subsec:corona}). For our simulations, based on prototype \extp\ response files (version 4, released in 2015 October), we assumed a typical non-PRE (i.e. constant emitting area) burst with a duration of 20 s, a peak temperature of $kT_{\mathrm{bb}}=2.5$~keV, and a 2--10 keV peak flux of $F^{\mathrm{peak}}_{\mathrm{X}}=2.4\times10^{-8}~\flux$. Furthermore, we take the corona parameters during the X-ray burst as determined by \citet{ji2015_gs1862} as input. We then fit the synthetic data with the pre-burst spectrum for the persistent emission subtracted. The results are shown in Figure~\ref{fig:gs1826_comparison}, where we compare simulations for \extp\ (left) with simulations for \rxte\ (right). From the residuals of the spectral fits (bottom panels), it is clear that there are significant model deviations in the simulated \extp\ data of an X-ray burst, while these are only marginal in the \rxte\ data. To quantify this; the null hypothesis probability (i.e., probability that the deviation between data and the model is only due to the statistical errors) for the fits shown in Figure~\ref{fig:gs1826_comparison} is $\approx 3\times 10^{-5}$ for the \extp\ data and $\approx$0.6 for the \rxte\ data. Changes in the persistent X-ray spectrum during a single X-ray burst should thus be significantly detectable with \extp.

\begin{figure}
\centering
\mbox{
\includegraphics[width=1.0\textwidth]{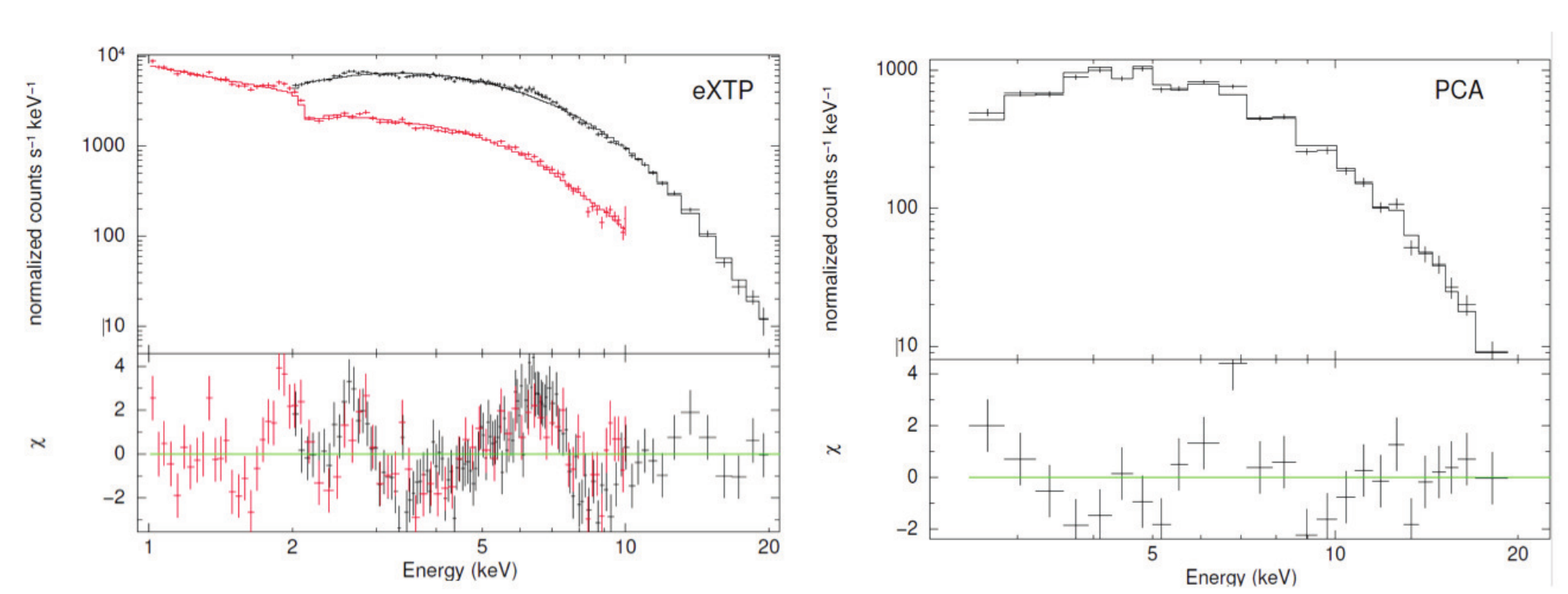}
}
\caption{Simulation of detecting disk reflection features during bursts with \extp\ (left panel; SFA in red and LAD in black) and \rxte\ (right panel) for an exposure time of 2~s. The bottom panels show the model residuals when reflection is not included in the fits. As a template for these simulations we used the superburst detected from 4U 1820--30 with \rxte\ \citep[][]{ballantyne2004}}
\label{fig:4u1820_comparison}
\end{figure}

The broad bandpass of 0.5--30 keV, good energy resolution, and large effective area of \extp\ also hold good prospects to detect disk reflection features \citep[][]{zhang2016_extp}. We therefore also explore the capabilities of detecting such features with \extp\ during X-ray bursts. There are only a few cases of reflection features during bursts observed by \rxte\ (and \swift), during long bursts of a few sources (see Section~\ref{subsec:specfeat}). So far, no disk reflection has been detected during a normal (i.e. short) X-ray bursts. This is likely due to the poor statistics of the current data which are limited by the small detection area of the telescopes. This difficulty will be largely overcome with \extp. Here we show simulations of how well the disk reflection during a single burst can be detected by \extp, and how this compares to \rxte. As a template, we take the spectrum and the reflection properties as observed by \rxte\ for the superburst of 4U 1820--30 \citep[][]{ballantyne2004}. We then consider an exposure of 2~s and generate a synthetic spectrum. We fit this synthetic spectrum with a model without a reflection component. As shown in Figure~\ref{fig:4u1820_comparison}, the residuals clearly demonstrate the requirement of an additional emission component in fitting the simulated \extp\ data, while this is much less clear for \rxte\ (mainly because the SFA have a better energy resolution, similar to CCD detectors).

\subsubsection{Missions concepts: {\it STROBE-X}}\label{subsubsec:loft}
The Large Observatory For x-ray Timing (\loft) was a mission concept proposed to ESA as M3 and M4 candidate in the framework of the Cosmic Vision 2015--2025 program \citep[][]{feroci2016}. Although it was highly ranked, it was eventually not selected. The \loft\ Coordination Team decided to concentrate its technical and programmatic efforts to further consolidate a coordinated European participation to \extp\ \citep[see Section~\ref{subsubsec:extp};][]{zhang2016_extp} and not to submit another \loft proposal to the ESA M5 call. 

The \loft\ Consortium also supports the study of the {\it STROBE-X} concept, which demonstrates the feasibility of a \loft-like mission within the context of the NASA Probe-class missions, potentially starting their development within the Decadal Survey 2020 context, for a launch in the late 2020s \citep[][]{wilsonhodge2016}. The {\it STROBE-X} concept involves a Large Area Detector (LAD) that consists of collimated arrays of silicon drift detectors, providing a 1$^{\circ}$ FOV and sensitivity in the 2--30 keV range. The peak effective area of $10$~m$^2$ near 10 keV is $>10$ larger than \rxte/PCA and optimized for fast X-ray timing. The {\it STROBE-X} concept also includes a Wide Field Monitor (WFM); a 2--50 keV coded mask imager that provides nearly continuous X-ray imaging over a 4 steradian FOV with a sensitivity $\gtrsim$20 times larger than that of the \rxte\ All-Sky Monitor (ASM). The concept has been enhanced relative to the original \loft\ concept by the addition of about 3m$^2$ of X-ray concentrator arrays, similar to those on NICER. This will allow better studies of soft X-ray continua and soft X-ray spectral lines than other timing missions.

Owing to the enormous effective area, the LAD holds great prospects for X-ray burst studies. Based on the original \loft\ design, \citet{peille2014} illustrated its value for detecting changes in kHz QPOs during bursts, whereas \citet{keek2016_future} demonstrated its great ability for detecting and studying reflection features during X-ray bursts. Furthermore, a \loft-like mission would enable the detection of hard X-ray deficits during single bursts rather than stacked bursts \citep{zand2015_loft}. The value of a \loft-like mission for X-ray burst studies in general is detailed in \citet{zand2015_loft}. In particular, the large sky-coverage of the WFM would allow the detection of thousands of bursts (including tens of rare superbursts). Furthermore, the enormous effective area of the LAD would allow the collection of an unprecedented amount of photons (free from pile-up issues) for each burst, thereby facilitating studies of all the different burst/accretion flow interactions reviewed in this work (see Table~\ref{tab:prospects}).

\subsection{New observational approaches}\label{subsec:studies}

\subsubsection{Coordinated X-ray/radio observations}\label{subsec:xradio}
Going forward, one of the things worth pursuing is to determine how radio jets respond to the input of extra photons from X-ray bursts (see Section~\ref{subsec:jet}). In the scenario that the corona is blown apart by radiation pressure from the X-ray bursts, the mass supply for the jet would disappear with it. As a result, the jet should turn off in the radio shortly after the X-ray burst. The time delay for the start of the radio shut off at different radio frequencies could then be used to map out the velocity structure of the jet.  In the scenario where the disk merely collapses due to Compton cooling, it seems likely that what happens will vary from source to source, given that some NS LMXBs show jet quenching in soft states, while others do not \citep[][]{migliari2010}.

Studying the response of the jet to an X-ray burst is a challenging task; multiple burst need to be stacked and captured simultaneously in the radio and X-ray band, but these events typically recur on timescales of several hours and are not fully predictable. The clocked burster GS 1826--238 is an exception; it exhibits very regular and predictable bursting behavior. This source therefore seems the ideal target to search for the effect of X-ray bursts on the radio jet. Its X-ray bursts recur on a timescale of $\approx 5$~hr, last for $\approx 150$~s, and can be quite well predicted within a $\approx$30~min time window. Furthermore, GS 1826--238 is firmly detected in the radio band, with a flux density of $\approx 70~\mu$Jy at 8 GHz (van den Eijnden et al. in prep.). Accumulating 10 bursts (i.e. $\approx$1500~s of burst emission) would require 10 observing blocks of $\approx$30 min, i.e. 5~hr of radio observations in total. With such an exposure time, the {\it VLA} can go down to radio flux densities of $\approx 5~\mu$Jy at 10 GHz, which implies that a 50\% reduction in the radio flux during an X-ray burst could be detected at high ($\approx 7 \sigma$) significance. 

Looking into the future, the Square Kilometer Array (\textit{SKA}) will be completed soon. This will provide highly increased sensitivity in the radio band and the ability to scan the sky much more rapidly than ever before. Among its many science objectives, the SKA is going to be of great value for studying jets in LMXBs \citep[e.g.][]{fender2004}, including burst/jet interaction studies such as outlined above.

\subsubsection{Optical searches for X-ray bursts}\label{subsec:xopt}
It has long been known that X-ray bursts produce optical emission through heating the outer accretion disk and the donor star, which then reprocess the emission into the optical band \citep[e.g.][]{grindlay1978,hackwell1979,mcclintock1979,lawrence1983,matsuoka1984,kong2000,hynes2006,mescheryakov2011}. Typically, the optical emission brightens by $\approx 1-2$ magnitudes during an X-ray burst. The optical emission may well become one of our best tools to search for X-ray bursts and develop statistics in the near future. At the present time, {\it Evryscope}\footnote{http://evryscope.astro.unc.edu} has coverage of the entire Southern sky with two minute cadence. Its sensitivity limit is about $m_V = 18.2$ for a one hour integration, and about 1.8 magnitudes worse for a 2 minute integration \citep[][]{law2015}.  Given the relation between the orbital period, X-ray luminosity and optical brightness \citep[][]{vanparadijs1994}, and taking a 10 hour orbital period as typical, bursters should be detectable to distances of more than 10 kpc except in highly reddened parts of the Galaxy.  Superbursts, in particular, should be easily identifiable. Such optical systems are relatively inexpensive, and are of interest to a wide range of projects (planet transits, searches for gravitational wave source counterparts, etc), so it may be that we soon will have nearly continuous coverage of nearly the entire sky that will detect X-ray bursts out to quite substantial distances. For instance, the Large Synoptic Survey Telescope (\textit{LSST}) is expected to come online in 2019 and will continuously monitor 20\,000 square degrees of the sky at various observing cadences, hence it will be sensitive to transient events on timescales ranging from seconds to hours \citep[e.g.][]{ivezic2008}. The \textit{LSST} has a large FOV of nearly 10 square degrees and will provide color information in 6 bands with a typical sensitivity of $\sim$24.5~mag in 2x15-s long integrations.

\subsubsection{Developing analysis techniques}\label{subsec:techniques}
The wide application of the ``variable persistent emission' approach clearly revealed how a new analysis technique can yield valuable information about the interaction between X-ray bursts and the accretion flow (see Section~\ref{subsubsec:enhanced}). However, an X-ray burst can have an effect on the disk as well as the corona, and both the flux and the spectrum of these emission components may change. Therefore, this simple method does not capture the complexity of the X-ray burst/accretion flow interaction and more elaborate methods  need to be devised. The prospects of new and concept missions that have good sensitivity at energies $>$30 keV is key in the effort to determine the effect of X-ray bursts on the coronal emission \citep[see e.g.][]{kajava2017_hardX}. 
Even with these new missions, however, it remains a difficult task to disentangle the X-ray burst flux from the different emission components of the accretion flow. In particular during soft states there are three emission components with a more of less similar shape radiating in a similar energy band: i) the X-ray burst emission itself that approaches a black-body spectral shape with a temperature of $\approx$1--3~keV, ii) the boundary/spreading layer that has a Planckian spectral shape with a color temperature of $\approx$2.5~keV \citep[][]{suleimanov2006}, and iii) the accretion disk that emits as a multi-color black body with a temperature of $\approx$1~keV. There is thus a high level of spectral degeneracy.

A promising avenue to disentangle the X-ray burst and accretion spectrum is to apply mathematical methods such as the NMF (see Section~\ref{subsec:change}). Similar techniques were already successfully applied to break spectral degeneracies in the accretion emission of Active Galactic Nuclei (AGN) and LMXBs \citep[e.g.][]{mittaz1990,vaughan2004,malzac2006,koljonen2013,parker2015}. Three recent studies applied the NMF method to analyze X-ray bursts, finding hints for coronal cooling during a hard state burst \citep[][]{degenaar2015_burst}, and a changing boundary/spreading layer during soft state bursts \citep[][]{koljonen2016,kajava2017_4u1608}. These three initial studies demonstrate the applicability and potential diagnostic power of this technique to further our understanding of how bursts affect their immediate environment. It would be worth to apply NMF-like techniques to more different sources and burst populations. Employing such techniques more broadly, for instance to series of stacked bursts from the rich \rxte\ archive (e.g. from GS 1826--238) or to intermediate-duration bursts, may prove to be a valuable exercise to disentangle the X-ray burst and accretion emission, and to map out how the latter is varying.

\subsubsection{Promising targets for future studies}\label{subsec:targets}
The various observational studies reviewed in this work demonstrate the potential of X-ray bursts as a tool to further our understanding of the accretion flow properties of NS LMXBs. Below we briefly touch upon specific targets and projects that may further progress this field of research.

\begin{itemize}

\item {\it Bursters with radio counterparts:} As discussed in Section~\ref{subsec:xradio}, searching for changes in the radio emission during an X-ray burst can provide interesting new information about the response of the jet and the corona to a bright soft X-ray shower. Although GS 1826--238 is by far the best candidate (when it is in its hard state showing its regular bursting pattern\footnote{We note that GS 1826--238  transitioned to a soft X-ray state in 2016, which affected its regular bursting pattern \citep[][]{chenevez2016}. After a few months it returned to the hard X-ray state that it normally resides in \citep[][]{chenevez2016}, but recent reports suggests that it is again in a (prolonged) soft state \citep[][]{palmer2017}.}; see Section~\ref{subsec:xradio}), there are several other (persistently accreting) bursters with sufficiently bright radio emission (e.g. GX 17+2, GX 13+1, Cyg X-2, Ser X-1, 4U 1820--30, 4U 1728--34) that could serve as targets if a pilot study of GS 1826--238 would turn out to be successful (see also Figure~\ref{fig:lxlr}).

\item {\it Bursters with optical counterparts:} Many X-ray burst sources have optical counterparts identified. Studying the reprocessed optical emission in tandem with the X-ray burst emission itself can provide valuable information about the accretion flow. In particular, monitoring for color variations in the optical while simultaneously searching for column density changes in the X-ray band could give important information on the amount and evolution of any accretion disk material expelled by the X-ray burst. Instruments with multi-wavelength capabilities such as \swift, but in particular \astrosat\ (as opposed to \swift\ its optical/UV data is time-tagged and simultaneously obtained in several filters), can facilitate such studies (see Section~\ref{subsec:astrosat}). Due to their long duration, superbursts might provide the best conditions for a detailed study; superbursts have been observed from 15 different LMXBs \citep[][]{zand2017_superbursts} and these would therefore be promising targets for this study. However, superbursts are rare and the chances of catching one simultaneous at X-ray and optical wavelengths is not high. Intermediate-duration bursts are less rare than superbursts, and could potentially also be employed for this study; about two dozen of LMXBs have shown such long bursts.

\item {\it Z-sources:} Among the population of NS LMXBs, the sub-group of Z-sources accrete near the Eddington limit. Two of these persistent Z-sources are X-ray bursters: Cyg X-2 and GX 17+2. These are interesting targets to study if/how an Eddington accretion flow is affected by the brief energy injection of an X-ray burst. Of the two, Cyg X-2 is frequently bursting, but its X-ray bursts are relatively short \citep[tens of seconds long; e.g.][]{kuulkers1995,smale1998,titarchuk2002}. GX 17+2 is also a frequent burster and may be a more suitable target as it also exhibits longer bursts \citep[up to half an hour; e.g.][]{tawara1984,sztajno1986,kuulkers1997,kuulkers2002}. Particularly interesting is that the Z-sources also show X-ray flares, which possibly have a different origin than the X-ray bursts (accretion flow origin versus thermonuclear burning on the NS surface). It would be interesting to study the response of the accretion flow to X-ray bursts and flares, to look for similarities and differences.

\item {\it Very low $\dot{M}$ bursters:} At the other extreme end, we have a sub-population of NS LMXBs that accrete (transiently or persistently) at very low rates of $\sim 10^{-5}-10^{-3}$ of the Eddington limit. Historically, several objects were discovered through the detection of their X-ray bursts whereas the accretion emission could not be detected with older instruments \citep[e.g.][]{zand1999_burstonly,cocchi2001,cornelisse2002}. Therefore, these very low $\dot{M}$ bursters are also referred to as ``burst only sources'', although with the current generation of sensitive X-ray instruments their accretion emission is detected and studied \citep[e.g.][]{zand05,zand07,zand2008,zand09_J1718,delsanto2007,delsanto2010,degenaar2010_burst,degenaar2017_igrj1706,armas2013,vandeneijnden2017_igr}. Nevertheless, their faint accretion emission is almost negligible compared to the X-ray burst peak emission and hence a clear view of the X-ray burst can be obtained. Vice versa, studying the X-ray bursts of these objects can potentially shed more light on the properties of the accretion flow: are the disks in these very low $\dot{M}$ bursters truncated and/or are these objects deprived of a corona? Several (intermediate-duration) bursts from these low $\dot{M}$ bursters were detected with \beppo\ \citep[e.g.][]{zand1999_burstonly,zand05,cocchi2001,cornelisse2002}, \rxte\ \citep[e.g.][]{galloway2008}, \inte\ \citep[e.g.][]{chelovekov07,delsanto2007,chenevez2008} and \swift\ \citep[e.g.][]{degenaar2010_burst,degenaar2011_burst,degenaar2013_igrj1706,degenaar2014_xmmsource,bozzo2015,keek2016}. However, a detailed analysis of the X-ray burst/accretion flow interaction has not yet been performed for these objects except for two intermediate-duration bursts of one source that shows discrete spectral features \citep[][see also Section~\ref{subsec:specfeat}]{degenaar2013_igrj1706,keek2016}. Unfortunately, due to the low accretion rate, it takes a long time to accumulate sufficient fuel to ignite an X-ray burst, so many of the low $\dot{M}$ bursters have only single burst detections (i.e. stacking studies are not possible).

\item {\it Intermediate-duration bursts:} These bursts that last tens of minutes to hours are often detected from low $\dot{M}$ bursters, the X-ray bursts are bright, and are less rare than superbursts. The intermediate-duration bursts are pure He bursts, so the ignition happens fast, and the response of the disk and/or corona would likely be immediate. During the long tails we should be able to see the effect of the corona and/or accretion disk to settle back again. For instance, if the X-ray burst decays are not smooth (i.e. deviate from pure power-law like), there might be this effect. It is therefore certainly worthwhile to take a more detailed look at the changing accretion emission during intermediate-duration bursts.

\item {\it Dippers:} There are several bursters that are viewed at high inclination and display X-ray dips \citep[][]{white1985}, which are thought to be due to partial obscuration of the NS (and inner accretion flow) by a thickened region of the accretion disk \citep[e.g.][]{diaztrigo2006}. Examples of bursting X-ray dippers are EXO 0748--676, MXB 1659--298, AX J1745.6--2901, Swift J1749.4--2807, GRS 1747--312, GX 13+1, 4U 1915--05, 4U 1254--69, 4U 1746--37 (in NGC 6441), GRS 1747--312 (in Terzan 6), 4U 1323--62 and XTE J1710--281 \citep[see e.g.][for a recent overview]{galloway2016}. Dipping sources can be exploited to probe the basic parameters of the corona. For instance, timing the ingress of dips allows for a direct measurement of the radial extent of the corona and studying the broad-band spectrum of dippers can accurately constrain the high-energy cutoff of the Comptonized emission, hence the electron temperature of the corona. On occasion, bursts have been detected during dips \citep[e.g.][]{gottwald1986,wijnands2002,boirin2007,hyodo2009}. An in-depth study of bursts occuring during dips could provide another interesting avenue to learn more about the accretion flow properties of NS LMXBs.

\end{itemize}

\section{Summary and conclusions}\label{sec:summary}
Ever since the discovery of X-ray bursts in the 1970s, there have been hints that the intense radiation of thermonuclear flashes impacts the surrounding accretion flow. However, in the past $\approx$5~yr this research has taken a significant leap by studying these interactions systematically and in the best possible detail, using good-quality data. In this work we have reviewed the observational evidence for the effect of X-ray bursts on both the accretion disk and the corona. Particularly strong observational evidence for such interactions is provided by: i) variations in the (soft) persistent flux and spectrum during an X-ray burst, ii) a (state-dependent) reduction of the hard ($>$30 keV) X-ray flux during burst peaks, iii) the detection of (evolving) discrete spectral features during very energetic bursts and iv) changes in kHz QPOs induced by bursts. 

These studies are very valuable because X-ray bursts represent short, highly repeatably instances in which the accretion disk and corona (and perhaps also the jet) are disrupted and restored. This provides an exciting opportunity to shed new light on the properties of accretion disks and coronae in NS LMXBs, which is expected to have strong links to those in stellar-mass and supermassive BHs. We have attempted to explain some of the observational phenomena in a basic theoretical framework, although it is clear that the X-ray burst/accretion flow interactions are complex and much work remains to be done. Indeed, this research area is only just emerging.

In order to grasp the full complexity of the X-ray burst/accretion flow interaction, we need to further develop our analysis techniques. Continued study and obtaining new, high-quality data are instrumental to achieve this. Fortunately, there are several recently launched and concept X-ray missions that can be exploited for the kind of studies reviewed in this work. In particular, the recently launched Indian satellite \astrosat\ should be able to catch X-ray bursts simultaneously over a broad X-ray energy range as well as in several different optical/UV filters. Furthermore, the recently installed NASA mission \nicer\ is particularly well suited to study (evolving) reflection features during bursts. In addition, the recently launched Chinese mission \hxmt\ will be a powerful tool to study hard X-ray deficits during X-ray bursts in greater detail than has been possible so far. Furthermore, the concept mission \extp\ will allow for detailed study of both burst-disk and burst-corona interactions via the detection of reflection features and changes in the hard X-ray emission. Finally, similar to \extp, the design of {\it STROBE-X} is ideally suited to collect a huge number of X-ray bursts and carry out all the different types of studies reviewed in this work. 

We conclude that X-ray bursts are a useful tool to study different aspects of the accretion flow in NS LMXBs. There is still a large data archive that can be mined for such studies (e.g. \rxte, \inte, \nustar, \swift), analysis techniques are evolving, and current and new X-ray missions are expected to provide valuable data for this type of research. We therefore expect that studying burst/accretion flow interactions will further advance our understanding of accretion disks and coronae over the next decade.

\begin{acknowledgements}
We thank the International Space Sciences Institute in Beijing for hosting the meetings that initiated the writing of this review. We are most grateful to Jean in 't Zand, Wenfei Yu, and Philippe Peille for kindly providing plots for their publications for use in this review, and to Duncan Galloway, Jean in 't Zand and Valery Suleimanov for providing valuable comments on a draft version of the manuscript. We thank James Miller-Jones and Andrzej Zdziarski for useful discussions. We thank the anonymous referee for constructive comments that helped improve this manuscript. ND is supported by a Vidi grant from the Netherlands Organization for Scientific Research (NWO) and a Marie Curie Intra-European fellowship from the European Commission under contract no. FP-PEOPLE-2013-IEF-627148. TMB acknowledges financial contribution from the agreement ASI-INAF I/037/12/0. JM acknowledges financial support from the French National Research Agency (CHAOS project ANR-12- BS05-0009). ZSN, ZS and CYP thank support from XTP project XDA 04060604, the Strategic Priority Research Program ``The Emergence of Cosmological Structures'' of the Chinese Academy of Sciences, grant no XDB09000000, the National Key Research and Development Program of China (2016YFA0400800) and the Chinese NSFC 11473027, 11733009.
\end{acknowledgements}

\end{document}